\documentclass{aastex631}

\usepackage{natbib}
\usepackage{graphicx}
\usepackage{hyperref}
\usepackage{amsmath}
\usepackage{lipsum}
\usepackage{grffile}
\usepackage{hyphenat}
\usepackage{longtable}

\begin{document}

\title{Dartmouth Stellar Evolution Emulator (DSEE) 1: Generative Stellar Evolution Model Database}

\author{Jiaqi (Martin) Ying}
\affiliation{Department of Physics and Astronomy, Dartmouth College, 6127 Wilder Laboratory, Hanover, NH 03755, USA}

\author{Brian Chaboyer} 
\affiliation{Department of Physics and Astronomy, Dartmouth College, 6127 Wilder Laboratory, Hanover, NH 03755, USA}

\author{Phillip A. Cargile}
\affiliation{Center for Astrophysics $\vert$ Harvard \& Smithsonian, 60 Garden St, Cambridge, MA 02138, USA}

\author{Wenxin Du}
\affiliation{Department of Mathematics and Computer Science, College of the Holy Cross, Worcester, MA, 01610, USA}

\author{George Dufresne}
\affiliation{Department of Physics and Astronomy, Dartmouth College, 6127 Wilder Laboratory, Hanover, NH 03755, USA}

\begin{abstract}
    We present the Dartmouth Stellar Evolution Emulator (DSEE), a flow-based stellar evolution model emulator trained on a comprehensive database comprising over eight million evolutionary tracks that vary across twenty input-physics dimensions and span broad ranges in mass and composition. DSEE learns phase-conditioned stellar state snapshots, unifying track and isochrone construction as marginals of one generative model. It delivers continuous interpolation across high-dimensional physics, probabilistic predictions with calibrated credible intervals, and orders-of-magnitude speedups over direct modeling. Validation against current stellar evolution models shows high fidelity across the HR diagrams, while distributional tests recover the full distributions obtained from brute-force Monte Carlo sampling. To broaden impact, DSEE is integrated into the open-source CONF1DENCE package, enabling fast, end-to-end creation of stellar tracks and isochrones. CONF1DENCE includes the ability to make uncertainty-aware age determinations for clusters taking into account observational effects. CONF1DENCE  replaces bespoke, fixed-physics grids with a generative, physics-marginalized emulator, setting a practical new standard for stellar modeling and enabling survey-scale analyses with rigorous uncertainty.
\end{abstract}

\section{Introduction}

Stellar evolution models are a cornerstone of modern astrophysics, underpinning our understanding of how stars eveolve and informing measurements of fundamental quantities like stellar ages, masses, and distances. From exoplanet host star characterization to determining the ages of the oldest star clusters, theoretical stellar evolution tracks and isochrones serve as indispensable reference points. Over the decades, several major stellar evolution databases have been developed and widely adopted including Dartmouth Stellar Evolution Database (for example, DSEP, \citealp{dotterACSSurveyGalactic2007,dotterDartmouthStellarEvolution2008}), MESA Isochrones and Stellar Tracks (MIST, \citealp{dotterMESAISOCHRONESStelLAR2016,choiMesaIsochronesStellar2016}), Padova Trieste Stellar Evolution Code (PARSEC, \citealp{bressanPARSECStellarTracks2012,nguyenPARSECV20Stellar2022}), Bag of Stellar Tracks and Isochrones (BaSTI, \citealp{pietrinferniLargeStellarEvolution2004,pietrinferniUpdatedBaSTIStellar2024}), Yonsei–Yale isochrones (${\textup{Y}^2}$, \citealp{yiBetterAgeEstimates2001,demarqueY2IsochronesImproved2004}), etc. These databases provide model predictions that have been extensively used to interpret observations. 

However, they all share a fundamental limitation: the grid-based approach samples only a limited set of parameters and holds many other physics inputs fixed at standard values. Typically, only 3–4 dimensions (stellar mass, age, composition, etc.) are varied in public grids. At the same time, properties such as the mixing-length parameter, convective overshoot, diffusion efficiencies, and key reaction rates remain fixed. Interpolation within a fixed-physics grid can therefore understate real model uncertainty and bias derived quantities when those underlying physics choices matter. 
For instance, a number of detailed studies have demonstrated that the mixing length parameter needed to match observations can vary (e.g.\ \citealp{Tayar2017,joyceClassicallyAsteroseismicallyConstrained2018a,joyceNotAllStars2018,vianiInvestigatingMetallicityMixinglengthRelation2018, Stokholm2019,LiY2024})
and is not fixed at a single value as it is currently done in existing stellar evolution databases. 
In an era of increasingly precise and numerous stellar observations, this poses a serious challenge: traditional static grids cannot easily expand to cover the entire high-dimensional space of stellar physics without becoming impractically large or sparse.

Our recent work addressed this by explicitly marginalizing over uncertain physics. By varying over $20$ stellar evolution parameters and constructing $10,000$ sets of isochrones for each of the 10 Milky Way globular clusters, we show that uncertainties in stellar evolution parameters such as metallicity, $\alpha$-enhancement, mixing length parameter, and atomic diffusion have a significant contribution to the uncertainty in age estimation \citep{yingAbsoluteAgeM922023,yingAbsoluteAgeNGC2024,yingAbsoluteAgeMilky2025}. The cost, however, was steep, as this approach required evolving over half a million stellar evolution models per globular cluster to explore the high-dimensional parameter space fully. This fundamental limit prohibits scalability and makes cluster-specific analyses hard to generalize to the broader survey landscape. 

Advances in data-driven modeling offer a path forward. Machine-learning emulators can learn from precomputed tracks to act as rapid, differentiable interpolators. Early studies used neural networks and ensemble regressors to infer stellar parameters directly from observables \citep[e.g.][]{vermaAsteroseismicDeterminationFundamental2016, bellingerFundamentalParametersMainSequence2016}. Subsequent work introduced dedicated emulators that approximate grid interpolation \citep[e.g.][]{honAsteroseismicInferenceSubgiant2020, garraffoStelNetHierarchicalNeural2021, honFlowBasedGenerativeEmulation2024}. Hierarchical approaches have further treated initial helium abundance ($Y$) and convective mixing length parameter $\alpha_{\textup{MLT}}$ as population-level random variables, replacing repeated calls to evolution codes with learned surrogates and yielding precise posteriors \citep{lyttleHierarchicallyModellingKepler2021}. These efforts demonstrate that well-trained networks can capture the information content of high-dimensional grids and return uncertainty-aware inferences at a fraction of the computational cost.

In this paper, we present the Dartmouth Stellar Evolution Emulator (DSEE), a normalizing–flow–based emulator trained on a multi-million-model DSEP database spanning broad ranges in mass, composition, and input physics. By learning a bijective map from a simple latent distribution to the space of observable stellar states, DSEE provides continuous interpolation across a high-dimensional physics space,  probabilistic predictions that naturally encode model-physics uncertainty, and orders-of-magnitude speedups relative to direct simulation. DSEE learns the correlations between stellar evolution parameters representing the initial stellar evolution model, and stellar state parameters at different evolutionary phases as a function of stellar age, producing tracks and isochrones on demand with DSEP-level accuracy and credible intervals that capture physics variability. Within the training domain, DSEE interpolates smoothly and with high precision. Once near the edges, it degrades gracefully, offering modest, well-behaved extrapolation accompanied by wider uncertainty bands that signal limited support. In practice, this turns what once required bespoke, large grids into a fast, uncertainty-aware workflow suitable for large surveys and interactive analyses.

This is the first of a series of papers on the Dartmouth Stellar Evolution Emulator. The primary focus of this paper is on the methodology behind DSEE and an in-depth validation of its performance. In \S \ref{sec:uncertainty}, we describe the creation of the Monte Carlo stellar model training dataset, detailing the $20+$ dimensional parameter space and how DSEP was used to generate millions of evolutionary tracks. In \S \ref{sec:Architecture Overview}, we outline the architecture of our normalizing flow emulator and the training procedure adopted to optimize it. In \S \ref{sec:validation}, we rigorously validate DSEE’s outputs against conventional stellar models: we compare emulator-generated tracks to actual DSEP calculations and to other standard model sets (e.g., DSED), and we quantify the interpolation errors across the HR diagram. In \S \ref{sec:confidence}, we integrate DSEE into the open-source CONF1DENCE package to enhance functionality and broaden accessibility to the community. Finally, in \S \ref{sec:conclusion}, we provide our conclusions and a brief outlook. Subsequent papers will build on the foundation laid here, employing DSEE to tackle astrophysical problems (such as globular cluster analyses and probabilistic isochrone fitting) that were previously intractable with static grids. Through DSEE, we aim to enable the community to fully exploit the flood of high-quality stellar data, marrying it with equally sophisticated stellar models in a fast, flexible, and physically comprehensive framework.

\section{Uncertainty in Stellar Evolution Parameters}\label{sec:uncertainty}

The interplay between theory and observation in astronomy has been very successful in the field of stellar evolution \citep{eggenObservationalAspectsStellar1965}. However, the uncertainty introduced by the complexity of stellar evolution models is often overlooked. Although modern stellar evolution models were all built from first principles and solved for stellar evolution structures, the choice of numerical techniques and input physics, such as opacity and the treatment of convection, can lead to different results. Most models are tuned to match the Sun's properties to ensure consistency at solar conditions, yet stars that are unlike the Sun can yield discrepant results across codes. It is therefore crucial to quantify how variations in these assumed physics affect stellar evolution outcomes and to incorporate those uncertainties into our analyses.

In this section, we will explore the different sources of uncertainties in the Dartmouth Stellar Evolution Program (DSEP), summarized in Table \ref{tab:DSEE_MC}.

\begin{table}[!htbp]
\caption{Dartmouth Stellar Evolution Emulator training set Monte Carlo input parameters \label{tab:DSEE_MC} }
\begin{tabular}{rllll}
\hline
Index &
Variable                               & Distribution & Range        & Source                                               \\
\hline
1 &Mass & Stepwise-Uniform & $0.1 \sim 4.0$ &  N/A  \\
2 & {[}Fe/H{]} & Stepwise-Uniform  & $-4.0 \sim 0.5$ &  N/A  \\
3 & {[}$\alpha$/Fe{]}  & Stepwise-Uniform  & $-0.2 \sim 0.8$ &  N/A  \\
4 & $\Delta Y/\Delta Z$ & Uniform & $1.75 \sim 2.5$ & see text\\
5 & Helium abundance   & Uniform      & $0.2465(25) + \left(\Delta Y/\Delta Z\right) Z$   & \citet{averEffectsHeTextbackslashuplambda108302015}                                       \\
6 & Mixing length                 & Uniform      & $1.0 \sim 2.5$      & see text                                         \\
7 & Heavy element diffusion       & Uniform      & $0.5 \sim 1.3$      & \citet{thoulElementDiffusionSolar1994}                                         \\
8 & Helium diffusion              & Uniform      & $0.5 \sim 1.3$    & \citet{thoulElementDiffusionSolar1994}                                           \\
9 & Surface boundary condition &   Binary          &     1/2; 1/2             & \citet{eddingtonInternalConstitutionStars1926}\\
  &             &                     && \citet{hauschildtNextGenModelAtmosphere1999} \\
10 & Low temperature opacities     & Uniform      & $0.7 \sim 1.3$        & \citet{fergusonLowTemperatureOpacities2005}                                      \\
11 & High temperature opacities    & Normal       & $1.0 \pm 0.03$           & \citet{iglesiasUpdatedOpalOpacities1996}     \\
12 & Plasma neutrino loses         & Normal       & $1.0 \pm 0.05$     & \citet{haftStandardNonstandardPlasma1994}     \\
13 & Conductive opacities          & Uniform       & $0.9 \sim 1.1$       & 
\citet{blouinNewConductiveOpacities2020}   \\
    &             &                     && \citet{cassisiElectronConductionOpacities2021}   \\
14 & Convective envelope overshoot & Uniform      & $0 \sim 0.2$      & see text                                          \\
15 & Convective core overshoot & Uniform & $0 \sim 0.2$ & see text \\
16 & $p + p \to H_2 + e + \nu$   
& Normal       & $\left(4.07 \pm 0.04 \right)\times 10^{-22}$ & \citet{acharyaUncertaintyQuantificationProtonproton2016}\\
 & & && \citet{marcucciProtonProtonWeakCapture2013}\\
17 &${ }^{3}\mathrm{He} + { }^{3}\mathrm{He} \to { }^{4}\mathrm{He} + \mathrm{p} + \mathrm{p}$                 & Normal       & $5150 \pm 500$& \citet{adelbergerSolarFusionCross2011}\\
18 & ${ }^{3}\mathrm{He} + { }^{4}\mathrm{He} \to { }^{7}\mathrm{Be} + \gamma$                  & Normal       & $0.54 \pm 0.03$&\citet{deboerMonteCarloUncertainty2014}\\
19 & Triple-$\alpha$ & Normal & $1.0 \pm 0.15$ & \citet{sunoPreciseCalculationTripleAlpha2016}\\
20 & ${ }^{12}\mathrm{C} + \mathrm{p} \to { }^{13}\mathrm{N} + \gamma$                & Normal       & $1.45 \pm 0.50$ & \citet{xuNACREIIUpdate2013}\\
21 & ${ }^{12} \mathrm{C} + { }^{4}\mathrm{He} \to  { }^{16} \mathrm{O} + \gamma$& Normal & $1.0 \pm 0.15$ & \citet{deboer12Ca16OReaction2017} \\
22 & ${ }^{13}\mathrm{C} + \mathrm{p} \to { }^{14}\mathrm{N} + \gamma$              & Normal       & $5.50 \pm 1.20$& \citet{chakrabortySystematicRmatrixAnalysis2015}\\
23 & ${ }^{14}\mathrm{N} + \mathrm{p} \to { }^{15}\mathrm{O} + \gamma$             & Normal       & $3.32 \pm 0.11$ & \citet{martaN14pensuremathgammaO15ReactionStudied2011}\\
24 & ${ }^{16}\mathrm{N} + p \to { }^{17}\mathrm{F} + \gamma$               & Normal       & $9.40 \pm 0.80$ & \citet{adelbergerSolarFusionCross2011}\\
\hline
\end{tabular}
\end{table}

\subsection{Abundance}
The primordial helium abundance $Y_{\textup{prim}}$ is the mass fraction of helium produced in the first few minutes after the Big Bang, during the Big Bang Nucleosynthesis (BBN). Thus, $Y_{\textup{prim}}$ is the initial helium abundance $Y$ for the first generation of stars. For stars formed later, the first generation of stars enriches the environment, and their initial helium abundance should be adjusted by their metallicity $\frac{\Delta Y}{\Delta Z}$ as:
\begin{equation}
    Y = Y_{\textup{prim}} + \frac{\Delta Y}{\Delta Z} \cdot Z.
\end{equation}
As stellar evolution codes require the initial chemical composition to construct the pre-main-sequence model, the accuracy of both $Y_{\textup{prim}}$ and $\frac{\Delta Y}{\Delta Z}$ is crucial. 

Historically, $Y_{\textup{prim}}$ is determined by observing helium and hydrogen recombination lines in metal-poor extragalactic HII regions (dwarf galaxies). 
Classic analyses used only visible helium lines (e.g., HeI 4471, 5876 \AA) and hydrogen lines, but these suffered from systematic uncertainties due to degeneracies in nebular temperature, density, and radiative transfer effects. 
\citet{peimbertPrimordialHeliumAbundance2016} use new, high-precision atomic physics calculations for helium recombination and find $Y_{\textup{prim}} = 0.2446 \pm 0.0029$, and \citet{averEffectsHeTextbackslashuplambda108302015} include the HeI $\lambda 10830$ infrared line in helium abundance analyses and find $Y_{\textup{prim}} = 0.2449 \pm 0.0040$. Combine with $Y_{\textup{prim}} = 0.2470 \pm 0.0002$ derived from the BBN model based on the Planck determination of the baryon density \citep{planckcollaborationPlanck2018Results2020}, we choose to vary $Y_{\textup{prim}}$ uniformly from  $0.2440$ to $0.2490$. 

There is a wide range of the linear enrichment coefficient $\frac{\Delta Y}{\Delta Z}$ reported in the literature. For example, \citet{Balser2006} found $\frac{\Delta Y}{\Delta Z} = 1.41\pm 0.62$ based upon the measured helium abundance in two Galactic H II regions.  \citet{Serenelli2010} estimated for the Sun that  $ 1.7 \le \frac{\Delta Y}{\Delta Z} \le 2.2$,  based upon the choice of the solar composition.  Fitting observations of nearby K dwarfs, \citet{casagrandeHeliumAbundanceDY2007} found $\frac{\Delta Y}{\Delta Z} = 2.1\pm 0.9$. \citet{Verma2019} analyzed the seismic glitch signatures
caused by the ionization of helium in 38 Galactic stars to find $\frac{\Delta Y}{\Delta Z} = 1.3\pm 0.8$. The Galactic chemical evolution models of \citet{Weller2025} find $ 1.4 \le \frac{\Delta Y}{\Delta Z} \le 1.8$  \citet{Tognelli2021} determined $\frac{\Delta Y}{\Delta Z} = 2.0\pm 0.83$ from fitting stellar models to the Hyades main sequence. 
As a result, we choose to vary $\frac{\Delta Y}{\Delta Z}$ uniformly from  $1.75$ to $2.50$. 

\subsection{Convection} \label{sec:convection}

Convection describes the bulk motion of materials driven by a steep temperature gradient within the star. When the temperature gradient becomes sufficiently large, hotter and less dense material rises towards the stellar surface, while cooler, denser material sinks inward, establishing a circulating flow. This process is analogous to water boiling in a pot, where heated water near the bottom rises and cooler water at the surface descends, creating a continuous convective circulation. In stars, convection is a crucial mechanism for energy transport, efficiently redistributing heat and influencing both stellar structure and evolution.

In reality, convection within stars is inherently a three-dimensional, turbulent process, involving complex fluid motions that transport energy and mix stellar material across convective boundaries. Ideally, to fully characterize convection and accurately track the motion of stellar material, three-dimensional (3D) hydrodynamic simulations would be employed. These simulations directly model convective flows, turbulence, and the detailed physics of convective boundaries, thereby providing physically realistic insights into convective transport and mixing processes.

However, 3D hydrodynamic simulations are computationally far too expensive to integrate directly into 1D stellar evolution codes, which require calculations spanning billions of years of stellar lifetimes. Even though pioneer studies have been conducted in the area of using 3D hydrodynamic simulations for certain stages of evolution \citep[e.g.][]{rizzuti3DStellarEvolution2023} or utilizing a pre-calculated 3D atmosphere model grid \citep[e.g.][]{zhouCoupling1DStellar2025}, none of the existing methods are sufficiently versatile to cover a wide range of parameters, and striking a balance between accuracy and efficiency remains a significant challenge. Consequently, stellar evolution models rely on simplified, parameterized descriptions of convection. The mixing length theory (MLT) and convective overshooting prescriptions are widely used approximations that reduce the inherently complex 3D convective phenomena to manageable, empirical parameters. MLT uses a simplified physical framework to represent the efficiency of convective energy transport through a single adjustable parameter—the mixing length parameter. Similarly, convective overshooting treatments introduce parameters to model how far convective motions extend beyond their classical boundaries.

\subsubsection{Mixing Length}\label{sec:MLT}
MLT is the classical framework for convective heat transport in stellar models. In this theory, convective eddies are assumed to travel a mean free path – the mixing length – before dissolving and releasing their heat. The mixing length $\ell$ is usually expressed as a multiple of the local pressure scale height $H_p$:
\begin{equation}
    \ell = \alpha_{\textup{MLT}} \cdot H_p, \label{eq:mlt}
\end{equation}
where $\alpha_{\textup{MLT}}$ is a dimensionless free parameter. 

The choice of $\alpha_{\textup{MLT}}$ has a significant influence on model predictions of effective temperature, radius, and, to a lessr extent the age for a given mass. A higher $\alpha_{\textup{MLT}}$ leads to more efficient convective energy transport, which tends to flatten the temperature gradient in convective envelopes, and leads to a higher temperature and a smaller radius \citep{joyceReviewMixingLength2023}. 

Since the mixing length theory is just a 1D convection model trying to address an intrinsic 3D hydrodynamical problem, one cannot determine $\alpha_{\textup{MLT}}$ from first principles.  Instead, it is calibrated using observations. For example, \citet{joyceNotAllStars2018} calibrates a solar model by adjusting the mixing length, initial helium abundance, and initial heavy element abundance (which affect the chemical evolution degenerately) until the model reproduces the observed solar radius, luminosity, and surface abundance at the solar age to better than $0.1\%$ accuracy. However, \citet{joyceNotAllStars2018} also suggests that the choice of mixing length varies a lot for non-solar types of stars. Other studies have reinforced this idea that a single mixing length does not apply to all stars (e.g.\ \citealp{joyceClassicallyAsteroseismicallyConstrained2018a,vianiInvestigatingMetallicityMixinglengthRelation2018, Stokholm2019,joyceReviewMixingLength2023,Joyce2023,LiY2024}).
As a result, we choose to vary $\alpha_{\textup{MLT}}$ uniformly from  $1.0$ to $2.50$. 

\subsubsection{Overshooting}\label{sec:overshoot}
MLT is used to describe convection within formally unstable regions (as given by the Schwarzschild or Ledoux criteria) of a star. Convective overshooting refers to the extension of convective motion and mixing into adjacent regions which are formally stable by the Schwarzschild or Ledoux criteria.  In real stars, convective eddies have inertia and can travel beyond the boundary of the convective zone, depositing heat and mixing material into the stable zone. DSEP accounts for this phenomenon with a parametric overshooting model. There are two distinct cases to consider: convective core overshoot (from a convective core into the overlying radiative zone) and convective envelope overshoot (from a convective envelope into the underlying radiative zone). These are treated somewhat differently and have different consequences for stellar evolution.

Various studies have calibrated the amount of convective overshoot during central hydrogen burning by comparing to observations \citep[e.g.][]{demarqueY2IsochronesImproved2004, claretNewGridsStellar2004, pietrinferniLargeStellarEvolution2004, mowlaviStellarMassAge2012,acharyaUncertaintyQuantificationProtonproton2016,Valle2017,Mombarg2019,Claret2019,Anders2023,Valle2025}. These studies have generally found a fairly small value of $0.0$ to $0.2$ pressure scale heights. As a result, we choose to vary both core convective overshoot and envelope convective overshoot uniformly from  $0.0$ to $0.2$, in all evolutionary phases.

\subsection{Surface Boundary Condition}\label{sec:SBC}
Solving the stellar structure equations not only requires the trivial boundary condition at the center of the star but also an appropriate surface boundary condition at the outermost layer of the stellar surface. The model atmosphere is usually formulated by providing the temperature $T$ as a function of optical depth $\tau$, and matches the interior solution with these conditions at the matching point. Here we introduce two the two model atmosphere used in our new work.  
We choose not to include the \citep{krishnaswamyProfilesStrongLines1966} model atmosphere, as it was the least favored model in our previous studies \citep{yingAbsoluteAgeM922023,yingAbsoluteAgeNGC2024,yingAbsoluteAgeMilky2025}, and is only calibrated to the Sun.

\subsubsection{Eddington Grey Atmosphere}\label{sec:eddington}
The Eddington grey atmosphere \citep{eddingtonInternalConstitutionStars1926} is the classical analytical model for stellar atmospheres. It assumes a grey atmosphere with opacity independent of wavelength and radiative equilibrium in a plane-parallel atmosphere. Under these assumptions and using Eddington’s approximation that specific intensity is constant over the upward and downward hemispheres, we get an atmosphere model that is in local thermodynamic equilibrium with temperature structure:
\begin{equation}
    T^4 (\tau) = \frac{3}{4} T^4_{\textup{eff}} \left( \tau + \frac{2}{3}\right). \label{eq:eddington}
\end{equation}
Because of its simplicity, the Eddington grey atmosphere model is widely used in stellar evolution codes as a baseline outer boundary condition. However, since real stellar atmospheres are not truly grey and have more complex temperature profiles in their outer layers, calibrations are required to correct for the oversimplification.

\subsubsection{PHOENIX Model Atmosphere}\label{sec:phoenix}
The PHOENIX atmosphere model \citep{hauschildtNextGenModelAtmosphere1999} is a detailed model atmosphere that can be used as the outer boundary condition for stellar models. PHOENIX assumes a 1D, hydrostatic equilibrium atmosphere in radiative-convective equilibrium and solves the full radiative transfer equations for a stellar atmosphere with thousands or millions of frequency points, incorporating extensive atomic and molecular line opacities (“line blanketing”), and including convection where relevant. It can produce atmospheric structures in either plane-parallel or spherical geometry. It also features very detailed physics: opacity databases with millions of lines, and can include NLTE (non-local thermodynamic equilibrium) treatment for certain species. The PHOENIX model atmospheres have a maximum effective temperature of $10^4\,$K.  Above this temperature, the \cite{Kurucz1993} model atmospheres are used.

\subsection{Diffusion}\label{sec:diffusion}
Atomic diffusion describes the process by which heavier elements settle faster in the stellar interior than relative to hydrogen. As a result, the composition of a star like the Sun changes gradually over its billions of years of its lifetime. The calculation of atomic diffusion involves solving for diffusion velocities resulting from concentration gradients, gravity, and thermal gradients for both helium and heavy elements. DSEP \citep{chaboyerHeavyElementDiffusionMetalpoor2001} uses the formulation described by \citet{thoulElementDiffusionSolar1994}:
\begin{equation}
    \frac{\partial X_s}{\partial t}=-\frac{1}{\rho r^2} \frac{\partial}{\partial r}\left[r^2 X_s T^{5 / 2} \xi_s(r)\right], \label{eq:thoul_diff}
\end{equation}
where the rate of change of element mass fraction $\frac{\partial X_s}{\partial t}$ due to diffusion is expressed in radius $r$, mass fraction $X_s$, temperature $T$ and diffusion coefficient $\xi_s(r)$. The diffusion coefficient is related to internal gradients of pressure $p$, temperature $T$, and concentration in stars $C$:
\begin{equation}
    \xi_s(r)=A_p(s) \frac{\partial \ln p}{\partial r}+A_T(s) \frac{\partial \ln T}{\partial r}+\sum_{t \neq e, 2} A_t(s) \frac{\partial \ln C_t}{\partial r}. \label{eq:diff_coeff}
\end{equation}
\citet{thoulElementDiffusionSolar1994} suggested that their expression for the diffusion velocity is expected to be accurate to $\sim 15\%$. 
Comparisons with observations of iron abundances in cluster stars suggest that the \citet{thoulElementDiffusionSolar1994} velocities may be an overestimate as it neglects turbulent mixing processes, which may also be occurring in stars  \citep[e.g.][]{chaboyerHeavyElementDiffusionMetalpoor2001, Semenova2020,Moedas2022}. As a result, we choose to multiply the default turbulent diffusion coefficient $\xi_s$ from \citet{thoulElementDiffusionSolar1994} with a coefficient that varies uniformly from  $0.5$ to $1.3$.   Due to its long timescale, diffusion is turned-off in our models at the end of central hydrogen fusion.

\subsection{Opacity}\label{sec:opal}
Opacity describes the ability of radiation to pass through a material. It plays a crucial role in stellar structure and evolution because it dictates how effectively energy produced in the stellar core can be transported outward. Therefore, accurate modeling of stellar opacities is fundamental for understanding and predicting the evolution of stars across their lifetimes. 

Since opacity strongly depends on temperature, density, and composition, the convention in the stellar evolution modeling community is to adapt the extensive opacity tables for diverse physical conditions. Different opacity regimes dominate in distinct regions and evolution stages of a star; therefore, a combination of opacity tables is utilized in stellar evolution models. For DSEP, we mainly use three sets of opacity tables: low-temperature radiative opacity tables \citep{fergusonLowTemperatureOpacities2005}, high-temperature radiative opacity tables \citep{iglesiasUpdatedOpalOpacities1996}, and conductive opacity tables \citep{hubbardThermalConductionElectrons1969, canutoElectricalConductivityConductive1970, blouinNewConductiveOpacities2020}.

More importantly, the choice of opacity tables and uncertainties in opacity measurement and interpolation must be considered when modeling stellar interiors.
\subsubsection{Low Temperature Opacity}\label{sec:low-temp_opal}
In the cool outer layers $(\log{T} \lesssim 4)$, molecules and even dust grains form and contribute significantly to the opacity. \citet{fergusonLowTemperatureOpacities2005} includes these molecular line and grain absorption effects, providing opacities for stellar atmospheres and envelopes; therefore, it is used as a low-temperature opacity table in DSEP. \citet{fergusonLowTemperatureOpacities2005} has a good agreement with other opacity tables at higher temperatures (e.g., OPAL \citep{iglesiasUpdatedOpalOpacities1996}, Opacity Project \citep{seatonOpacitiesStellarEnvelopes1994}, and AF94 \citep{jorgensenMoleculesStellarEnvironment1994,alexanderLowTemperatureRosselandOpacities1994}). Due to differences in molecular and grain physics included in different calculations, the uncertainty at extremely low temperatures ($\log{T} < 3.2$ when grains dominate the opacity) becomes large. As a result, we choose to multiply the low-temperature opacities from \citet{fergusonLowTemperatureOpacities2005} with a coefficient that varies uniformly from  $0.7$ to $1.3$, see \citet{chaboyerLowerLimitAge1996} for further details.

\subsubsection{High Temperature Opacity}\label{sec:high-temp_opal}
At higher temperatures in the stellar interior $(\log{T} \gtrsim 4)$, most atoms are ionized and opacity is governed by atomic processes like bound-free and free-free transitions and electron scattering. High-temperature tables such as OPAL opacities \citep{iglesiasUpdatedOpalOpacities1996} tabulate these radiative opacities for various compositions, covering conditions deep within stellar cores. The OPAL opacity tables agree with Opacity Project \citep{seatonOpacitiesStellarEnvelopes1994}, and AF94 \citep{jorgensenMoleculesStellarEnvironment1994,alexanderLowTemperatureRosselandOpacities1994} well in high temperatures. However, there are some experimental suggestions that the current opacity calculation are too low around $T \simeq 6.2\times 10^6\,$K \citep{Nagayama2019,Mayes2025} although other experiments have not confirmed these results \citep{Hoarty2023}.   Therefore, we choose to multiply the high-temperature opacities from \citet{iglesiasUpdatedOpalOpacities1996} with a coefficient that samples from a normal distribution with $\sigma = 0.03$, see \citet{Chaboyer2002} for further details

It is worth noting that the more up-to-date OPLIB high-temperature radiative opacities \citep{colganNEWGENERATIONALAMOS2016} offer a much finer and denser grid, which improves interpolation, particularly in low-metallicity and hydrogen-poor regimes. \citet{boudreauxUpdatedHightemperatureOpacities2023} found that OPLIB opacities are systematically lower than OPAL opacities for high temperatures ($\log{T} > 5$). These generally lower opacities will decrease the radiative temperature gradient throughout most of the model's radius. \citet{faragExpandedSetAlamos2024} found that the radiative opacities differ $20 \sim 80\%$ between OPLIB and OPAL in various temperature–density–composition regimes. Using different opacity tables will result in measurable changes in internal profiles: core density, core temperature, and the base of the convection zone, and produce significant structural differences, even for well-studied stars like the Sun. More importantly, models using OPLIB still require solar metallicities higher than the observed solar abundances \citep[AGSS09 or MB22][]{asplundChemicalCompositionSun2009,Magg2022,Buldgen2025} to better match helioseismic and neutrino constraints. As a result, we need to conduct more studies on the choice of high-temperature opacity tables on stellar evolution models, and we do not include OPLIB in this work.

\subsubsection{Conductive Opacity}\label{sec:condopac}

\begin{figure}
\centering
    \includegraphics[width=0.6\textwidth]{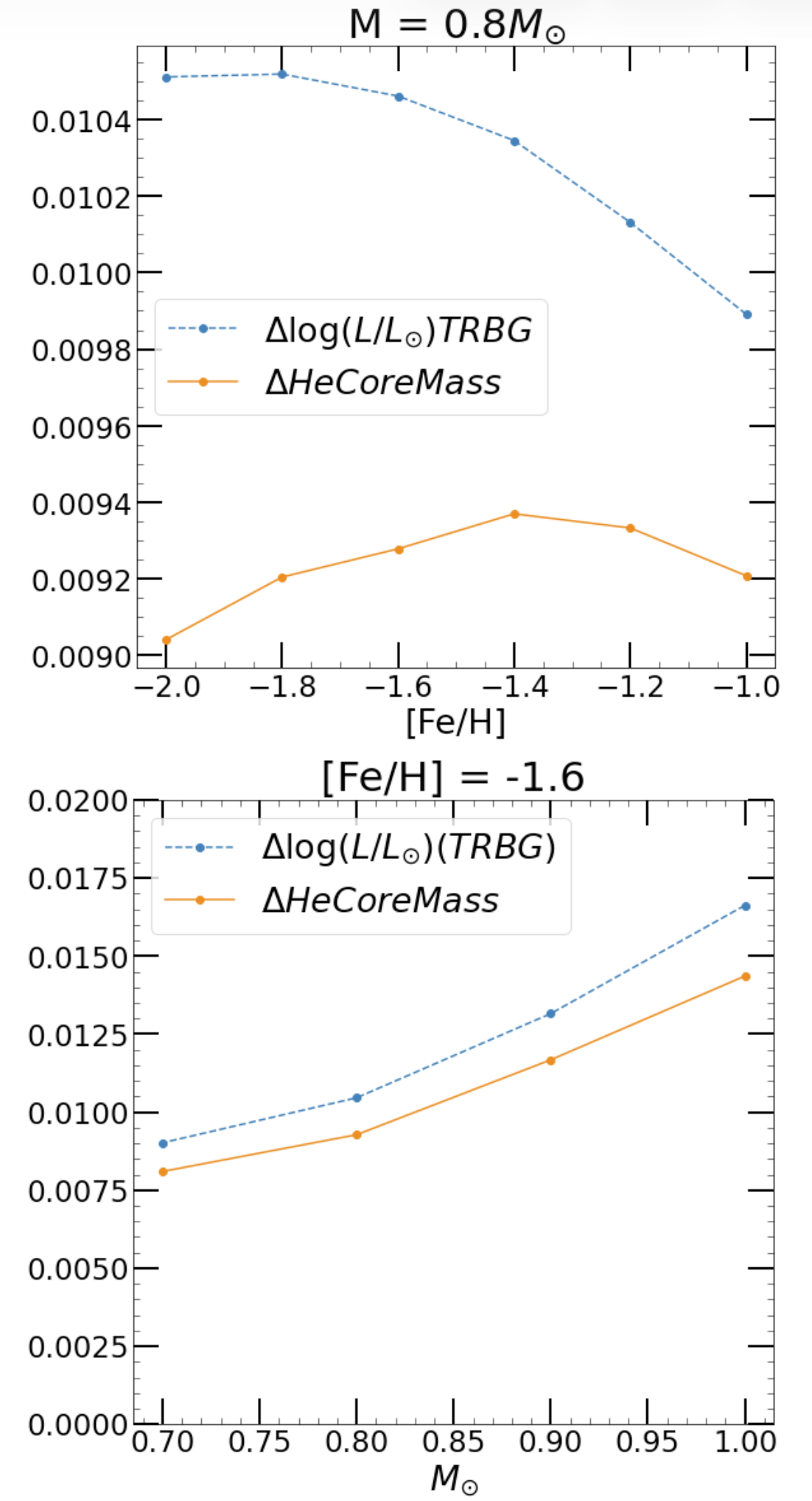}
\caption{\textit{Top}: Plot of change in luminosity at the tip of red-giant branch (TRGB) and the Helium core mass against initial metallicity with $M = 0.8\, M_{\odot}$ with minimal luminosity of the star required for the Helium flash to happen as a function of mass. \textit{Bottom}: Plot of change in luminosity at the tip of red-giant branch (TRGB) and the Helium core mass against initial mass with $[\mathrm{[Fe/H]} = -1.6$ with minimal luminosity of the star required for the Helium flash to happen as a function of metallicity.}
\label{fig:condopac_ldiff}
\end{figure}

Conductive opacities are important within stars in the high density, relatively cool temperature regime.  This occurs in the inert helium cores of lower mass ($M \la 1.5\,M_{\odot}$) red giant branch stars or in white dwarfs.  Uncertainties in the conductive opacities can impact the predicted properties of stars on the upper giant branch, including the tip of the red giant branch (TRGB). 
The conductive opacity previously adopted by DSEP \citep{chaboyerTestingMetalPoorStellar2017} was from \citet{hubbardThermalConductionElectrons1969} and \citet{canutoElectricalConductivityConductive1970}, with the latter one taking the relativistic effect at high density ($\log{\rho} > 6$) into consideration. The conductive opacity is updated by \citet{blouinNewConductiveOpacities2020}. The updated conductive opacities are evaluated from the mean force quantum Landau–Fokker–Planck (qLFP) plasma kinetic theory. In the weakly coupled limit, the effective Coulomb logarithms and opacities of qLFP are in good agreement with the analytic expressions in \citet{hubbardThermalConductionElectrons1969}. At higher Coulomb coupling, due to the use of mean-force scattering potentials, the qLFP calculations are more accurate than \citet{hubbardThermalConductionElectrons1969} \citep{blouinNewConductiveOpacities2020}.

The updated conductive opacities only have significant ($\ga 10\%$) differences   in the high temperature ($T\ga 10^7\,$K) and high density ($\rho \ga 10^4 \textrm{\,gm\,cm}^{-3}$) regimes.
The updated conductive opacity is implemented into DSEP through the interpolation program developed by \citet{cassisiElectronConductionOpacities2021}.  Multiple runs of DSEP with different initial parameters are performed. Figure  \ref{fig:condopac_ldiff} compares the old conductive opacity and the updated conductive opacity from two perspectives: change in luminosity  and the change in Helium core mass both at the TRGB. Figure \ref{fig:condopac_ldiff} shows that overall, the updated conductive opacity only has a minor effect on these two parameters.

\citet{cassisiElectronConductionOpacities2021} implement updated conductive opacity calculations from \citet{blouinNewConductiveOpacities2020}, which provide improved accuracy in regimes of moderate electron degeneracy $(\theta \equiv T/T_F \sim 1)$, where $T_F$ is the Fermi temperature. Because these opacities become less reliable in the strongly degenerate limit ($\theta \ll 1$), \citet{cassisiElectronConductionOpacities2021} introduces three damping prescriptions to transition smoothly between the \citet{blouinNewConductiveOpacities2020} values and \citet{cassisiUpdatedElectronConductionOpacities2007} opacities, which remain valid in the highly degenerate regime using the damping function:
\begin{equation}
    f(\theta) = \left[ 1 + (\theta/a)^b \right]^{-1},
\end{equation}
where the weak damping approach adopts a slow and gradual convergence, using parameters $(a = 0.01, b = 2)$, and the strong damping approach imposes a much faster suppression of the \citet{blouinNewConductiveOpacities2020} opacity using parameters $(a = 0.1, b = 3)$. Since \citet{cassisiElectronConductionOpacities2021} suggests that the true conductive opacity is between the weak-damping and strong-damping results, we chose to implement the updated conductive opacity in the following steps:
\begin{enumerate}
    \item Randomly pick a number $w$ between $-1$ and $+1$ from a uniform distribution.
    \item Calculate the conductive opacity with the weak damping, or strong damping correction, and then determine the difference.
    \item If the difference is $< 10\%$, which suggests a strongly degenerate case, and we calculate the conductive opacity by averaging the two opacities, and then multiply them by the uncertainty correction of $1 + 0.1 \cdot w$, where $0.1$ is the chosen conductive opacity uncertainty coefficient.
    \item If the difference is more than $10\%$, then we convert $w$ to a weighting function between $0$ and $1$, where $0$ would mean only use the weak damping, and $1$ would mean you are using the strong damping case. The conductive opacity is the weighted sum of opacities from two damping approaches.
\end{enumerate}
\subsection{Plasma Neutrino Energy Loss}\label{sec:neutrino}
Plasma neutrino emission is a key cooling mechanism in stellar interiors at high densities and temperatures. Unlike photon radiation, neutrinos can freely escape the stellar core, carrying away energy and cooling the star’s interior. In regions of moderate to high degeneracy and temperatures ($\log{T} \sim 7 - 8$), the plasma neutrino process becomes an important source of energy loss in stellar evolution models \citep{haftStandardNonstandardPlasma1994}. 

The contribution of plasma neutrino emission to stellar evolution has been studied for decades, and its inclusion in models is accompanied by some theoretical uncertainty. \citet{haftStandardNonstandardPlasma1994} revisited the standard plasma neutrino loss rates and highlighted the uncertainties in earlier calculations. They found that many widely used analytic approximations (from authors like \citet{braatenEmissivityHotPlasma1991} and \citet{itohNeutrinoEnergyLoss1992}) were not sufficiently accurate in the regimes where plasma neutrinos dominate (degenerate, semi-degenerate cores). \citet{haftStandardNonstandardPlasma1994} derived a new fitting formula for the neutrino emission by comparing against exact quantum electrodynamics plasma dispersion calculations \citep{braatenNeutrinoEnergyLoss1993}. This new formula is accurate to within $\sim 4\%$ across the relevant temperature-density range where the plasma process is important. As a result, we choose to multiply the plasma neutrino energy loss rate from \citet{haftStandardNonstandardPlasma1994} with a coefficient that samples from a normal distribution with $\sigma = 0.05$.
\subsection{Nuclear Reaction Rates}\label{sec:nuclear_rate}
\subsubsection{Triple-alphaReaction Rate}\label{sec:talpha}
\begin{figure}[h]
\centering
\includegraphics[width=0.45\textwidth]{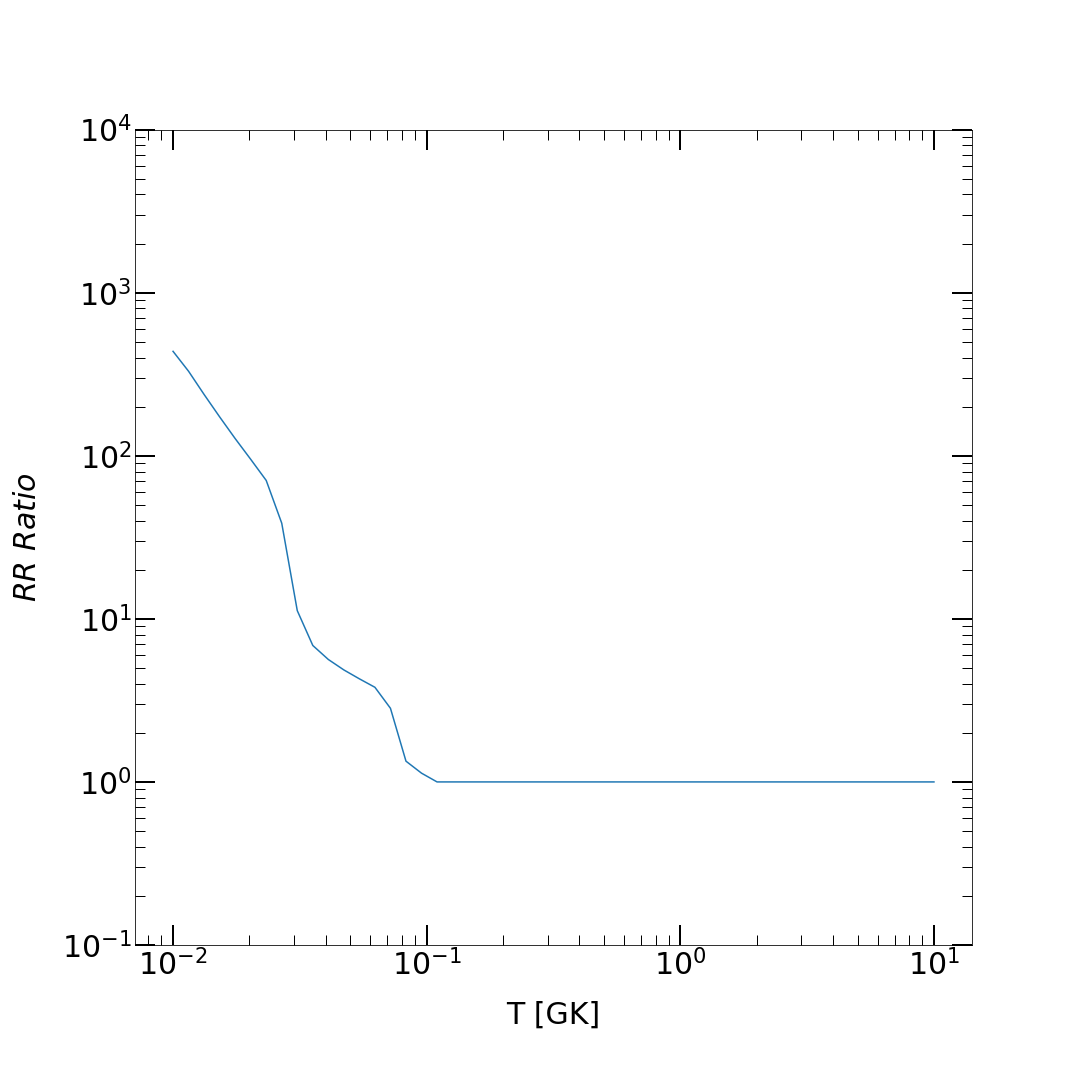}
\caption{Ratio of the new triple-$\alpha$ reaction rate \citep{sunoPreciseCalculationTripleAlpha2016} to the NACRE rate \citep{anguloCompilationChargedparticleInduced1999} as a function of temperature.}
\label{fig:talpha_ratio}
\end{figure}
During the evolution of a star, as the temperature of the hydrogen-burning shell increases and the degenerate core builds in mass, the temperature eventually reaches approximately $10^8 K$. This will trigger the helium fusion via the triple-$\alpha$ reaction \citep{collinsFundamentalsStellarAstrophysics1989}. The triple-$\alpha$ reaction rate previuosly adopted by DSEP \citep{chaboyerTestingMetalPoorStellar2017} was from \citet{anguloCompilationChargedparticleInduced1999}, the nuclear astrophysics compilation of reactions (NACRE). Using the transmission-free complex absorbing potential method, the triple-$\alpha$ reaction rate was recalculated by \citet{sunoPreciseCalculationTripleAlpha2016}. \citet{sunoPreciseCalculationTripleAlpha2016} showed that the new reaction rate agrees with the NACRE rate at high temperatures $T \geq 0.1$ GK. However, at low temperature $0.01$ GK$ \leq T < 0.1$ GK, the new rate can be larger than the NACRE rate by orders of magnitude, as shown in Fig. \ref{fig:talpha_ratio}. This difference can play an important role in DSEP. 

Under normal conditions, helium burning could begin, allowing for an orderly transition of nuclear energy generation processes. However, in lower mass stars the helium core is degenerate, and the electron pressure resulting from quantum degeneracy is not strongly temperature-dependent. This means the hot core will continue to receive more energy from helium burning, despite being unable to cool through core expansion, which is supported by the electron pressure. Because the triple-$\alpha$ reaction is highly temperature-dependent, higher temperatures will lead to faster reactions, and faster reactions will, in turn, lead to even higher temperatures. This positive feedback mechanism will dramatically increase helium luminosity in a short time \citep{collinsFundamentalsStellarAstrophysics1989}. This is called the helium flash and occurs when the star reaches the TRGB.
\begin{figure}[h]
\centering
\includegraphics[width=0.55\textwidth]{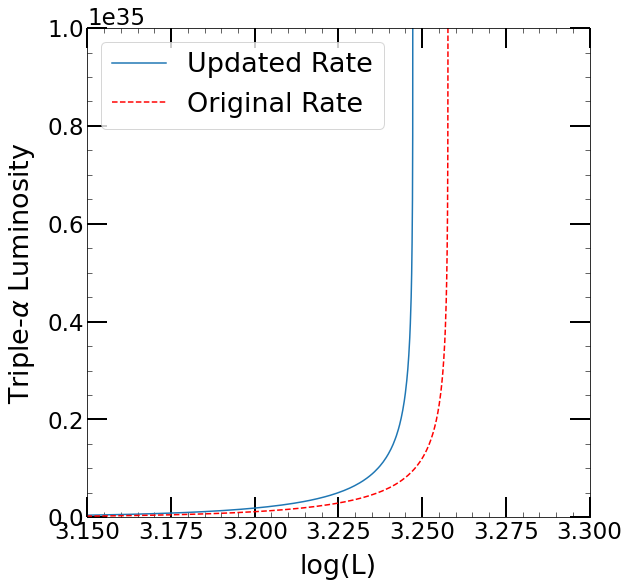}
\caption{Helium flash of both updated and old triple-$\alpha$ reaction rate with mass $= 0.7 M_{\odot}$ and [Fe/H] $= -1.6$.}
\label{fig:talpha_heliumflash}
\end{figure}

Fig. \ref {fig:talpha_heliumflash} shows the triple-$\alpha$ luminosity as a function of the luminosity of the star. Helium flash represents the moment when the He of the star is ignited and the triple-$\alpha$ luminosity increases dramatically. Because the updated triple-$\alpha$ reaction rate is much higher than the old one at lower temperatures, the Helium flash is expected to occur earlier with the updated triple-$\alpha$ reaction rate. In other words, the star's luminosity required for the He ignition is less for the updated triple-$\alpha$ reaction rate. 

\begin{figure}
\centering
    \includegraphics[width=0.6\textwidth]{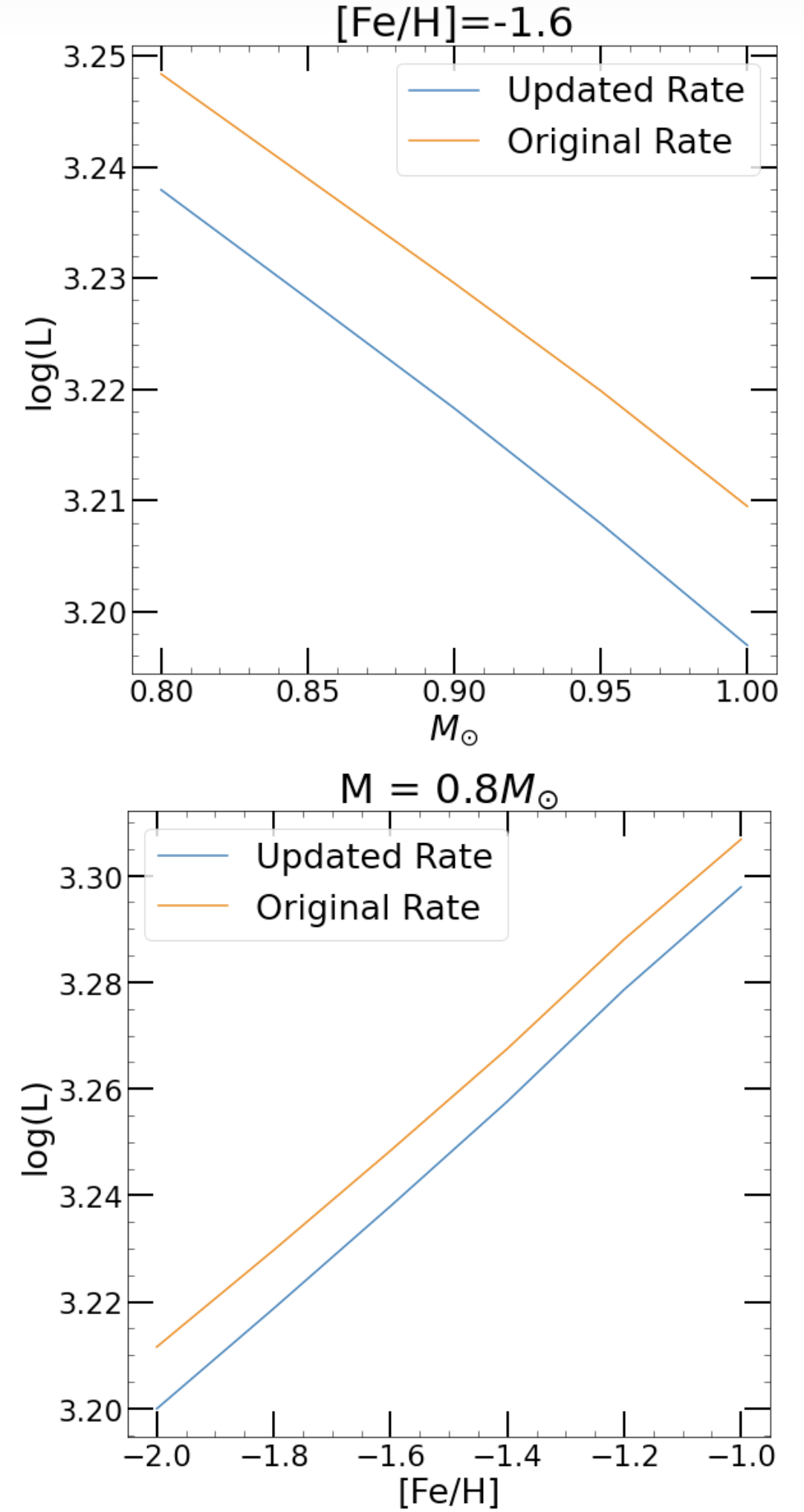}
\caption{\textit{Top}: Plot of luminosity against initial mass with [Fe/H] = $-1.6$ with minimal luminosity of the star required for the Helium flash to happen as a function of mass. \textit{Bottom}: Plot of luminosity against initial metallicity with mass = $0.8 M_{\odot}$ with minimal luminosity of the star required for the Helium flash to happen as a function of metallicity.}

\label{fig:talpha_flashtiming}
\end{figure}

The minimal luminosity of the star required for the Helium flash is calculated for both old and updated triple-$\alpha$ reaction rates for models with mass from $0.8 M_{\odot}$ to $1 M_{\odot}$ and [Fe/H] from $-2.0$ to $-1.0$. Fig. \ref{fig:talpha_flashtiming} demonstrates that the same conclusion can be generalized: for the same initial condition (a fixed mass or metallicity), the updated triple-$\alpha$ reaction rate, compared to the old triple-$\alpha$ reaction rate, will trigger the helium flash at a lower luminosity. Fig. \ref{fig:talpha_flashtiming} also suggests that the difference between the minimal luminosity required for the updated triple-$\alpha$ reaction rate and the old triple-$\alpha$ reaction rate is almost invariant to the change in initial mass or metallicity.  Overall, the updated triple-$\alpha$ reaction rate has only a minor impact on the properties of the TRGB.

\citet{sunoPreciseCalculationTripleAlpha2016} does not discuss the uncertainty in the Triple-$\alpha$ reaction rate. However, \citet{togniniImpactUncertainties3a2023} reports an average $12\%$ of uncertainty in the Triple-$\alpha$ reaction rate while \citet{kibediRadiativeWidthHoyle2020} found a surprising $34\%$ difference in the reaction adopted in models of stellar evolution and nucleosynthesis. We choose to multiply the Triple-$\alpha$ reaction rate from \citet{sunoPreciseCalculationTripleAlpha2016} with a coefficient that samples from a normal distribution with $\sigma = 0.15$.

\subsubsection{Carbon to Oxygen Reaction Rate}\label{sec:c12o16}
The ${ }^{12} \mathrm{C}(\alpha, \gamma)^{16} \mathrm{O}$ reaction rate previously adopted by DSEP \citep{chaboyerTestingMetalPoorStellar2017} was from \citet{kunzAstrophysicalReactionRate2002}. This rate was updated by \citet{deboerMonteCarloUncertainty2014}. The calculation was based on the updated triple-$\alpha$ reaction rate \citep{sunoPreciseCalculationTripleAlpha2016}.

The updated ${ }^{12} \mathrm{C}(\alpha, \gamma)^{16} \mathrm{O}$ reaction rate is implemented into DSEP. Note that the ${ }^{12} \mathrm{C}(\alpha, \gamma)^{16} \mathrm{O}$ reaction rate is only important for helium fusion after a reasonable amount of ${ }^{12} \mathrm{C}$ has been created in the star though the triple-alpha reaction. Therefore, it is reasonable to assume that this update will not have a significant impact on the TRGB and this was confirmed by model calculations.  

\begin{figure}
\centering
    \includegraphics[width=0.6\textwidth]{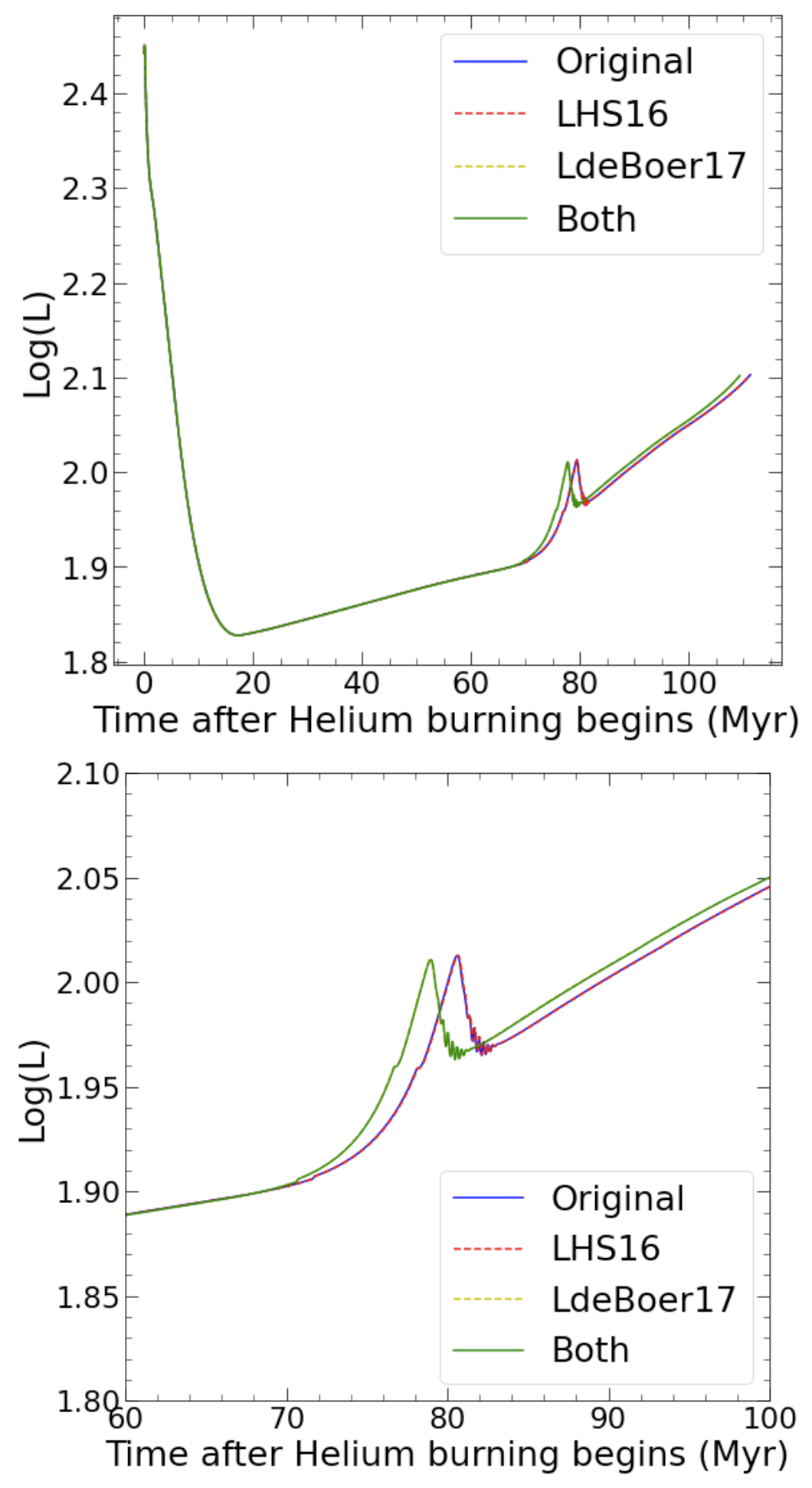}
\caption{Plots of luminosity as a function of time after the ignition  of helium fusion for a $M=3\,M_\odot$ model. Four combinations were chosen: ``Original'' DSEP (blue), updated triple-$\alpha$ reaction rate (red dotted, LHS16), updated ${ }^{12} \mathrm{C}(\alpha, \gamma)^{16} \mathrm{O}$ reaction rate (yellow dotted, LdeBoer17), and both updated triple-$\alpha$ reaction rate and updated ${ }^{12} \mathrm{C}(\alpha, \gamma)^{16} \mathrm{O}$ reaction rate (green). \textit{Bottom}: Zoomed in version of the star's luminosity as a funciton of time after the helium burning starts. The updated ${ }^{12} \mathrm{C}(\alpha, \gamma)^{16} \mathrm{O}$ reaction rate leads to a slightly higher luminosity at the same age.}

\label{fig:he_burn}
\end{figure}

We evolved a higher-mass ($M = 3\, M_{\odot}$),star to investigate the effect of these updated He fusion reaction rates on stellar evolution. Figure \ref{fig:he_burn} shows that the updated triple-$\alpha$ reaction rate will have little effect on the luminosity. However, the updated  ${ }^{12} \mathrm{C}(\alpha, \gamma)^{16} \mathrm{O}$ reaction rate leads to a slight difference in total luminosity. This is expected since the updated  ${ }^{12} \mathrm{C}(\alpha, \gamma)^{16} \mathrm{O}$ reaction rate is on the same order of magnitude (within $\sim 20\%) $ as the previous rate. The updated triple-$\alpha$ reaction rate had little effect on higher mass stars. As is well known, higher-mass stars will reach a higher core temperature much faster than lower-mass stars. As a result, higher mass stars will usually have a core temperature that is outside the range where the updated triple-$\alpha$ reaction differs significantly from the previous rate. 

\citet{deboer12Ca16OReaction2017} suggests an uncertainty of $15 - 20\%$ due to interpolation and \citet{pepperImpactUncertainties12Ca2022} suggests an uncertainty of $5 - 10\%$ in the reaction rate ${ }^{12} \mathrm{C}(\alpha, \gamma)^{16} \mathrm{O}$. We choose to multiply the ${ }^{12} \mathrm{C}(\alpha, \gamma)^{16} \mathrm{O}$ reaction rate from \citet{deboer12Ca16OReaction2017} with a coefficient that samples from a normal distribution with $\sigma = 0.15$.

\section{DSEE Overview: Flow-Based Model Architecture} \label{sec:Architecture Overview}

In \S \ref{sec:uncertainty}, we identify the source of uncertainties in stellar evolution models. In our previous studies, we used a Monte Carlo approach to study the impact of those uncertainties on the age estimation of Milky Way globular clusters \citep{yingAbsoluteAgeM922023,yingAbsoluteAgeNGC2024,yingAbsoluteAgeMilky2025}. Although we successfully estimated the absolute age of Milky Way globular clusters and created the error budget for the uncertainty in age, this approach required evolving over half a million stellar evolution models per globular cluster to fully explore the high-dimensional parameter space. As a result, the prohibitively high computational cost and the parameter space specific to individual globular clusters make it challenging to keep pace with the rapidly expanding volume of observational data. 

To address these challenges, we propose developing a new stellar evolution model grid with the following features:
\begin{enumerate}
    \item Comprehensive coverage of composition and ages.
    \item User-friendly design. 
    \item Inclusion of non-standard stellar evolution parameters to explicitly account for model-dependent uncertainties. 
\end{enumerate}

Our previous studies and \S \ref{sec:uncertainty} have demonstrated the feasibility of constructing a new stellar evolution model grid that meets the first two criteria. However, achieving the third criterion is challenging because the spacing between grid points grows rapidly as dimensionality increases, complicating comprehensive coverage of the parameter space. Consequently, the ideal number of stellar evolution models required by the database expands exponentially with the inclusion of non-standard stellar evolution parameters, making storage and retrieval of specific models increasingly impractical. Furthermore, most stellar evolution databases employ interpolation methods to generate models that are not precisely located on grid points. In high-dimensional spaces, even linear interpolation becomes computationally demanding, and accuracy significantly diminishes because high-dimensional datasets typically appear sparse and scattered \citep{linKernelInterpolationHigh2020}. 

Modern machine learning (ML) advances have enabled rapid and efficient utilization of pre-computed stellar model grids for inferring stellar properties. Instead of manually interpolating grids or running time-consuming fits for each star, ML models can learn the complex mapping between observable stellar quantities and the underlying stellar parameters from large grids of evolutionary models. In \S \ref{sec:isochrones}, we unify track building and isochrone construction by treating both as marginalized stellar output variables (hereafter referred to as stellar snapshots). In \S \ref{sec:ingredient} we outline the key components for constructing the Dartmouth Stellar Evolution Emulator (DSEE), and in \S \ref{sec:training}, we detail the training and optimization process.

\subsection{Isochrones: marginalized stellar evolutionary snapshots} \label{sec:isochrones}

As ML methods demonstrate a superior computational performance in interpolation between stellar evolution models, they can also be used to construct isochrones. A stellar isochrone represents a population of stars with the same age, and it is a fundamental tool to estimate the age of a population of stars. However, stars with different initial masses will have very different lifetimes and timesteps. To make the comparison, we need to find a uniform basis: equivalent evolutionary phases \citep[EEPs;][]{Simpson1970,dotterMESAISOCHRONESStelLAR2016}) EEPs are certain points on the evolution track of a star that can be easily identified through certain physical features. For example, the zero-age main sequence (ZAMS) is an ideal EEP since it represents the first time a star joins the main sequence and can be characterized as the point at which hydrogen-burning luminosity dominates the total luminosity of the star. The Terminal-age main sequence (TAMS), when a star burns out its central hydrogen, is another ideal EEP. EEPs, such as ZAMS and TAMS, form the basis for generating isochrones. In other words, isochrones are stellar evolution snapshots marginalized by time. The theory-driven EEP-based isochrone construction methods \cite[e.g.][]{dotterMESAISOCHRONESStelLAR2016} have demonstrated significant success in various subfields of astronomy. However, the interpolation methods introduce external uncertainties and can only be applied when a complete set of stellar evolution models exists, differing only in mass. 

\citet{honFlowBasedGenerativeEmulation2024} used conditional normalizing flows to emulate grids of stellar evolutionary models. They treat the grid’s input parameters and output properties collectively as a multi-dimensional probability distribution, and train the ML model to learn the complex joint relationships between inputs and outputs. The resulting flow can generate a continuous range of stellar evolutionary tracks, not just at the discrete grid points, by sampling from this learned distribution. As a result, isochrones are naturally constructed by putting a condition on age. Moreover, the model acts as a smooth interpolator/extrapolator that understands the covariances in the grid (for instance, how changes in mass and metallicity simultaneously affect luminosity and radius over time). 

Similarly, \citet{van-laneNovelApplicationConditional2023} showed the power of probabilistic ML in a different context – gyrochronology – where the physics of stellar spin-down is not fully understood. They trained a conditional normalizing flow on the rotation periods of cluster stars to learn the relationship between rotation, color, and age directly from the data, without relying on an explicit spin-down law. The result was a data-driven gyrochronology model that achieves age precisions comparable to traditional methods and successfully recovers known cluster ages. 

These studies illustrate how machine learning (ML) models can effectively bridge the gap between theoretical stellar evolution models and observational data in the era of big data astronomy. ML techniques demonstrate superior computational efficiency and scalability compared to grid based interpolation techniques enabling robust analyses of large astronomical datasets. The inherent flexibility of ML methods facilitates rigorous statistical assessments, making them well-suited to automated analaysis of large datasets.  

\subsection{Bigger, Better, Stronger: Enhancing Models through Improved Architecture, Richer Parameters, and Expanded Data}\label{sec:ingredient}

The key ingredients of the Dartmouth Stellar Evolution Emulator are:
\begin{enumerate}
    \item The updated Dartmouth Stellar Evolution Program with the stability and flexibility to vary a large selection of stellar evolution parameters
    \item A compilation of evolved stellar evolution models that covers a wide range in stellar evolution parameter space
    \item A sophisticated machine learning architecture with the capability to learn and generate stellar evolution models quickly and accurately
\end{enumerate}

\subsubsection{Stellar Evolution Model Grid} \label{sec:grid}

A robust and comprehensive training dataset is essential to the successful development of an accurate and efficient stellar evolution emulator. Traditionally, the Dartmouth Stellar Evolution Database (DSED) has provided stellar evolution models varying in metallicity ([Fe/H]), alpha-element enhancement ([$\alpha$/Fe]), and helium abundance (Y), comprising a total of $67$ distinct parameter combinations (see Table \ref{tab:DSED}). $57$ stellar evolution models with different masses were included for each combination, leading to a total of $3,819$ stellar evolution models in the database. However, to ensure the effectiveness of the Dartmouth Stellar Evolution Emulator (DSEE), it is crucial to significantly expand the coverage of this parameter.

\begin{table*}[htbp]
\caption{The Current Dartmouth Stellar Evolution Database}
\centering
\begin{tabular}{|c|cccccc|cc|cc|}
\hline
\multicolumn{11}{|c|}{\textbf{COMPOSITION}} \\
\hline
 & \multicolumn{6}{c|}{Y=0.245+1.6Z} & \multicolumn{2}{c|}{Y=0.33} & \multicolumn{2}{c|}{Y=0.40} \\
\hline
\textbf{[Fe/H]} & \multicolumn{6}{c|}{[$\alpha$/Fe]} & \multicolumn{2}{c|}{[$\alpha$/Fe]} & \multicolumn{2}{c|}{[$\alpha$/Fe]} \\
\hline
 & -0.2 & 0.0 & +0.2 & +0.4 & +0.6 & +0.8 & 0.0 & +0.4 & 0.0 & +0.4 \\
\hline
-2.5 & -0.2 & 0.0 & +0.2 & +0.4 & +0.6 & +0.8& 0.0 & +0.4 & 0.0 & +0.4 \\
-2.0 & -0.2 & 0.0 & +0.2 & +0.4 & +0.6 & +0.8& 0.0 & +0.4 & 0.0 & +0.4 \\
-1.5 & -0.2 & 0.0 & +0.2 & +0.4 & +0.6 & +0.8& 0.0 & +0.4 & 0.0 & +0.4 \\
-1.0 & -0.2 & 0.0 & +0.2 & +0.4 & +0.6 & +0.8& 0.0 & +0.4 & 0.0 & +0.4 \\
-0.5 & -0.2 & 0.0 & +0.2 & +0.4 & +0.6 & +0.8& 0.0 & +0.4 & 0.0 & +0.4 \\
0.0 & -0.2 & 0.0 & +0.2 & +0.4 & +0.6 & +0.8& 0.0 & +0.4 & 0.0 & +0.4 \\
+0.15 &  & 0.0 &  & &&&&&&\\
+0.3 & -0.2 & 0.0 & +0.2 & &&&&&&\\
+0.5 & -0.2 & 0.0 & +0.2 & &&&&&&\\
\hline
\end{tabular}
\label{tab:DSED}
\end{table*}

To address this necessity, we have substantially broadened our parameter space, encompassing over $20$ critical stellar evolution parameters, each spanning extensive and physically relevant ranges. By employing a high-dimensional parameter exploration approach, we have evolved approximately $8.4$ million ($2^{23}$) individual stellar evolution models using the Dartmouth Stellar Evolution Program (DSEP). For low-mass stars, each stellar model was evolved either until a stellar age of $20$ Gyr or until it reached the helium flash event. For high-mass stars that do not undergo the helium flash, each stellar model was evolved through the end of core helium fusion. This method ensured the coverage of critical evolutionary stages across the parameter space. Table \ref{tab:DSEE_MC} summarizes the selection of stellar evolution parameters being varied in the database and their corresponding distribution.

\begin{figure*}[htp]
    \centering
    \includegraphics[width=0.9\textwidth]{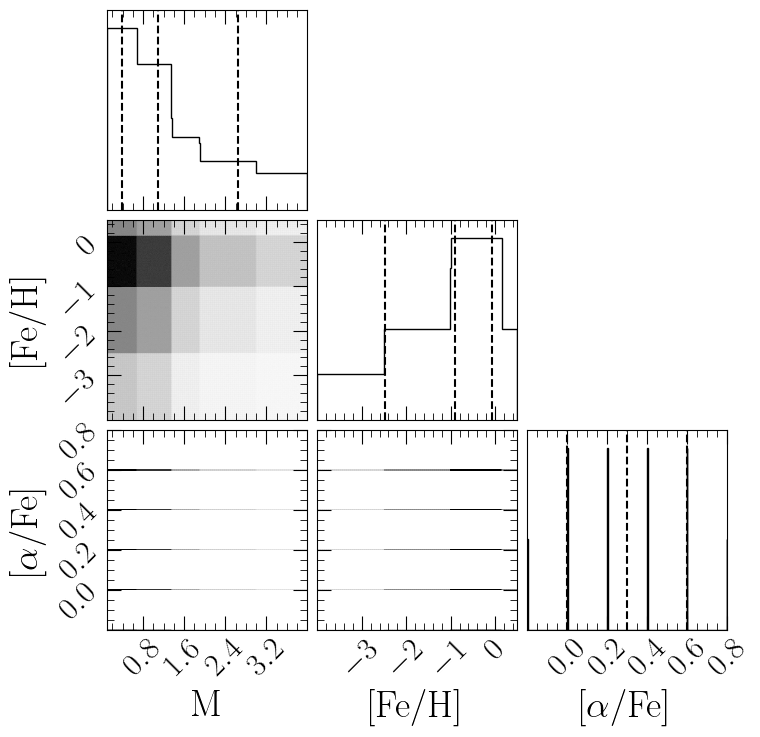}
    \caption{Corner plot showing the distribution of mass, [Fe/H], and [$\alpha$/Fe] for stellar evolution models used to train the Dartmouth Stellar Evolution Emulator.}
    \label{fig:3param}
\end{figure*}

We choose to follow the mass distribution used to construct isochrones in previous studies \citep{dotterDartmouthStellarEvolution2008,yingAbsoluteAgeM922023,yingAbsoluteAgeNGC2024,yingAbsoluteAgeMilky2025}.  Thesse studies  have shown success in capturing the essential morphological changes for isochrone construction using a stepwise-uniform distribution with $\sim 32\%$ of low-mass stars from $0.1 M_{\odot}$ to $0.68 M_{\odot}$, $\sim 29\%$ of intermediate-low-mass stars from $0.68 M_{\odot}$ to $1.35 M_{\odot}$, $\sim 12\%$ of intermediate-high-mass stars from $1.35 M_{\odot}$ to $1.9 M_{\odot}$, $\sim 16\%$ of vintermediate-high-mass stars from $1.90 M_{\odot}$ to $3.0 M_{\odot}$, and $\sim 11\%$ of high-mass stars from $3.0 M_{\odot}$ to $4.0 M_{\odot}$. We choose a stepwise-uniform distribution of metallicity based on the distribution of metallicity in the galaxy \citep{nessMetallicityDistributionMilky2016}, the change in morphology on the CMD per change in metallicity, and our interest in metal-poor Milky Way Globular Clusters. $\sim 15\%$ of stars have $-4.0 <$ [Fe/H] $<-2.5$, $\sim 31\%$ of stars have $-2.5 <$ [Fe/H] $<-1.0$, $\sim 47\%$ of stars have $-1.0 <$ [Fe/H] $< -0.15$, and $\sim 7\%$ of stars have $-0.15 <$ [Fe/H] $<0.5$. Due to the limited choice of 
PHOENIX model atmospheres, we can only choose [$\alpha$/Fe] $=-0.2, 0.0, 0.2, 0.4, 0.6, 0.8$. Therefore, we use a stepwise-uniform distribution for [$\alpha$/Fe]. $20\%$ of stars have [$\alpha$/Fe] $=-0.2$ or $0.8$, and $80\%$ of stars have [$\alpha$/Fe] $=0.0, 0.2, 0.4$ or $0.6$. The distribution of mass, [Fe/H], and [$\alpha$/Fe] are show in Figure~\ref{fig:3param}. 

\begin{figure*}[htp]
    \includegraphics[width=\textwidth]{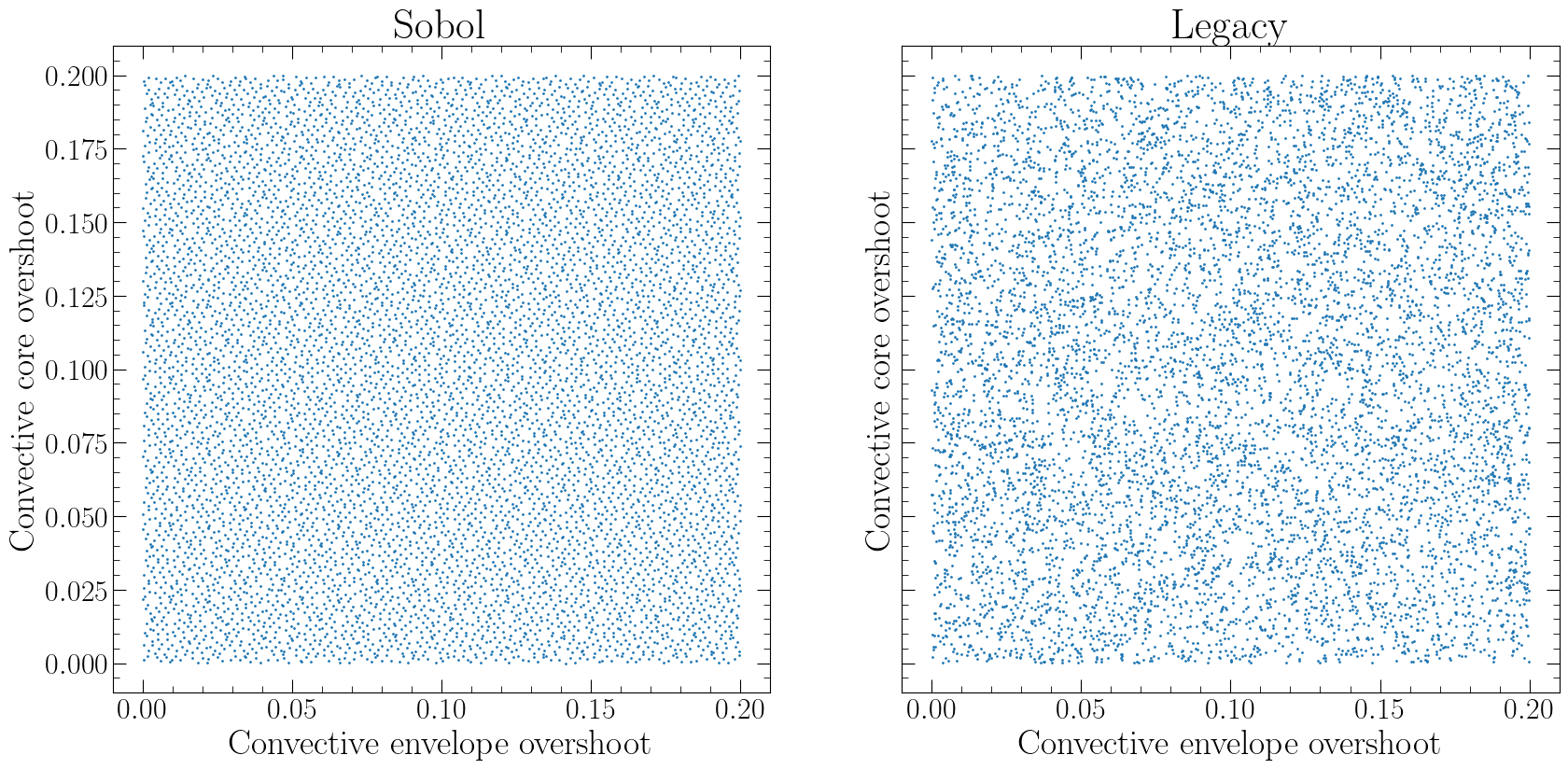}
    \caption{Comparing the samples generated using the Sobol sequence(left) and the traditional Monte Carlo approach(right) for a uniform distribution.}
    \label{fig:sobol_uni}
\end{figure*}

\begin{figure*}[htp]
    \centering
    \includegraphics[width=0.9\textwidth]{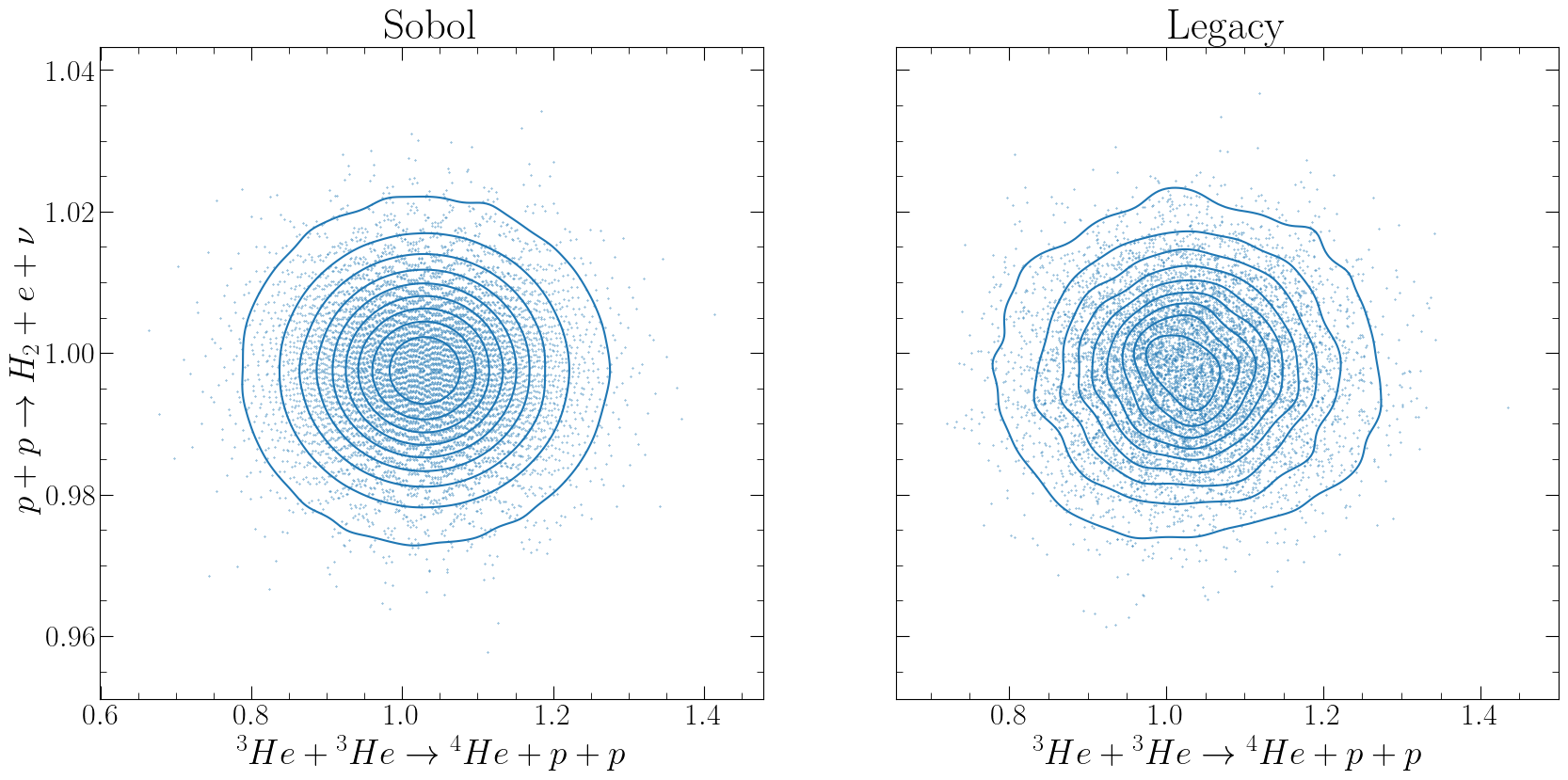}
    \caption{Comparing the samples generated using the Sobol sequence(left) and the traditional Monte Carlo approach(right) for a normal distribution.}
    \label{fig:sobol_normal}
\end{figure*}

To construct the high-dimensional input parameter space necessary for training our machine learning model, we employed a Sobol sequence \citep{sobol1967distribution}, a type of quasi-random, low-discrepancy sampling technique. The primary motivation for adopting Sobol sequences is their ability to provide a more uniform and evenly distributed coverage of the parameter space compared to purely random sampling methods, as shown in Figure ~\ref{fig:sobol_uni} and Figure ~\ref{fig:sobol_normal}. Ensuring a low-discrepancy distribution minimizes potential sparse regions in the input parameter space, which, if present, could cause inadequate information representation and consequently reduce the predictive accuracy and reliability of the trained emulator. Thus, employing Sobol sequences enhances the density and uniformity of sampling, significantly improving the overall quality and robustness of the resulting training dataset. We generate $2^{23}$ sets of stellar evolution parameters corresponding to $23$ stellar evolution parameters\footnote{$\Delta Y/\Delta Z$ is absorbed in the initial helium abundance} in Table \ref{tab:DSEE_MC}. Each set of stellar evolution parameters was used to evolve a stellar evolution model using the updated Dartmouth Stellar Evolution Program. Each model takes $5 \sim 60$ minutes of CPU time and is represented in an evolution track with $1,000 \sim 10,000$ timesteps, depending on the evolutionary trajectory of the stellar model given the initial condition. The result is a total of $\sim 2 \times 10^{10}$ evolutionary snapshots that take about $2.4$ TB of storage space.

\begin{figure*}[tp]
    \centering
    \includegraphics[width=0.9\textwidth]{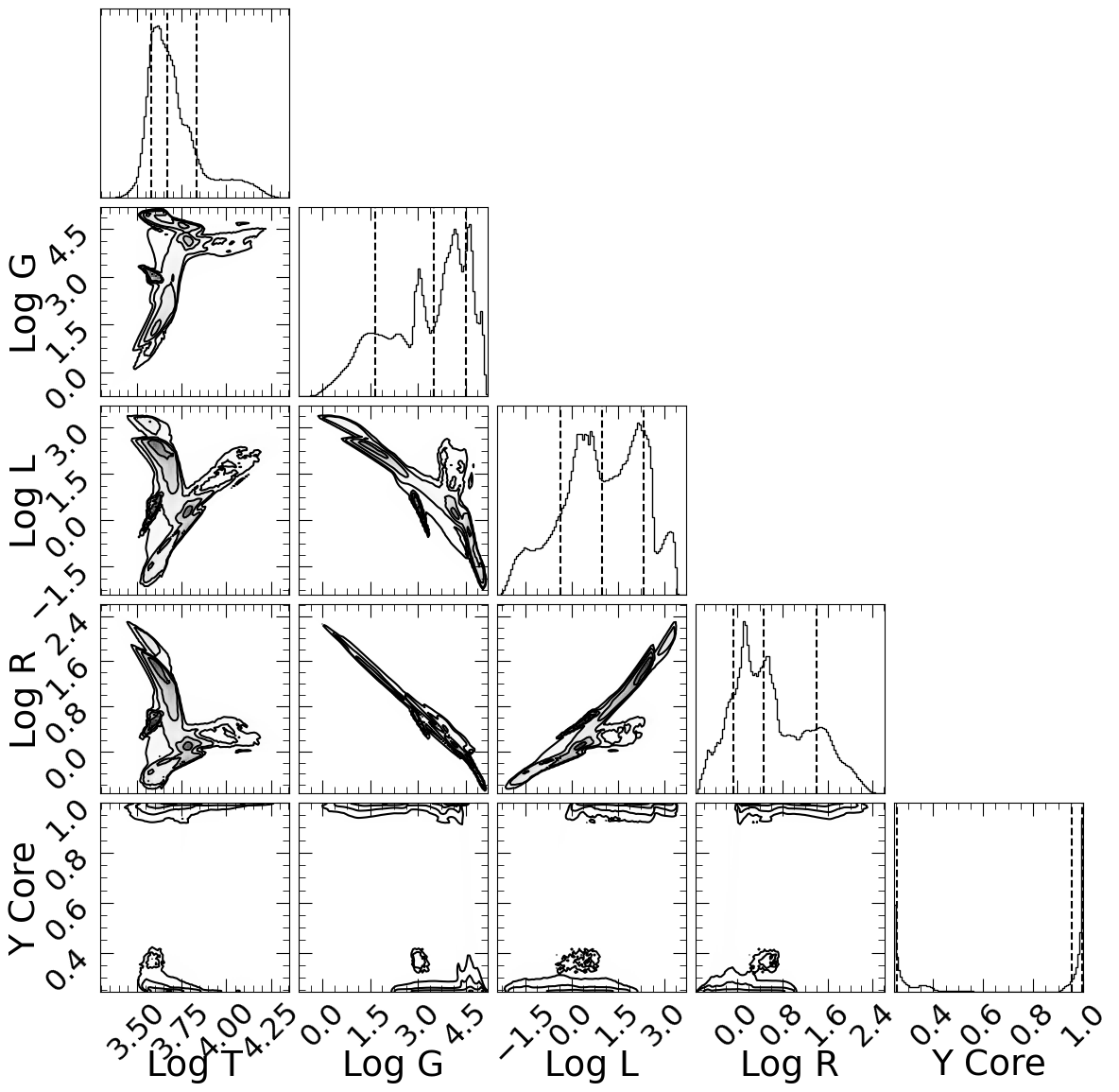}
    \caption{Corner plot showing the 5-dimensional stellar evolution model output in the DSEE training set.}
    \label{fig:DSEE_output_dist}
\end{figure*}

\begin{figure}[htp]
    \centering
    \includegraphics[width=0.45\textwidth]{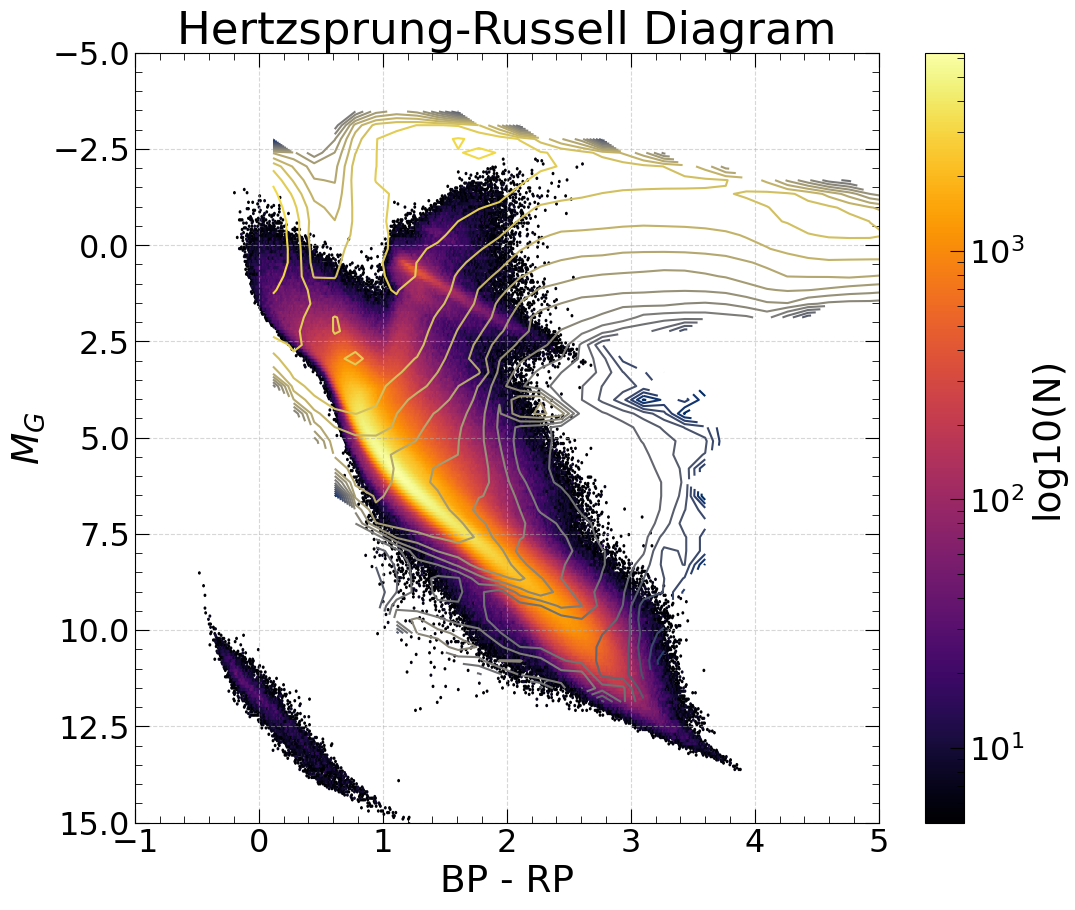}
    \caption{Comparing the distribution of evolutionary snapshots in the database with over $60$ million stars within $5,000$ ly from the Sun on a Hertzsprung-Russell diagram in Gaia DR3 absolute magnitude. The color indicates the number density of Gaia DR3 sources on the HR diagram and the contour represents the number density level of evolutionary snapshots in the database.}
    \label{fig:gaia_cmd}
\end{figure}

Figure \ref{fig:DSEE_output_dist} shows the distribution of evolutionary snapshots in 5-dimensional stellar evolution model output space, and Figure \ref{fig:gaia_cmd} shows the distribution of evolutionary snapshots from our database plotted over $60$ million stars within $5,000$ ly from the Sun from Gaia DR3 \citep{gaiacollaborationGaiaDataRelease2023}. The evolutionary snapshots in our database comprehensively span the regions of the Hertzsprung–Russell (HR) diagram predominantly occupied by main-sequence and red giant branch stars as observed in Gaia DR3. However, we note several intentional limitations of this database: it does not include evolutionary phases corresponding to white dwarfs, high-mass stars, or helium-burning stars with masses below approximately $1.5 M_{\odot}$. Nevertheless, this extensive coverage ensures that our stellar evolution database accurately represents the parameter space relevant to a large majority of stars.  This newly generated dataset represents the largest and most comprehensive collection of stellar evolution models currently available. This extensive database provides a solid foundation, enabling the Dartmouth Stellar Evolution Emulator to accurately capture the intricate relationships between various stellar parameters and their corresponding evolutionary outcomes. Consequently, it facilitates rapid, reliable, and continuous modeling across a wide range of astrophysical applications.

\subsubsection{Flow-based Generative Model} \label{sec:NF}
As discussed, there are significant uncertainties associated with the calculation of stellar models. To address these uncertainties in a comprehensive way, we adopt a normalizing flow architecture for our emulator. Normalizing flows offer a probabilistic and bijective framework that explicitly models uncertainties, making them particularly suitable for representing the complex interdependencies and parameter sensitivities inherent in stellar evolution modeling.

Normalizing flows are a class of generative models that define complex probability densities through invertible transformations of simple base distributions. In a flow, one starts with an easy-to-sample base density (e.g., a standard Gaussian) and applies a sequence of invertible, differentiable mappings $f$ to obtain a target data distribution. By the change-of-variables formula, the resulting density $(p(x))$ can be evaluated exactly as: 
\begin{equation}
    p(x) = \pi_u(f^{-1}(x)) \left|\det \frac{\partial f^{-1}}{\partial x}\right|, \label{eq:cov}
\end{equation}
which is tractable if $f^{-1}$ and its Jacobian determinant is easy to compute. Flows therefore permit explicit density evaluation (exact log-likelihood) and direct sampling in contrast to Variational Autoencoders (VAEs, \citealp{kingmaIntroductionVariationalAutoencoders2019}) or Generative Adversarial Networks (GANs, \citealp{goodfellowGenerativeAdversarialNetworks2014}), which either approximate the likelihood or lack a tractable density.

Two widely used design families are autoregressive and coupling-based flows. Autoregressive flows (e.g., IAF, \citealp{kingmaImprovingVariationalInference2016}, MAF, \citealp{papamakariosMaskedAutoregressiveFlow2017}) transform dimensions sequentially, yielding triangular Jacobians and exact likelihoods. With Gaussian conditionals,

\begin{equation}
    x_i=\mu_i\left(x_{<i}\right)+\sigma_i\left(x_{<i}\right) u_i, \label{eq:maf}
\end{equation}
with $\mu_i, \sigma_i$ given by neural nets that take previous $x$’s as input. The Jacobian of MAF is triangular, making its log-determinant straightforward:
\begin{equation}
    \log \left|\operatorname{det} \frac{\partial f^{-1}}{\partial x}\right|=-\sum_i \log \sigma_i\left(x_{<i}\right). \label{eq:jacobian}
\end{equation}

Coupling-layer flows (e.g., NICE, Real NVP, Glow; \citealp{dinhNICENonlinearIndependent2014, dinhDensityEstimationUsing2016, kingmaGlowGenerativeFlow2018}) split variables and transform one subset conditioned on the other, allowing parallel forward and inverse passes; historically, however, they often relied on simple element-wise (additive/affine) transforms that can limit flexibility.

To increase expressivity without sacrificing tractability, we adopt Neural Spline Flows (NSF;\citealp{durkanNeuralSplineFlows2019}), which replace the simple additive/affine elementwise transforms used in many coupling or autoregressive flows with monotonic rational–quadratic splines. These spline transforms are exactly invertible and have closed-form log-Jacobians, so likelihood evaluation and sampling remain efficient, and yet each layer can implement highly nonlinear, localized warps of the density. With sufficiently many segments, the splines act as universal approximators of differentiable monotone functions on bounded intervals, meaning a single NSF layer can capture curvature and sharp changes that would otherwise require stacking many affine layers.

In practice, this added flexibility yields smoother, more faithful fits to the non-Gaussian, skewed, and occasionally multimodal conditional densities that arise in phase-conditioned stellar states, while preserving numerical stability and exact maximum-likelihood training.

In DSEE, we model the stellar state as a conditional density:
\begin{equation}
    p\!\left(\mathbf{M}_{\mathbf{\theta}} \mid \mathbf{\theta}\right), \label{eq:DSEE_cond_p}
\end{equation}
where stellar evolution parameters $\mathbf{\theta}$ are consist of widely used initial conditions (e.g., $M_{\rm init}, [{\rm Fe/H}], [\alpha/{\rm Fe}]$), and DSEE specific input-physics settings (e.g., $\alpha_{\rm MLT}$, overshoot, diffusion, opacities), and age of the star $t$. A conditioning network maps $\mathbf{\theta}$ to the NSF parameters. The flow provides exact log-likelihoods for maximum-likelihood estimation, calibrated credible intervals from sampling, and smooth interpolation across the high-dimensional physics space. Tracks come from sweeping $t$ at fixed initial conditions; isochrones from fixing $t$ and other initial parameters but marginalizing over $M_{\rm init}$. We will discuss more details in \S \ref{sec:training} and \S \ref{sec:validation}.

\subsection{Training and Optimization}\label{sec:training}
We follow a similar approach as \citet{honFlowBasedGenerativeEmulation2024} and employ Neural Spline Flows \citep{durkanNeuralSplineFlows2019}, implemented in the Zuko library\citep{rozet2022zuko} to train our model. Specifically, our architecture consists of $10$ NSF transforms, each parameterized by a $10$-layer multi-layer perceptron (MLP) with $256$ neurons per layer. In each flow layer, this neural network acts as a conditioner: given the context (initial conditions, input-physics vector, and time), it outputs the knot widths/heights and boundary slopes that parameterize the NSF’s monotonic spline. These outputs are constrained to ensure valid, invertible transforms. During maximum-likelihood training, gradients update the MLP so that, for each context, it emits spline parameters that reallocate probability mass toward the theoretical stellar states. Training is performed using an AMD 7950X CPU paired with an RTX 5090 GPU. After extensive hyperparameter experimentation, we selected a batch size of $40,000$. This batch size represents the minimum value required to ensure that CPU-GPU communication does not become a bottleneck, thus maximizing GPU parallel computation efficiency. We utilize the Adam optimizer \citep{kingmaAdamMethodStochastic2017} with an initial learning rate of $1 \times 10^{-4}$, combined with a learning rate scheduler to decrease the learning rate whenever the validation metric plateaus. Due to the substantial size of our training dataset, each training epoch requires approximately $40$ hours of wall time. The emulator is trained for a total of $100$ epochs, achieving rapid convergence facilitated by the extensive and well-sampled training data.

Since theoretical isochrones are usually converted to magnitudes to be compared with observational data, we also trained a simple $3$-layer MLP with $200$ neurons in each layer to perform bolometric corrections. We use bolometric correction tables from the YBC database \citep{chenYBCStellarBolometric2019} as the training data. Each set of bolometric correction tables was used to train a bolometric correction model for $1,000$ epochs with a $128$ batch size. The light-weighted bolometric correction model can be stored in a GPU and perform bolometric correction with no additional synchronization between the GPU and CPU. 

The artificial star test is an essential tool to understand the photometric uncertainty. For each artificial star test, we train two neural-network models for photometric uncertainty and completeness, respectively. For photometric uncertainty, we use a smaller NSF model with $5$ NSF transforms, each parameterized by a $5$-layer multi-layer perception (MLP) with $64$ neurons per layer. For completeness, we use a $3$-layer MLP with $200$ neurons in each layer. Similar to the bolometric correction model, photometric uncertainty and completeness models can be stored in GPU memory for the rapid construction of simulated color-magnitude diagrams (sCMDs).

\begin{figure}[htp]
    \centering
    \includegraphics[width=0.45\textwidth]{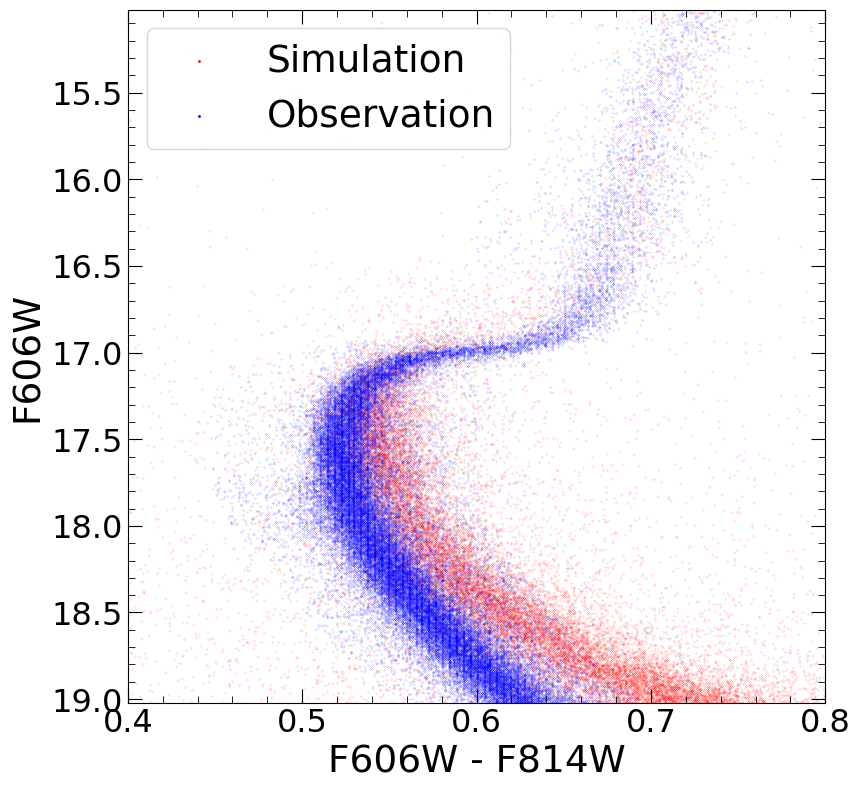}
    \caption{Comparing a simulated color-magnitude diagram (in red) with the observed color-magnitude diagram for 47Tuc (in blue). 
    This particular simulated CMD was chosen at random from our database for a particular set of physics inputs and is not a good fit to the data.  Our database contains a wide range of physics input parameters, and, as discussed in \cite{yingAbsoluteAgeMilky2025} it is possible to find a range of parameters which would provide a good fit to the observed color-magnitude diagram. } 
    
    \label{fig:sCMD_emulate}
\end{figure}

Our previous studies demonstrate the time-consuming and complicated process of constructing a sCMD from evolving stellar models, constructing isochrones, sampling stars given a mass distribution, and injecting photometric uncertainty based on an artificial star test. With the help of ML, we can emulate a sCMD, as shown in Figure \ref{fig:sCMD_emulate}, almost instantly, which leads to the possibility of performing large-scale analysis.

\section{Validation and Performance}\label{sec:validation}

\subsection{Validation Through Stellar Evolution Models} \label{sec:evolion_model_validation}

To validate the performance of the Dartmouth Stellar Evolution Emulator (DSEE), we construct a comprehensive test set comprising $185$ stellar evolution models. These models were generated by randomly sampling parameters from the stellar evolution parameter space and evolving each model up to either $20$ Gyr or the onset of the helium flash event. In total, our test set consists of over $300,000$ evolutionary snapshots, representing distinct stages of stellar evolution.

\begin{figure}[htp]
    \centering
    \includegraphics[width=0.45\textwidth]{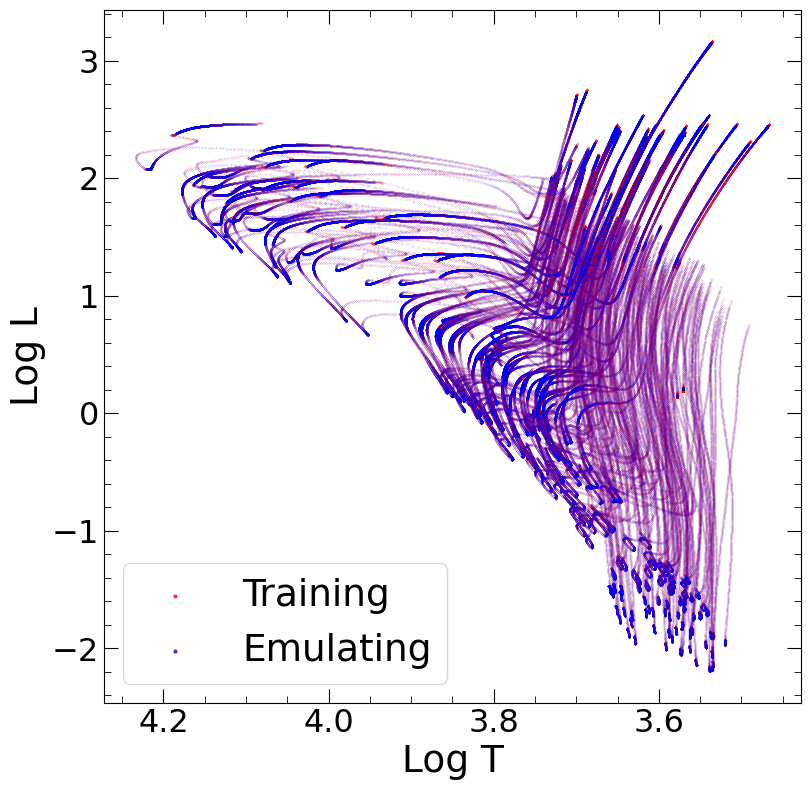}
    \caption{Comparing $185$ stellar models evolved using DSEP with stellar models emulated using DSEE. Each red dot represents an evolutionary snapshot from the DSEP models, and each blue dot represents an evolutionary snapshot generated by DSEE.}
    \label{fig:validate_tracks}
\end{figure}

Utilizing the DSEE, we emulate the stellar evolution tracks corresponding to the identical set of randomly sampled input parameters. For each evolutionary snapshot in the test set, we generated $10,000$ emulator predictions and adopted the median of these samples as the emulator’s representative output. 
To evaluate emulator accuracy, we compared these outputs directly against the test set evolutionary tracks on the Hertzsprung–Russell (HR) diagram, as depicted in Figure \ref{fig:validate_tracks}. Visually, one can see that the emulator is doing a reasonable job of representing the actual stellar evolution tracks.

To validate the accuracy of the age prediction, we sample $10,000$ models using DSEE with the same physical parameters and varied ages. We determine the best-fit DSEE model by finding the nearest neighbor in the $5$-dimensional output space. Because stellar ages span a vast range, from millions to tens of billions of years, we choose to quantify model accuracy in relative terms, reporting errors as percentages rather than absolute values. 

The comparison between the computed and emulated stellar models is quantified in Table~\ref{tab:validate_accuracy} which reports the $90$th-percentile uncertainties (in absolute terms) for effective temperature, radius, luminosity, and surface gravity, and in relative terms for age. For example $90\%$ of the emulator evolutionary snapshots have effective temperature uncertainties within the range $-0.006 < \log{\textup{T}} < 0.006$ of the computed stellar evolutionary tracks.

Our validation demonstrates that the emulator achieves remarkably high accuracy particularly for main-sequence and main-sequence turn-off stars. In contrast, a higher degree of uncertainty is observed at the tip of the Red Giant Branch (TRGB) and the Hertzsprung gap (the subgiant phase for higher mass stars). This increased uncertainty arises primarily due to the rapid evolutionary changes characteristic of the RGB and Hertzsprung gap phase. We anticipate that the emulator's accuracy within these evolutionary stages can be improved by strategically enhancing the training set to increase data density, specifically within the RGB and Hertzsprung gap regions. 

\begin{table}[]
\caption{Validation Accuracy \label{tab:validate_accuracy}}
\centering
\begin{tabular}{lllll}
\hline
$\Delta$ Log T & $\Delta$ Log R & $\Delta$ Log L & $\Delta$ Log G & $\Delta$ Age ($\%$)\\
\hline
0.006 &0.048 & 0.075 & 0.096 & 1.06 \\
\hline
\end{tabular}
\end{table}

\subsection{Validation Through Dartmouth Stellar Evolution Database}

\begin{figure}[htp]
    \centering
    \includegraphics[width=0.9\textwidth]{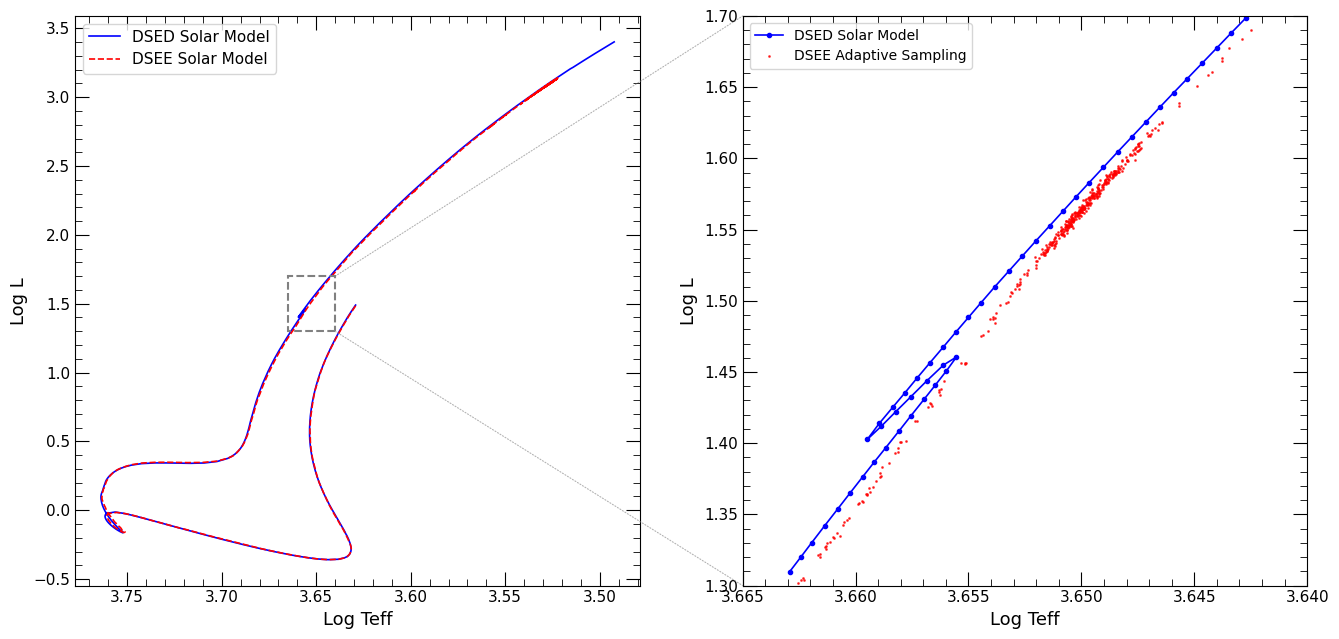}
    \caption{Comparing the solar model from the Dartmouth Stellar Evolution Database (DSED) with the solar model generated by the DSEE emulator using the same set of parameters. Left: the blue solid line is the evolutionary track of the DSED solar model and the red dashed line is the track generated by DSEE. The dashed gray rectangle marks the RGB bump region. Right: zoomed view of the RGB bump region. Blue circles connected by a solid line show the DSED track, which exhibits the characteristic luminosity reversal of the RGB bump. Red points show the DSEE track generated with adaptive sampling, which iteratively concentrates evaluation points in regions of rapid change along the HR diagram. The visible clustering of DSEE points at the bump location indicates that the emulator has learned a softened version of this feature — preserving the rapid change in slope at the correct evolutionary phase, though smoothing the sharp luminosity reversal intrinsic to the physical models.}
    \label{fig:DSED_solar}
\end{figure}

We compared our model with the Dartmouth Stellar Evolution Database (DSED, \citealp{dotterDartmouthStellarEvolution2008}). DSED employs the PHOENIX model atmosphere \citep{husserNewExtensiveLibrary2013} with a solar-calibrated mixing length parameter, $\alpha_{\textup{MLT}} = 1.938$. The DSED models incorporate convective core overshoot based on the formalism where the extent of overshoot scales with the size of the convective core \citep{demarqueY2IsochronesImproved2004}. Specifically, the overshoot amount is parameterized in terms of the local pressure scale height ($\lambda_p$) and varies according to stellar mass and composition. Models at the minimum stellar mass incorporate an overshoot of $0.05 \lambda_p$, those exceeding the minimum by $0.1 M_{\odot}$ incorporate $0.1 \lambda_p$, and those exceeding the minimum mass by $0.2 M_{\odot}$ or more receive $0.2 \lambda_p$ of overshoot. Additionally, DSED incorporates atomic diffusion following the prescription of \citet{thoulElementDiffusionSolar1994}, with inhibition in the outermost $0.01 M_{\odot}$. Standard nuclear reaction rates and opacity tables are adopted in DSED.

Leveraging the inherent flexibility of our Dartmouth Stellar Evolution Emulator (DSEE), we adjusted input parameters to replicate stellar evolution models provided by DSED. Our analysis demonstrates that DSEE effectively reproduces most features of these reference models. Figure \ref{fig:DSED_solar} shows an example of this capability through a direct comparison of the solar model from DSED and its emulation by DSEE.  DSEE faithfully replicates the overall morphology of the evolutionary track across the main sequence, subagent branch, and red giant branch. For the Red Giant Branch (RGB) bump, a luminosity anomaly occurring when the hydrogen-burning shell encounters the chemically homogeneous region left by the receding convective envelope \citep{joyceINVESTIGATINGCONSISTENCYStelLAR2015}, DSEE produces a softened representation of this feature rather than a sharp luminosity reversal. As shown in the zoomed panel of Figure \ref{fig:DSED_solar}, adaptive sampling of the emulator reveals a clustering of evaluation points close to the bump location of the DSED solar model with the emulator exhibiting $\sim 40\%$ higher uncertainty in $T_{\mathrm{eff}}$ and $\sim 34\%$ higher uncertainty in luminosity relative to the surrounding RGB. This elevated uncertainty and point concentration indicate that the neural network has learned the rapid morphological transition at this evolutionary phase, though the intrinsically smooth nature of the network prevents it from reproducing the discontinuous luminosity reversal present in the physical models. Similarly, DSEE model does not evolve to the Tip of Red Giant Branch (TRGB) as this region lies near the boundary of the training set where the emulator has limited coverage and is therefore less constrained.

\subsection{Validation Through Monte-Carlo Isochrones}

\begin{figure*}[htp]
    \centering
    \includegraphics[width=0.9\textwidth]{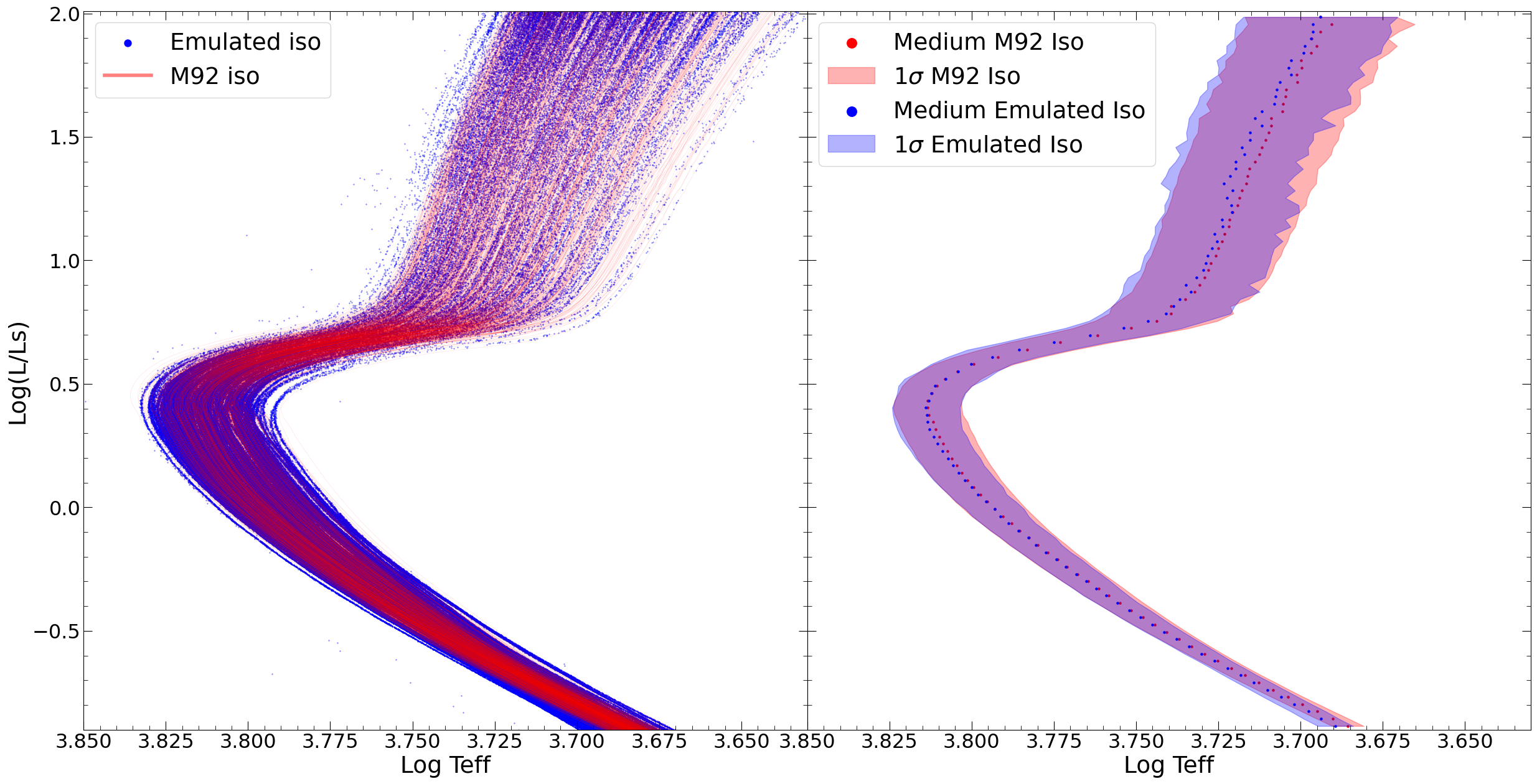}
    \caption{Comparing the $1,000$ $13$ Gyr Monte-Carlo isochrones generated in the M92 study \citep{yingAbsoluteAgeM922023} with $1,000$ emulated isochrones from DSEE. Left: Emulated isochrones, each consisting of $3,000$ evolutionary snapshots with different masses (in blue), are plotted over the $1,000$ $13$ Gyr Monte-Carlo isochrones (in red). Right: Corresponding median effective temperatures at given luminosities, with shaded regions indicating $1\sigma$ around the median effective temperature.}
    \label{fig:M92_iso_remake}
\end{figure*}

The capability of DSEE extends beyond emulating single stellar evolution models or individual isochrones. Figure \ref{fig:M92_iso_remake} compares the $1,000$ $13$ Gyr Monte-Carlo isochrones generated from the M92 study \citep{yingAbsoluteAgeM922023}, each constructed using a distinct set of stellar evolution parameters varied through a Monte Carlo method (as detailed in Table \ref{tab:DSEE_MC}). In our new approach, we generate $1,000$ set of stellar evolution parameters using the Sobol sequence within the same parameter ranges for [Fe/H], [$\alpha$/Fe], and Helium abundance. Using DSEE, we then sampled $3,000$ evolutionary snapshots per parameter set, uniformly distributed in mass. Figure \ref{fig:M92_iso_remake} illustrates that DSEE successfully reproduces the distribution of the Monte-Carlo isochrones while significantly reducing computational complexity.

\subsection{Uncertainty Quantification and Calibration} \label{sec:calibration}

A central advantage of the Dartmouth Stellar Evolution Emulator (DSEE) is its probabilistic design: by learning the full conditional probability density of stellar states given input physics and evolutionary phase, it captures both the mean behavior and the intrinsic dispersion of stellar models. 
This enables DSEE to provide calibrated credible intervals that reflect not only the variability of the training data but also the inherent uncertainty of stellar physics.

To evaluate these uncertainty estimates, we conducted a calibration experiment. 
For each randomly selected set of stellar parameters, we generated 10,000 realizations by resampling the emulator’s latent space while holding the conditioning variables fixed. 
The resulting distributions of luminosity, radius, effective temperature, and surface gravity were compared against those from Monte Carlo DSEP models with perturbed input physics (Table~\ref{tab:DSEE_MC}). 
Across all evolutionary phases, $91\%$ of DSEP models fall within DSEE’s $95\%$ credible interval, indicating robust statistical calibration.

DSEE performs exceptionally well along the main sequence and near the main-sequence turn-off (MSTO), where uncertainty widths track the expected physical variance. 
Beyond the MSTO, as stars evolve toward the subgiant and red giant branches (RGBs), the credible intervals broaden systematically. 
This reflects two effects: (1) interpolation accuracy declines outside the densest regions of the training set (\S ~\ref{sec:evolion_model_validation}), and (2) the MSTO–RGB phases are intrinsically more sensitive to variations in mixing length, opacity, and diffusion, leading to genuine increases in model uncertainty. 

Despite this, DSEE preserves a smooth, physically consistent median track from the MSTO through the tip of the red giant branch (TRGB), demonstrating that the learned manifold remains coherent even where uncertainty grows. 
The broader intervals near the RGB likely arise from two additional sources:
\begin{enumerate}
    \item \textbf{Training set density.} Although the database includes over eight million stellar models, the MSTO and RGB stages are brief and therefore underrepresented. Denser sampling could reduce stochastic spread, though limited coverage alone cannot fully explain the observed uncertainty.
    \item \textbf{Finite temporal resolution.} 
    Because stellar tracks are computed at discrete time steps, the mapping between mass, age, and evolutionary phase is not perfectly continuous. 
    As noted by \citet{dotterMESAISOCHRONESStelLAR2016}, small discontinuities in age resolution can break the monotonic mass–phase relation near the MSTO and RGB tip. 
    DSEE, which learns from these patterns, encodes this non-monotonicity as increased uncertainty rather than imposing an artificial smoothness.
\end{enumerate}

Overall, DSEE's predicted uncertainty reflects a realistic combination of data sparsity, intrinsic model sensitivity, and the discrete nature of stellar evolution calculations. 
Its credible intervals are physically meaningful and self-consistent—expanding naturally where input physics or training coverage limits predictive confidence. 
Users should therefore interpret broader intervals not as emulator failure, but as an honest quantification of the limits of current stellar model knowledge.

An interactive demonstration of this uncertainty propagation experiment, including stochastic track and isochrone generation, is available in the \href{https://github.com/200k33p3r/CONF1DENCE}{CONF1DENCE repository}. A frozen version (which corresponds to the version of CONF1DENCE as it exists upon the completion of peer review of this paper) is available on Zenodo
\href{https://doi.org/10.5281/zenodo.18331872}{https://doi.org/10.5281/zenodo.18331872} \citep{CONF1DENCE_zenodo}.

\section{CONF1DENCE} \label{sec:confidence}
CONF1DENCE (COnditional Normalizing Flow 1D stellar Evolution acCElerator) is a Python package\footnote{Available at \url{https://github.com/200k33p3r/CONF1DENCE}} designed to streamline the entire process of stellar parameter inference from training the normalizing flow, emulating stellar evolutionary snapshots, to inferring stellar parameters. Additionally, CONF1DENCE includes integrated functionality for training neural networks to compute bolometric corrections, photometric uncertainties, completeness corrections, and simulated observational data in order to compare to cluster observations. 

\begin{figure*}[htp]
    \centering
    \includegraphics[width=\textwidth]{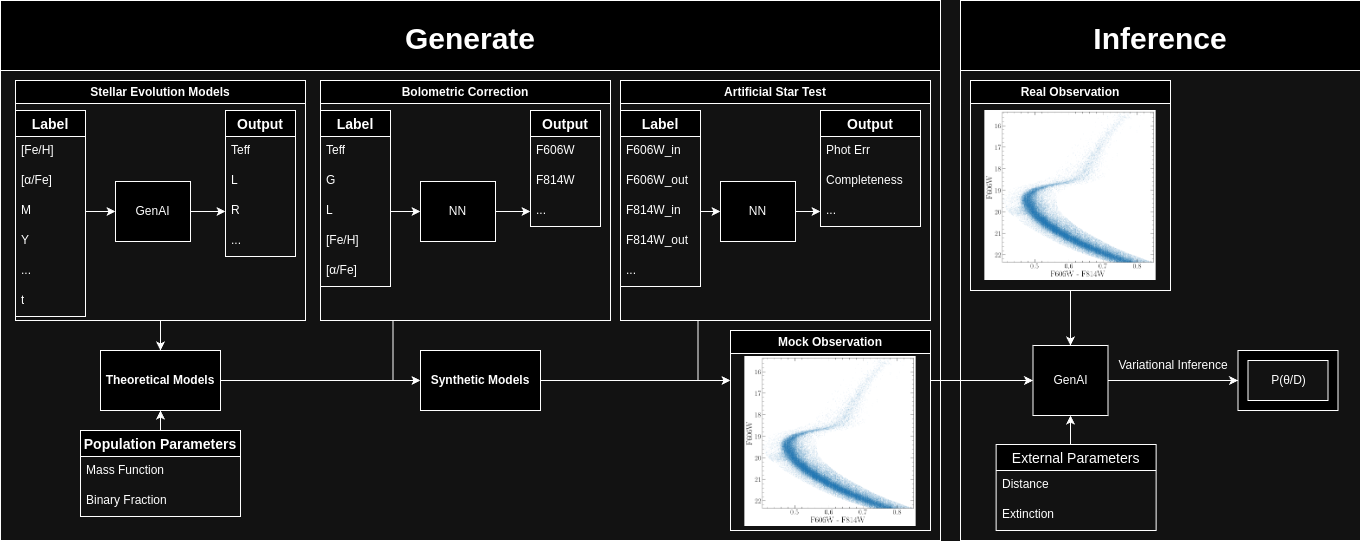}
    \caption{Illusitration of the workflow using CONF1DENCE to inference stellar evolution parameters for globular clusters}
    \label{fig:CONF1DENCE}
\end{figure*}

Figure \ref{fig:CONF1DENCE} illustrates the workflow for inferring stellar evolution parameters of globular clusters using CONF1DENCE. Given an initial mass function, a distribution of stellar parameters can be sampled and emulated using DSEE. Additional stellar populations can be simulated to represent binary stars based on the binary fraction. The resulting physical parameters are converted into observable quantities using a bolometric correction neural network tailored to the specific photometric filters (we trained over 100 sets of filters covers for all major telescopes). Observational uncertainties and completeness corrections are subsequently applied to these observables. When a sufficient number of stellar members are present, an additional normalizing flow model can be trained directly on the observational data. Due to the computational efficiency of normalizing flows as density estimators, the resulting synthetic CMD (sCMD) allows for rapid comparison with observed CMDs. Variational inference-based statistical methods are then employed to infer precise stellar parameters.

CONF1DENCE enables the creation of custom-designed inference pipelines to handle diverse types of data for various tasks. It can also be used to easily interact with DSEE to emulate stellar evolution models or isochrones with robust uncertainty estimations. For further examples and more information, please visit the \href{https://github.com/200k33p3r/CONF1DENCE}{CONF1DENCE repository}.

\section{Conclusion and Future Work} \label{sec:conclusion}

In this work, we introduced the Dartmouth Stellar Evolution Emulator (DSEE), a powerful new generative tool for stellar evolution modeling. DSEE uses a normalizing flow architecture to emulate continuous stellar evolution tracks and isochrones, and it was trained on a Monte Carlo grid of over eight million Dartmouth Stellar Evolution Program (DSEP) models spanning $23$ varied input physics parameters. This unprecedented training set and modern machine-learning approach enable DSEE to overcome the limitations of traditional fixed stellar model grids, allowing for the fast on-demand generation of stellar models across a high-dimensional parameter space while accounting for uncertainties in the input physics. Our emulator thereby provides a flexible, uncertainty-aware alternative to classical stellar model libraries, built to meet the needs of today’s precise and extensive astronomical datasets.

DSEE achieves high accuracy in reproducing standard stellar evolution results. We validated the emulator extensively and found that its predictions closely match those from direct DSEP code integrations across the HR diagram, particularly for the main-sequence and turn-off phases. The emulator’s interpolated tracks are essentially as precise as the original DSEP models, yet can be generated in a fraction of a second.

A unique advantage of our emulator is its ability to encode uncertainties and produce probabilistic outputs. Because DSEE is built on normalizing flows, it returns a full probability distribution for stellar model quantities given any set of input parameters. In practice, this means that uncertainties arising from uncertain physics inputs are naturally captured as generative variability in the model’s output. We demonstrated this capability by showing that ensembles of isochrones generated with DSEE can reproduce the same spread of outcomes as those from brute-force Monte Carlo simulations. This probabilistic aspect is crucial for rigorous uncertainty quantification in stellar modeling and is a key improvement over traditional static grids, which lack such flexibility.

To facilitate broader use of these advances, we also release an open-source pipeline named CONF1DENCE for probabilistic cluster parameter inference. CONF1DENCE leverages DSEE to perform full forward modeling of observed color–magnitude diagrams in a Bayesian or likelihood framework. By rapidly sampling synthetic stellar populations (including single and binary stars) through the emulator and applying observational models (bolometric corrections, photometric errors, completeness) on-the-fly, the pipeline can efficiently compare against observed cluster data. The CONF1DENCE tool highlights the practical application of DSEE by reducing the barrier to complex, uncertainty-aware analysis of star clusters and other stellar populations using sophisticated models.
CONF1DENCE will continue to be developed and is available on Github \href{https://github.com/200k33p3r/CONF1DENCE}{https://github.com/200k33p3r/CONF1DENCE}.  A frozen version of CONF1DENCE is available on Zenodo
\href{https://doi.org/10.5281/zenodo.18331872}{https://doi.org/10.5281/zenodo.18331872} \citep{CONF1DENCE_zenodo}.

Future extension of this work will proceed along two main directions: further refinements and improvements in stellar evolution models, and expanded applications of these models to address broader questions across various subfields of astronomy. On the modeling side, we will refine DSEE by densifying the training set in rapid phases (e.g., the turn-off and RGB tip) and by using targeted sensitivity studies/active learning to place new models where the emulator’s uncertainty is largest. The physics vector will be expanded to include additional processes and extend coverage in mass and composition.We can use DSEE to further explore the influence of one or a combination of stellar evolution parameters, such as calibrating the mixing-length parameter for well-observed stars, or studying the correlation between parameters like the mixing length parameter, convective overshoot, and surface boundary conditions. Those studies can guide us toward the development of the next generation of stellar evolution models. 

A significant limitation of the current database is that it does not include the helium burning phases of evolution for lower mass stars that undergo the helium flash. This is due to numerical limitations in DSEP.  We are currently exploring  overcoming this limitation and anticipate that we will be able to generate a large database of stellar evolution models for lower mass core helium burning stars and will publish an enlarged database and improved version of CONF1DENCE in 2027.

On the applications side, we will deploy DSEE to problems that demand fast, uncertainty-aware modeling, such as exoplanet host characterization, cluster CMD fitting at survey scale, stellar-population and SSP modeling for galaxy studies, and related inferences where theoretical systematics currently dominate the error budget. To lower the barrier to entry, we will continue to develop the open-source CONF1DENCE package, improving performance, expanding photometric/observational modules, and offering streamlined, scalable inference routines, so that researchers without deep stellar-modeling expertise or large computational resources can still perform rigorous analyses. We will collaborate with domain experts across various subfields to tailor inference strategies to different data types (photometry, spectroscopy, and asteroseismology) and selection  functions.

\clearpage

\section*{Acknowledgment}

\begin{acknowledgments}
We thank the anonymous referee, whose insightful comments have improved the manuscript. 
This material is based upon work supported by the National Science Foundation under Award No.\ 2007174 and by NASA through AR 17043 from the Space Telescope Science Institute (STScI), which is operated by AURA, Inc., under NASA contract NAS5-26555. 
\end{acknowledgments}

\bibliography{references}

@ARTICLE{Kurucz1993,
       author = {{Kurucz}, Robert},
        title = "{ATLAS9 Stellar Atmosphere Programs and 2 km/s grid.}",
      journal = {Robert Kurucz CD-ROM},
         year = 1993,
        month = jan,
       volume = {13},
       adsurl = {https://ui.adsabs.harvard.edu/abs/1993KurCD..13.....K}
}

@misc{CONF1DENCE_zenodo,
  doi = {10.5281/ZENODO.18331872},
  url = {https://zenodo.org/doi/10.5281/zenodo.18331872},
  author = {Ying, Martin and Chaboyer, Brian and Cargile, Phillip and Dufresne, George},
  title = {CONF1DENCE},
  version = {1.0},
  publisher = {Zenodo},
  year = {2026},
  copyright = {Creative Commons Attribution 4.0 International}
}

@ARTICLE{Simpson1970,
       author = {{Simpson}, Erik and {Hills}, R.~E. and {Hoffman}, Wilson and {Kellman}, Sanford A. and {Morton}, Jr., Erwin and {Paresce}, Francesco and {Peterson}, Charles},
        title = "{Studies in Stellar Evolution.IX. Theoretical Isochrones for Early-Type Clusters}",
      journal = {\apj},
         year = 1970,
        month = mar,
       volume = {159},
        pages = {895},
          doi = {10.1086/150366},
       adsurl = {https://ui.adsabs.harvard.edu/abs/1970ApJ...159..895S}
}

@ARTICLE{Magg2022,
       author = {{Magg}, Ekaterina and {Bergemann}, Maria and {Serenelli}, Aldo and {Bautista}, Manuel and {Plez}, Bertrand and {Heiter}, Ulrike and {Gerber}, Jeffrey M. and {Ludwig}, Hans-G{\"u}nter and {Basu}, Sarbani and {Ferguson}, Jason W. and {Gallego}, Helena Carvajal and {Gamrath}, S{\'e}bastien and {Palmeri}, Patrick and {Quinet}, Pascal},
        title = "{Observational constraints on the origin of the elements. IV. Standard composition of the Sun}",
      journal = {\aap},
         year = 2022,
        month = may,
       volume = {661},
          eid = {A140},
        pages = {A140},
          doi = {10.1051/0004-6361/202142971},
archivePrefix = {arXiv},
       eprint = {2203.02255},
 primaryClass = {astro-ph.SR},
       adsurl = {https://ui.adsabs.harvard.edu/abs/2022A&A...661A.140M}
}

@ARTICLE{Buldgen2025,
       author = {{Buldgen}, Ga{\"e}l and {Canocchi}, Gloria and {Le Saux}, Arthur and {Baturin}, Vladimir A. and {Trampedach}, Regner and {Oreshina}, Anna V. and {Ayukov}, Sergey V. and {Pradhan}, Anil and {Pain}, Jean-Christophe and {Kunitomo}, Masanobu and {Appourchaux}, Thierry and {Garc{\'\i}a}, Rafael A. and {Deal}, Morgan and {Grevesse}, Nicolas and {Noels}, Arlette and {Christensen-Dalsgaard}, J{\o}rgen and {Guillot}, Tristan and {Nandal}, Devesh and {B{\'e}trisey}, J{\'e}r{\^o}me and {Blancard}, Christophe and {Colgan}, James and {Coss{\'e}}, Philippe and {Fontes}, Christopher J. and {Petitdemange}, Ludovic and {Pin{\c{c}}on}, Charly},
        title = "{The Future of Solar Modelling: Requirements for a New Generation of Solar Models}",
      journal = {\solphys},
         year = 2025,
        month = jul,
       volume = {300},
       number = {7},
          eid = {97},
        pages = {97},
          doi = {10.1007/s11207-025-02508-x},
archivePrefix = {arXiv},
       eprint = {2506.14514},
 primaryClass = {astro-ph.SR},
       adsurl = {https://ui.adsabs.harvard.edu/abs/2025SoPh..300...97B}
}

@ARTICLE{Mombarg2019,
       author = {{Mombarg}, J.~S.~G. and {Van Reeth}, T. and {Pedersen}, M.~G. and {Molenberghs}, G. and {Bowman}, D.~M. and {Johnston}, C. and {Tkachenko}, A. and {Aerts}, C.},
        title = "{Asteroseismic masses, ages, and core properties of {\ensuremath{\gamma}} Doradus stars using gravito-inertial dipole modes and spectroscopy}",
      journal = {\mnras},
         year = 2019,
        month = may,
       volume = {485},
       number = {3},
        pages = {3248-3263},
          doi = {10.1093/mnras/stz501},
archivePrefix = {arXiv},
       eprint = {1902.06746},
 primaryClass = {astro-ph.SR},
       adsurl = {https://ui.adsabs.harvard.edu/abs/2019MNRAS.485.3248M}
}

@ARTICLE{Moedas2022,
    author = {{Moedas}, Nuno and {Deal}, Morgan and {Bossini}, Diego and {Campilho}, Bernardo},
    title = "{Atomic diffusion and turbulent mixing in solar-like stars: Impact on the fundamental properties of FG-type stars}",
   journal = {\aap},
     year = 2022,
    month = oct,
    volume = {666},
     eid = {A43},
    pages = {A43},
     doi = {10.1051/0004-6361/202243210},
archivePrefix = {arXiv},
    eprint = {2207.02779},
 primaryClass = {astro-ph.SR},
    adsurl = {https://ui.adsabs.harvard.edu/abs/2022A&A...666A..43M}
}

@ARTICLE{Semenova2020,
    author = {{Semenova}, Ekaterina and {Bergemann}, Maria and {Deal}, Morgan and {Serenelli}, Aldo and {Hansen}, Camilla Juul and {Gallagher}, Andrew J. and {Bayo}, Amelia and {Bensby}, Thomas and {Bragaglia}, Angela and {Carraro}, Giovanni and {Morbidelli}, Lorenzo and {Pancino}, Elena and {Smiljanic}, Rodolfo},
    title = "{The Gaia-ESO survey: 3D NLTE abundances in the open cluster NGC 2420 suggest atomic diffusion and turbulent mixing are at the origin of chemical abundance variations}",
   journal = {\aap},
     year = 2020,
    month = nov,
    volume = {643},
     eid = {A164},
    pages = {A164},
     doi = {10.1051/0004-6361/202038833},
archivePrefix = {arXiv},
    eprint = {2007.09153},
 primaryClass = {astro-ph.SR},
    adsurl = {https://ui.adsabs.harvard.edu/abs/2020A&A...643A.164S}
}

@ARTICLE{Joyce2023,
       author = {{Joyce}, Meridith and {Johnson}, Christian I. and {Marchetti}, Tommaso and {Rich}, R. Michael and {Simion}, Iulia and {Bourke}, John},
        title = "{The Ages of Galactic Bulge Stars with Realistic Uncertainties}",
      journal = {\apj},
         year = 2023,
        month = mar,
       volume = {946},
       number = {1},
          eid = {28},
        pages = {28},
          doi = {10.3847/1538-4357/acb692},
archivePrefix = {arXiv},
       eprint = {2205.07964},
 primaryClass = {astro-ph.SR},
       adsurl = {https://ui.adsabs.harvard.edu/abs/2023ApJ...946...28J}
}

@ARTICLE{Tayar2017,
       author = {{Tayar}, Jamie and {Somers}, Garrett and {Pinsonneault}, Marc H. and {Stello}, Dennis and {Mints}, Alexey and {Johnson}, Jennifer A. and {Zamora}, O. and {Garc{\'\i}a-Hern{\'a}ndez}, D.~A. and {Maraston}, Claudia and {Serenelli}, Aldo and {Allende Prieto}, Carlos and {Bastien}, Fabienne A. and {Basu}, Sarbani and {Bird}, J.~C. and {Cohen}, R.~E. and {Cunha}, Katia and {Elsworth}, Yvonne and {Garc{\'\i}a}, Rafael A. and {Girardi}, Leo and {Hekker}, Saskia and {Holtzman}, Jon and {Huber}, Daniel and {Mathur}, Savita and {M{\'e}sz{\'a}ros}, Szabolcs and {Mosser}, B. and {Shetrone}, Matthew and {Silva Aguirre}, Victor and {Stassun}, Keivan and {Stringfellow}, Guy S. and {Zasowski}, Gail and {Roman-Lopes}, A.},
        title = "{The Correlation between Mixing Length and Metallicity on the Giant Branch: Implications for Ages in the Gaia Era}",
      journal = {\apj},
         year = 2017,
        month = may,
       volume = {840},
       number = {1},
          eid = {17},
        pages = {17},
          doi = {10.3847/1538-4357/aa6a1e},
archivePrefix = {arXiv},
       eprint = {1704.01164},
 primaryClass = {astro-ph.SR},
       adsurl = {https://ui.adsabs.harvard.edu/abs/2017ApJ...840...17T}
}

@article{Nagayama2019,
  title = {Systematic Study of $L$-Shell Opacity at Stellar Interior Temperatures},
  author = {Nagayama, T. and Bailey, J. E. and Loisel, G. P. and Dunham, G. S. and Rochau, G. A. and Blancard, C. and Colgan, J. and Coss\'e, Ph. and Faussurier, G. and Fontes, C. J. and Gilleron, F. and Hansen, S. B. and Iglesias, C. A. and Golovkin, I. E. and Kilcrease, D. P. and MacFarlane, J. J. and Mancini, R. C. and More, R. M. and Orban, C. and Pain, J.-C. and Sherrill, M. E. and Wilson, B. G.},
  journal = {Phys. Rev. Lett.},
  volume = {122},
  issue = {23},
  pages = {235001},
  numpages = {7},
  year = {2019},
  month = {Jun},
  publisher = {American Physical Society},
  doi = {10.1103/PhysRevLett.122.235001},
  url = {https://link.aps.org/doi/10.1103/PhysRevLett.122.235001}
}

@article{Hoarty2023,
    author = {Hoarty, D. J. and Morton, J. and Rougier, J. C. and Rubery, M. and Opachich, Y. P. and Swatton, D. and Richardson, S. and Heeter, R. F. and McLean, K. and Rose, S. J. and Perry, T. S. and Remington, B.},
    title = {Radiation burnthrough measurements to infer opacity at conditions close to the solar radiative zone–convective zone boundary},
    journal = {Physics of Plasmas},
    volume = {30},
    number = {6},
    pages = {063302},
    year = {2023},
    month = {06},
    issn = {1070-664X},
    doi = {10.1063/5.0141850},
    url = {https://doi.org/10.1063/5.0141850},
    eprint = {https://pubs.aip.org/aip/pop/article-pdf/doi/10.1063/5.0141850/17987260/063302_1_5.0141850.pdf},
}

@article{Mayes2025,
title = {Overview of oxygen opacity experiments at the National Ignition Facility and investigation of potential systematic errors},
journal = {High Energy Density Physics},
volume = {55},
pages = {101177},
year = {2025},
issn = {1574-1818},
doi = {https://doi.org/10.1016/j.hedp.2025.101177},
url = {https://www.sciencedirect.com/science/article/pii/S1574181825000059},
author = {D.C. Mayes and B.A. Hobbs and R.F. Heeter and T.S. Perry and H.M. Johns and Y.P. Opachich and M. Hohenberger and P.A. Bradley and E.C. Dutra and C.J. Fontes and E. Gallardo-Diaz and M.H. Montgomery and H.F. Robey and M.S. Wallace and D.E. Winget}
}

@ARTICLE{Valle2025,
       author = {{Valle}, G. and {Dell'Omodarme}, M. and {Prada Moroni}, P.~G. and {Degl'Innocenti}, S.},
        title = "{Tests and calibrations of stellar models with two triply eclipsing triple systems}",
      journal = {\aap},
         year = 2025,
        month = feb,
       volume = {694},
          eid = {A305},
        pages = {A305},
          doi = {10.1051/0004-6361/202452487},
archivePrefix = {arXiv},
       eprint = {2502.05484},
 primaryClass = {astro-ph.SR},
       adsurl = {https://ui.adsabs.harvard.edu/abs/2025A&A...694A.305V}
}

@ARTICLE{Chaboyer2002,
       author = {{Chaboyer}, Brian and {Krauss}, Lawrence M.},
        title = "{Theoretical Uncertainties in the Subgiant Mass-Age Relation and the Absolute Age of {\ensuremath{\omega}} Centauri}",
      journal = {\apjl},
         year = 2002,
        month = mar,
       volume = {567},
       number = {1},
        pages = {L45-L48},
          doi = {10.1086/339898},
archivePrefix = {arXiv},
       eprint = {astro-ph/0201443},
 primaryClass = {astro-ph},
       adsurl = {https://ui.adsabs.harvard.edu/abs/2002ApJ...567L..45C}
}

@ARTICLE{Anders2023,
       author = {{Anders}, Evan H. and {Pedersen}, May G.},
        title = "{Convective Boundary Mixing in Main-Sequence Stars: Theory and Empirical Constraints}",
      journal = {Galaxies},
         year = 2023,
        month = apr,
       volume = {11},
       number = {2},
          eid = {56},
        pages = {56},
          doi = {10.3390/galaxies11020056},
archivePrefix = {arXiv},
       eprint = {2303.12099},
 primaryClass = {astro-ph.SR},
       adsurl = {https://ui.adsabs.harvard.edu/abs/2023Galax..11...56A}
}

@ARTICLE{Claret2019,
       author = {{Claret}, Antonio and {Torres}, Guillermo},
        title = "{The Dependence of Convective Core Overshooting on Stellar Mass: Reality Check and Additional Evidence}",
      journal = {\apj},
         year = 2019,
        month = may,
       volume = {876},
       number = {2},
          eid = {134},
        pages = {134},
          doi = {10.3847/1538-4357/ab1589},
archivePrefix = {arXiv},
       eprint = {1904.02714},
 primaryClass = {astro-ph.SR},
       adsurl = {https://ui.adsabs.harvard.edu/abs/2019ApJ...876..134C}
}

@ARTICLE{Valle2017,
       author = {{Valle}, G. and {Dell'Omodarme}, M. and {Prada Moroni}, P.~G. and {Degl'Innocenti}, S.},
        title = "{Statistical errors and systematic biases in the calibration of the convective core overshooting with eclipsing binaries. A case study: TZ Fornacis}",
      journal = {\aap},
         year = 2017,
        month = apr,
       volume = {600},
          eid = {A41},
        pages = {A41},
          doi = {10.1051/0004-6361/201628240},
archivePrefix = {arXiv},
       eprint = {1612.07066},
 primaryClass = {astro-ph.SR},
       adsurl = {https://ui.adsabs.harvard.edu/abs/2017A&A...600A..41V}
}

@ARTICLE{Weller2025,
       author = {{Weller}, Miqaela K. and {Weinberg}, David H. and {Johnson}, James W.},
        title = "{Modelling the Galactic Chemical Evolution of Helium}",
      journal = {\mnras},
         year = 2025,
        month = apr,
       volume = {538},
       number = {3},
        pages = {1517-1534},
          doi = {10.1093/mnras/staf373},
archivePrefix = {arXiv},
       eprint = {2404.08765},
 primaryClass = {astro-ph.GA},
       adsurl = {https://ui.adsabs.harvard.edu/abs/2025MNRAS.538.1517W}
}

@ARTICLE{Verma2019,
       author = {{Verma}, Kuldeep and {Raodeo}, Keyuri and {Basu}, Sarbani and {Silva Aguirre}, V{\'\i}ctor and {Mazumdar}, Anwesh and {Mosumgaard}, Jakob R{\o}rsted and {Lund}, Mikkel N. and {Ranadive}, Pritesh},
        title = "{Helium abundance in a sample of cool stars: measurements from asteroseismology}",
      journal = {\mnras},
         year = 2019,
        month = mar,
       volume = {483},
       number = {4},
        pages = {4678-4694},
          doi = {10.1093/mnras/sty3374},
archivePrefix = {arXiv},
       eprint = {1812.02751},
 primaryClass = {astro-ph.SR},
       adsurl = {https://ui.adsabs.harvard.edu/abs/2019MNRAS.483.4678V}
}

@ARTICLE{Balser2006,
       author = {{Balser}, Dana S.},
        title = "{The Chemical Evolution of Helium}",
      journal = {\aj},
         year = 2006,
        month = dec,
       volume = {132},
       number = {6},
        pages = {2326-2332},
          doi = {10.1086/508515},
archivePrefix = {arXiv},
       eprint = {astro-ph/0608436},
 primaryClass = {astro-ph},
       adsurl = {https://ui.adsabs.harvard.edu/abs/2006AJ....132.2326B}
}

@ARTICLE{Serenelli2010,
       author = {{Serenelli}, Aldo M. and {Basu}, Sarbani},
        title = "{Determining the Initial Helium Abundance of the Sun}",
      journal = {\apj},
         year = 2010,
        month = aug,
       volume = {719},
       number = {1},
        pages = {865-872},
          doi = {10.1088/0004-637X/719/1/865},
archivePrefix = {arXiv},
       eprint = {1006.0244},
 primaryClass = {astro-ph.SR},
       adsurl = {https://ui.adsabs.harvard.edu/abs/2010ApJ...719..865S}
}

@ARTICLE{LiY2024,
       author = {{Li}, Yaguang and {Bedding}, Timothy R. and {Huber}, Daniel and {Stello}, Dennis and {van Saders}, Jennifer and {Zhou}, Yixiao and {Crawford}, Courtney L. and {Joyce}, Meridith and {Li}, Tanda and {Murphy}, Simon J. and {Sreenivas}, K.~R.},
        title = "{Realistic Uncertainties for Fundamental Properties of Asteroseismic Red Giants and the Interplay between Mixing Length, Metallicity, and {\ensuremath{\nu}}max}",
      journal = {\apj},
         year = 2024,
        month = oct,
       volume = {974},
       number = {1},
          eid = {77},
        pages = {77},
          doi = {10.3847/1538-4357/ad6c3e},
archivePrefix = {arXiv},
       eprint = {2407.09967},
 primaryClass = {astro-ph.SR},
       adsurl = {https://ui.adsabs.harvard.edu/abs/2024ApJ...974...77L}
}

@ARTICLE{Stokholm2019,
       author = {{Stokholm}, Amalie and {Nissen}, Poul Erik and {Silva Aguirre}, V{\'\i}ctor and {White}, Timothy R. and {Lund}, Mikkel N. and {Mosumgaard}, Jakob R{\o}rsted and {Huber}, Daniel and {Jessen-Hansen}, Jens},
        title = "{The subgiant HR 7322 as an asteroseismic benchmark star}",
      journal = {\mnras},
         year = 2019,
        month = oct,
       volume = {489},
       number = {1},
        pages = {928-940},
          doi = {10.1093/mnras/stz2222},
archivePrefix = {arXiv},
       eprint = {1908.03232},
 primaryClass = {astro-ph.SR},
       adsurl = {https://ui.adsabs.harvard.edu/abs/2019MNRAS.489..928S}
}

@ARTICLE{Tognelli2021,
    author = {{Tognelli}, E. and {Dell'Omodarme}, M. and {Valle}, G. and {Prada Moroni}, P.~G. and {Degl'Innocenti}, S.},
    title = "{Bayesian calibration of the mixing length parameter {\ensuremath{\alpha}}$_{ML}$ and of the helium-to-metal enrichment ratio {\ensuremath{\Delta}}Y/{\ensuremath{\Delta}}Z with open clusters: the Hyades test-bed}",
   journal = {\mnras},
     year = 2021,
    month = jan,
    volume = {501},
    number = {1},
    pages = {383-397},
     doi = {10.1093/mnras/staa3686},
archivePrefix = {arXiv},
    eprint = {2012.08193},
 primaryClass = {astro-ph.GA},
    adsurl = {https://ui.adsabs.harvard.edu/abs/2021MNRAS.501..383T}
}

@article{vianiInvestigatingMetallicityMixinglengthRelation2018,
	title = {Investigating the {Metallicity}-{Mixing}-length {Relation}},
	volume = {858},
	issn = {0004-637X},
	url = {https://ui.adsabs.harvard.edu/abs/2018ApJ...858...28V},
	doi = {10.3847/1538-4357/aab7eb},
	urldate = {2025-07-13},
	journal = {\apj},
	author = {Viani, Lucas S. and Basu, Sarbani and Ong J., M. Joel and Bonaca, Ana and Chaplin, William J.},
	month = may,
	year = {2018},
	note = {Publisher: IOP
ADS Bibcode: 2018ApJ...858...28V},
	pages = {28},
}

@article{joyceClassicallyAsteroseismicallyConstrained2018a,
	title = {Classically and {Asteroseismically} {Constrained} {1D} {Stellar} {Evolution} {Models} of α {Centauri} {A} and {B} {Using} {Empirical} {Mixing} {Length} {Calibrations}},
	volume = {864},
	issn = {0004-637X},
	url = {https://dx.doi.org/10.3847/1538-4357/aad464},
	doi = {10.3847/1538-4357/aad464},
	language = {en},
	number = {1},
	urldate = {2025-07-13},
	journal = {\apj},
	author = {Joyce, M. and Chaboyer, B.},
	month = sep,
	year = {2018},
	note = {Publisher: The American Astronomical Society},
	pages = {99},
}

@article{chenYBCStellarBolometric2019,
	title = {{YBC}: a stellar bolometric corrections database with variable extinction coefficients - {Application} to {PARSEC} isochrones},
	volume = {632},
	copyright = {© ESO 2019},
	issn = {0004-6361, 1432-0746},
	shorttitle = {{YBC}},
	url = {https://www.aanda.org/articles/aa/abs/2019/12/aa36612-19/aa36612-19.html},
	doi = {10.1051/0004-6361/201936612},
	language = {en},
	urldate = {2025-07-05},
	journal = {\aap},
	author = {Chen, Yang and Girardi, Léo and Fu, Xiaoting and Bressan, Alessandro and Aringer, Bernhard and Tio, Piero Dal and Pastorelli, Giada and Marigo, Paola and Costa, Guglielmo and Zhang, Xing},
	month = dec,
	year = {2019},
	note = {Publisher: EDP Sciences},
	pages = {A105},
}

@misc{kingmaAdamMethodStochastic2017,
	title = {Adam: {A} {Method} for {Stochastic} {Optimization}},
	shorttitle = {Adam},
	url = {http://arxiv.org/abs/1412.6980},
	doi = {10.48550/arXiv.1412.6980},
	urldate = {2025-07-05},
	publisher = {arXiv},
	author = {Kingma, Diederik P. and Ba, Jimmy},
	month = jan,
	year = {2017},
	note = {arXiv:1412.6980 [cs]},
}

@misc{rozet2022zuko,
	title = {Zuko: {Normalizing} flows in {PyTorch}},
	copyright = {MIT},
	url = {https://pypi.org/project/zuko},
	author = {Rozet, François and {others}},
	year = {2022},
	doi = {10.5281/zenodo.7625672},
}

@misc{linKernelInterpolationHigh2020,
	title = {Kernel {Interpolation} of {High} {Dimensional} {Scattered} {Data}},
	url = {https://ui.adsabs.harvard.edu/abs/2020arXiv200901514L},
	doi = {10.48550/arXiv.2009.01514},
	urldate = {2025-07-04},
	publisher = {arXiv},
	author = {Lin, Shao-Bo and Chang, Xiangyu and Sun, Xingping},
	month = sep,
	year = {2020},
	note = {ADS Bibcode: 2020arXiv200901514L},
}

@article{yiBetterAgeEstimates2001,
	title = {Toward {Better} {Age} {Estimates} for {Stellar} {Populations}: {The} {Y2} {Isochrones} for {Solar} {Mixture}},
	volume = {136},
	issn = {0067-0049},
	shorttitle = {Toward {Better} {Age} {Estimates} for {Stellar} {Populations}},
	url = {https://ui.adsabs.harvard.edu/abs/2001ApJS..136..417Y},
	doi = {10.1086/321795},
	urldate = {2025-07-04},
	journal = {\apjs},
	author = {Yi, Sukyoung and Demarque, Pierre and Kim, Yong-Cheol and Lee, Young-Wook and Ree, Chang H. and Lejeune, Thibault and Barnes, Sydney},
	month = oct,
	year = {2001},
	note = {Publisher: IOP
ADS Bibcode: 2001ApJS..136..417Y},
	pages = {417--437},
}

@article{pietrinferniLargeStellarEvolution2004,
	title = {A {Large} {Stellar} {Evolution} {Database} for {Population} {Synthesis} {Studies}. {I}. {Scaled} {Solar} {Models} and {Isochrones}},
	volume = {612},
	doi = {10.1086/422498},
	journal = {\apj},
	author = {Pietrinferni, A. and Cassisi, S. and Salaris, M. and Castelli, F.},
	month = sep,
	year = {2004},
	note = {\_eprint: arXiv:astro-ph/0405193},
	pages = {168--190},
}

@article{dotterACSSurveyGalactic2007,
	title = {The {ACS} {Survey} of {Galactic} {Globular} {Clusters}. {II}. {Stellar} {Evolution} {Tracks}, {Isochrones}, {Luminosity} {Functions}, and {Synthetic} {Horizontal}-{Branch} {Models}},
	volume = {134},
	issn = {0004-6256},
	url = {https://ui.adsabs.harvard.edu/abs/2007AJ....134..376D},
	doi = {10.1086/517915},
	urldate = {2025-07-04},
	journal = {\aj},
	author = {Dotter, Aaron and Chaboyer, Brian and Jevremović, Darko and Baron, E. and Ferguson, Jason W. and Sarajedini, Ata and Anderson, Jay},
	month = jul,
	year = {2007},
	note = {Publisher: IOP
ADS Bibcode: 2007AJ....134..376D},
	pages = {376--390},
}

@article{joyceINVESTIGATINGCONSISTENCYStelLAR2015,
	title = {{INVESTIGATING} {THE} {CONSISTENCY} {OF} {S}℡{LAR} {EVOLUTION} {MODELS} {WITH} {GLOBULAR} {CLUSTER} {OBSERVATIONS} {VIA} {THE} {RED} {GIANT} {BRANCH} {BUMP}},
	volume = {814},
	issn = {0004-637X},
	url = {https://dx.doi.org/10.1088/0004-637X/814/2/142},
	doi = {10.1088/0004-637X/814/2/142},
	language = {en},
	number = {2},
	urldate = {2025-07-03},
	journal = {\apj},
	author = {Joyce, M. and Chaboyer, B.},
	month = nov,
	year = {2015},
	note = {Publisher: The American Astronomical Society},
	pages = {142},
}

@article{gaiacollaborationGaiaDataRelease2023,
	title = {Gaia {Data} {Release} 3. {Summary} of the content and survey properties},
	volume = {674},
	issn = {0004-6361},
	url = {https://ui.adsabs.harvard.edu/abs/2023A&A...674A...1G},
	doi = {10.1051/0004-6361/202243940},
	urldate = {2025-07-02},
	journal = {\aap},
	author = {{Gaia Collaboration} and Vallenari, A. and Brown, A. G. A. and Prusti, T. and de Bruijne, J. H. J. and Arenou, F. and Babusiaux, C. and Biermann, M. and Creevey, O. L. and Ducourant, C. and Evans, D. W. and Eyer, L. and Guerra, R. and Hutton, A. and Jordi, C. and Klioner, S. A. and Lammers, U. L. and Lindegren, L. and Luri, X. and Mignard, F. and Panem, C. and Pourbaix, D. and Randich, S. and Sartoretti, P. and Soubiran, C. and Tanga, P. and Walton, N. A. and Bailer-Jones, C. A. L. and Bastian, U. and Drimmel, R. and Jansen, F. and Katz, D. and Lattanzi, M. G. and van Leeuwen, F. and Bakker, J. and Cacciari, C. and Castañeda, J. and De Angeli, F. and Fabricius, C. and Fouesneau, M. and Frémat, Y. and Galluccio, L. and Guerrier, A. and Heiter, U. and Masana, E. and Messineo, R. and Mowlavi, N. and Nicolas, C. and Nienartowicz, K. and Pailler, F. and Panuzzo, P. and Riclet, F. and Roux, W. and Seabroke, G. M. and Sordo, R. and Thévenin, F. and Gracia-Abril, G. and Portell, J. and Teyssier, D. and Altmann, M. and Andrae, R. and Audard, M. and Bellas-Velidis, I. and Benson, K. and Berthier, J. and Blomme, R. and Burgess, P. W. and Busonero, D. and Busso, G. and Cánovas, H. and Carry, B. and Cellino, A. and Cheek, N. and Clementini, G. and Damerdji, Y. and Davidson, M. and de Teodoro, P. and Nuñez Campos, M. and Delchambre, L. and Dell'Oro, A. and Esquej, P. and Fernández-Hernández, J. and Fraile, E. and Garabato, D. and García-Lario, P. and Gosset, E. and Haigron, R. and Halbwachs, J. -L. and Hambly, N. C. and Harrison, D. L. and Hernández, J. and Hestroffer, D. and Hodgkin, S. T. and Holl, B. and Janßen, K. and Jevardat de Fombelle, G. and Jordan, S. and Krone-Martins, A. and Lanzafame, A. C. and Löffler, W. and Marchal, O. and Marrese, P. M. and Moitinho, A. and Muinonen, K. and Osborne, P. and Pancino, E. and Pauwels, T. and Recio-Blanco, A. and Reylé, C. and Riello, M. and Rimoldini, L. and Roegiers, T. and Rybizki, J. and Sarro, L. M. and Siopis, C. and Smith, M. and Sozzetti, A. and Utrilla, E. and van Leeuwen, M. and Abbas, U. and Ábrahám, P. and Abreu Aramburu, A. and Aerts, C. and Aguado, J. J. and Ajaj, M. and Aldea-Montero, F. and Altavilla, G. and Álvarez, M. A. and Alves, J. and Anders, F. and Anderson, R. I. and Anglada Varela, E. and Antoja, T. and Baines, D. and Baker, S. G. and Balaguer-Núñez, L. and Balbinot, E. and Balog, Z. and Barache, C. and Barbato, D. and Barros, M. and Barstow, M. A. and Bartolomé, S. and Bassilana, J. -L. and Bauchet, N. and Becciani, U. and Bellazzini, M. and Berihuete, A. and Bernet, M. and Bertone, S. and Bianchi, L. and Binnenfeld, A. and Blanco-Cuaresma, S. and Blazere, A. and Boch, T. and Bombrun, A. and Bossini, D. and Bouquillon, S. and Bragaglia, A. and Bramante, L. and Breedt, E. and Bressan, A. and Brouillet, N. and Brugaletta, E. and Bucciarelli, B. and Burlacu, A. and Butkevich, A. G. and Buzzi, R. and Caffau, E. and Cancelliere, R. and Cantat-Gaudin, T. and Carballo, R. and Carlucci, T. and Carnerero, M. I. and Carrasco, J. M. and Casamiquela, L. and Castellani, M. and Castro-Ginard, A. and Chaoul, L. and Charlot, P. and Chemin, L. and Chiaramida, V. and Chiavassa, A. and Chornay, N. and Comoretto, G. and Contursi, G. and Cooper, W. J. and Cornez, T. and Cowell, S. and Crifo, F. and Cropper, M. and Crosta, M. and Crowley, C. and Dafonte, C. and Dapergolas, A. and David, M. and David, P. and de Laverny, P. and De Luise, F. and De March, R.},
	month = jun,
	year = {2023},
	note = {Publisher: EDP
ADS Bibcode: 2023A\&A...674A...1G},
	pages = {A1},
}

@article{nessMetallicityDistributionMilky2016,
	title = {The {Metallicity} {Distribution} of the {Milky} {Way} {Bulge}},
	volume = {33},
	issn = {1323-3580},
	url = {https://ui.adsabs.harvard.edu/abs/2016PASA...33...22N},
	doi = {10.1017/pasa.2015.51},
	urldate = {2025-07-02},
	journal = {Publications of the Astronomical Society of Australia},
	author = {Ness, M. and Freeman, K.},
	month = jun,
	year = {2016},
	note = {ADS Bibcode: 2016PASA...33...22N},
	pages = {e022},
}

@article{yingAbsoluteAgeMilky2025,
	title = {The {Absolute} {Age} of {Milky} {Way} {Globular} {Clusters}},
	volume = {987},
	issn = {0004-637X},
	url = {https://ui.adsabs.harvard.edu/abs/2025ApJ...987...52Y},
	doi = {10.3847/1538-4357/add471},
	urldate = {2025-07-02},
	journal = {\apj},
	author = {Ying, Jiaqi (Martin) and Chaboyer, Brian and Boylan-Kolchin, Michael and Weisz, Daniel R. and Goebel-Bain, Rowan},
	month = jul,
	year = {2025},
	note = {Publisher: IOP
ADS Bibcode: 2025ApJ...987...52Y},
	pages = {52},
}

@article{sobol1967distribution,
	title = {Distribution of points in a cube and approximate evaluation of integrals},
	volume = {7},
	language = {Russian},
	number = {4},
	journal = {Zhurnal Vychislitel'noi Matematiki i Matematicheskoi Fiziki},
	author = {Sobol', Ilya M.},
	year = {1967},
	pages = {784--802},
}

@misc{goodfellowGenerativeAdversarialNetworks2014,
	title = {Generative {Adversarial} {Networks}},
	url = {https://ui.adsabs.harvard.edu/abs/2014arXiv1406.2661G},
	doi = {10.48550/arXiv.1406.2661},
	urldate = {2025-07-01},
	publisher = {arXiv},
	author = {Goodfellow, Ian J. and Pouget-Abadie, Jean and Mirza, Mehdi and Xu, Bing and Warde-Farley, David and Ozair, Sherjil and Courville, Aaron and Bengio, Yoshua},
	month = jun,
	year = {2014},
	note = {ADS Bibcode: 2014arXiv1406.2661G},
}

@misc{kingmaIntroductionVariationalAutoencoders2019,
	title = {An {Introduction} to {Variational} {Autoencoders}},
	url = {https://ui.adsabs.harvard.edu/abs/2019arXiv190602691K},
	doi = {10.48550/arXiv.1906.02691},
	urldate = {2025-07-01},
	publisher = {arXiv},
	author = {Kingma, Diederik P. and Welling, Max},
	month = jun,
	year = {2019},
	note = {ADS Bibcode: 2019arXiv190602691K},
}

@misc{dinhDensityEstimationUsing2016,
	title = {Density estimation using {Real} {NVP}},
	url = {https://ui.adsabs.harvard.edu/abs/2016arXiv160508803D},
	doi = {10.48550/arXiv.1605.08803},
	urldate = {2025-07-01},
	publisher = {arXiv},
	author = {Dinh, Laurent and Sohl-Dickstein, Jascha and Bengio, Samy},
	month = may,
	year = {2016},
	note = {ADS Bibcode: 2016arXiv160508803D},
}

@misc{kingmaImprovingVariationalInference2016,
	title = {Improving {Variational} {Inference} with {Inverse} {Autoregressive} {Flow}},
	url = {https://ui.adsabs.harvard.edu/abs/2016arXiv160604934K},
	doi = {10.48550/arXiv.1606.04934},
	urldate = {2025-07-01},
	publisher = {arXiv},
	author = {Kingma, Diederik P. and Salimans, Tim and Jozefowicz, Rafal and Chen, Xi and Sutskever, Ilya and Welling, Max},
	month = jun,
	year = {2016},
	note = {ADS Bibcode: 2016arXiv160604934K},
}

@misc{kingmaGlowGenerativeFlow2018,
	title = {Glow: {Generative} {Flow} with {Invertible} 1x1 {Convolutions}},
	shorttitle = {Glow},
	url = {http://arxiv.org/abs/1807.03039},
	doi = {10.48550/arXiv.1807.03039},
	urldate = {2025-07-01},
	publisher = {arXiv},
	author = {Kingma, Diederik P. and Dhariwal, Prafulla},
	month = jul,
	year = {2018},
	note = {arXiv:1807.03039 [stat]},
}

@misc{dinhNICENonlinearIndependent2014,
	title = {{NICE}: {Non}-linear {Independent} {Components} {Estimation}},
	shorttitle = {{NICE}},
	url = {https://ui.adsabs.harvard.edu/abs/2014arXiv1410.8516D},
	doi = {10.48550/arXiv.1410.8516},
	urldate = {2025-07-01},
	publisher = {arXiv},
	author = {Dinh, Laurent and Krueger, David and Bengio, Yoshua},
	month = oct,
	year = {2014},
	note = {ADS Bibcode: 2014arXiv1410.8516D},
}

@misc{papamakariosMaskedAutoregressiveFlow2017,
	title = {Masked {Autoregressive} {Flow} for {Density} {Estimation}},
	url = {https://ui.adsabs.harvard.edu/abs/2017arXiv170507057P},
	doi = {10.48550/arXiv.1705.07057},
	urldate = {2025-07-01},
	publisher = {arXiv},
	author = {Papamakarios, George and Pavlakou, Theo and Murray, Iain},
	month = may,
	year = {2017},
	note = {ADS Bibcode: 2017arXiv170507057P},
}

@misc{durkanNeuralSplineFlows2019,
	title = {Neural {Spline} {Flows}},
	url = {https://ui.adsabs.harvard.edu/abs/2019arXiv190604032D},
	doi = {10.48550/arXiv.1906.04032},
	urldate = {2025-07-01},
	publisher = {arXiv},
	author = {Durkan, Conor and Bekasov, Artur and Murray, Iain and Papamakarios, George},
	month = jun,
	year = {2019},
	note = {ADS Bibcode: 2019arXiv190604032D},
}

@article{bellingerFundamentalParametersMainSequence2016,
	title = {Fundamental {Parameters} of {Main}-{Sequence} {Stars} in an {Instant} with {Machine} {Learning}},
	volume = {830},
	issn = {0004-637X},
	url = {https://ui.adsabs.harvard.edu/abs/2016ApJ...830...31B},
	doi = {10.3847/0004-637X/830/1/31},
	urldate = {2025-06-30},
	journal = {\apj},
	author = {Bellinger, Earl P. and Angelou, George C. and Hekker, Saskia and Basu, Sarbani and Ball, Warrick H. and Guggenberger, Elisabeth},
	month = oct,
	year = {2016},
	note = {Publisher: IOP
ADS Bibcode: 2016ApJ...830...31B},
	pages = {31},
}

@article{lyttleHierarchicallyModellingKepler2021,
	title = {Hierarchically modelling {Kepler} dwarfs and subgiants to improve inference of stellar properties with asteroseismology},
	volume = {505},
	issn = {0035-8711},
	url = {https://doi.org/10.1093/mnras/stab1368},
	doi = {10.1093/mnras/stab1368},
	number = {2},
	urldate = {2025-06-29},
	journal = {\mnras},
	author = {Lyttle, Alexander J and Davies, Guy R and Li, Tanda and Carboneau, Lindsey M and Leung, Ho-Hin and Westwood, Harry and Chaplin, William J and Hall, Oliver J and Huber, Daniel and Nielsen, Martin B and Basu, Sarbani and García, Rafael A},
	month = aug,
	year = {2021},
	pages = {2427--2446},
}

@misc{van-laneNovelApplicationConditional2023,
	title = {A {Novel} {Application} of {Conditional} {Normalizing} {Flows}: {Stellar} {Age} {Inference} with {Gyrochronology}},
	shorttitle = {A {Novel} {Application} of {Conditional} {Normalizing} {Flows}},
	url = {https://ui.adsabs.harvard.edu/abs/2023arXiv230708753V},
	doi = {10.48550/arXiv.2307.08753},
	urldate = {2025-06-29},
	publisher = {arXiv},
	author = {Van-Lane, Phil and Speagle, Joshua S. and Douglas, Stephanie},
	month = jul,
	year = {2023},
	note = {ADS Bibcode: 2023arXiv230708753V},
}

@article{vermaAsteroseismicDeterminationFundamental2016,
	title = {Asteroseismic determination of fundamental parameters of {Sun}-like stars using multilayered neural networks},
	volume = {461},
	issn = {0035-8711},
	url = {https://ui.adsabs.harvard.edu/abs/2016MNRAS.461.4206V},
	doi = {10.1093/mnras/stw1621},
	urldate = {2025-06-29},
	journal = {\mnras},
	author = {Verma, Kuldeep and Hanasoge, Shravan and Bhattacharya, Jishnu and Antia, H. M. and Krishnamurthi, Ganapathy},
	month = oct,
	year = {2016},
	note = {Publisher: OUP
ADS Bibcode: 2016MNRAS.461.4206V},
	pages = {4206--4214},
}

@article{honAsteroseismicInferenceSubgiant2020,
	title = {Asteroseismic inference of subgiant evolutionary parameters with deep learning},
	volume = {499},
	issn = {0035-8711},
	url = {https://doi.org/10.1093/mnras/staa2853},
	doi = {10.1093/mnras/staa2853},
	number = {2},
	urldate = {2025-06-29},
	journal = {\mnras},
	author = {Hon, Marc and Bellinger, Earl P and Hekker, Saskia and Stello, Dennis and Kuszlewicz, James S},
	month = oct,
	year = {2020},
	pages = {2445--2461},
}

@article{garraffoStelNetHierarchicalNeural2021,
	title = {{StelNet}: {Hierarchical} {Neural} {Network} for {Automatic} {Inference} in {Stellar} {Characterization}},
	volume = {162},
	issn = {1538-3881},
	shorttitle = {{StelNet}},
	url = {https://dx.doi.org/10.3847/1538-3881/ac0ef0},
	doi = {10.3847/1538-3881/ac0ef0},
	language = {en},
	number = {4},
	urldate = {2025-06-29},
	journal = {\aj},
	author = {Garraffo, Cecilia and Protopapas, Pavlos and Drake, Jeremy J. and Becker, Ignacio and Cargile, Phillip},
	month = sep,
	year = {2021},
	note = {Publisher: The American Astronomical Society},
	pages = {157},
}

@article{kibediRadiativeWidthHoyle2020,
	title = {Radiative {Width} of the {Hoyle} {State} from γ -{Ray} {Spectroscopy}},
	volume = {125},
	issn = {0031-9007},
	url = {https://ui.adsabs.harvard.edu/abs/2020PhRvL.125r2701K},
	doi = {10.1103/PhysRevLett.125.182701},
	urldate = {2025-06-20},
	journal = {\prl},
	author = {Kibédi, T. and Alshahrani, B. and Stuchbery, A. E. and Larsen, A. C. and Görgen, A. and Siem, S. and Guttormsen, M. and Giacoppo, F. and Morales, A. I. and Sahin, E. and Tveten, G. M. and Garrote, F. L. Bello and Campo, L. Crespo and Eriksen, T. K. and Klintefjord, M. and Maharramova, S. and Nyhus, H. -T. and Tornyi, T. G. and Renstrøm, T. and Paulsen, W.},
	month = oct,
	year = {2020},
	note = {Publisher: APS
ADS Bibcode: 2020PhRvL.125r2701K},
	pages = {182701},
}

@article{pepperImpactUncertainties12Ca2022,
	title = {The impact of the uncertainties in the {12C}(α, γ){16O} reaction rate on the evolution of low- to intermediate-mass stars},
	volume = {513},
	issn = {0035-8711},
	url = {https://ui.adsabs.harvard.edu/abs/2022MNRAS.513.1499P},
	doi = {10.1093/mnras/stac1016},
	urldate = {2025-06-20},
	journal = {\mnras},
	author = {Pepper, Ben T. and Istrate, A. G. and Romero, A. D. and Kepler, S. O.},
	month = jun,
	year = {2022},
	note = {Publisher: OUP
ADS Bibcode: 2022MNRAS.513.1499P},
	pages = {1499--1512},
}

@article{togniniImpactUncertainties3a2023,
	title = {Impact of the uncertainties of 3α and {12C}(α, γ){16O} reactions on the {He}-burning phases of low- and intermediate-mass stars},
	volume = {679},
	copyright = {© The Authors 2023},
	issn = {0004-6361, 1432-0746},
	url = {https://www.aanda.org/articles/aa/abs/2023/11/aa46382-23/aa46382-23.html},
	doi = {10.1051/0004-6361/202346382},
	language = {en},
	urldate = {2025-06-20},
	journal = {\aap},
	author = {Tognini, F. and Valle, G. and Dell’Omodarme, M. and Degl’Innocenti, S. and Moroni, P. G. Prada},
	month = nov,
	year = {2023},
	note = {Publisher: EDP Sciences},
	pages = {A75},
}

@article{nguyenPARSECV20Stellar2022,
	title = {{PARSEC} {V2}.0: {Stellar} tracks and isochrones of low- and intermediate-mass stars with rotation},
	volume = {665},
	issn = {0004-6361},
	shorttitle = {{PARSEC} {V2}.0},
	url = {https://ui.adsabs.harvard.edu/abs/2022A&A...665A.126N},
	doi = {10.1051/0004-6361/202244166},
	urldate = {2025-06-19},
	journal = {\aap},
	author = {Nguyen, C. T. and Costa, G. and Girardi, L. and Volpato, G. and Bressan, A. and Chen, Y. and Marigo, P. and Fu, X. and Goudfrooij, P.},
	month = sep,
	year = {2022},
	note = {ADS Bibcode: 2022A\&A...665A.126N},
	pages = {A126},
}

@article{braatenNeutrinoEnergyLoss1993,
	title = {Neutrino energy loss from the plasma process at all temperatures and densities},
	volume = {48},
	issn = {1550-79980556-2821},
	url = {https://ui.adsabs.harvard.edu/abs/1993PhRvD..48.1478B},
	doi = {10.1103/PhysRevD.48.1478},
	urldate = {2025-06-19},
	journal = {\prd},
	author = {Braaten, Eric and Segel, Daniel},
	month = aug,
	year = {1993},
	note = {Publisher: APS
ADS Bibcode: 1993PhRvD..48.1478B},
	pages = {1478--1491},
}

@article{itohNeutrinoEnergyLoss1992,
	title = {Neutrino {Energy} {Loss} in {Stellar} {Interiors}. {IV}. {Plasma} {Neutrino} {Process} for {Strongly} {Degenerate} {Electrons}},
	volume = {395},
	issn = {0004-637X},
	url = {https://ui.adsabs.harvard.edu/abs/1992ApJ...395..622I},
	doi = {10.1086/171682},
	urldate = {2025-06-19},
	journal = {\apj},
	author = {Itoh, Naoki and Mutoh, Haruhiko and Hikita, Atsushi and Kohyama, Yasuharu},
	month = aug,
	year = {1992},
	note = {Publisher: IOP
ADS Bibcode: 1992ApJ...395..622I},
	pages = {622},
}

@article{braatenEmissivityHotPlasma1991,
	title = {Emissivity of a hot plasma from photon and plasmon decay},
	volume = {66},
	url = {https://link.aps.org/doi/10.1103/PhysRevLett.66.1655},
	doi = {10.1103/PhysRevLett.66.1655},
	number = {13},
	urldate = {2025-06-19},
	journal = {\prl},
	author = {Braaten, Eric},
	month = apr,
	year = {1991},
	note = {Publisher: American Physical Society},
	pages = {1655--1658},
}

@article{bressanPARSECStellarTracks2012,
	title = {{PARSEC}: stellar tracks and isochrones with the {PAdova} and {TRieste} {Stellar} {Evolution} {Code}},
	volume = {427},
	issn = {0035-8711},
	shorttitle = {{PARSEC}},
	url = {https://doi.org/10.1111/j.1365-2966.2012.21948.x},
	doi = {10.1111/j.1365-2966.2012.21948.x},
	number = {1},
	urldate = {2024-10-09},
	journal = {\mnras},
	author = {Bressan, Alessandro and Marigo, Paola and Girardi, Léo and Salasnich, Bernardo and Dal Cero, Claudia and Rubele, Stefano and Nanni, Ambra},
	month = nov,
	year = {2012},
	pages = {127--145},
}

@article{deboer12Ca16OReaction2017,
	title = {The {12C}(α ,γ ){16O} reaction and its implications for stellar helium burning},
	volume = {89},
	issn = {0034-6861},
	url = {https://ui.adsabs.harvard.edu/abs/2017RvMP...89c5007D},
	doi = {10.1103/RevModPhys.89.035007},
	urldate = {2025-06-18},
	journal = {Reviews of Modern Physics},
	author = {deBoer, R. J. and Görres, J. and Wiescher, M. and Azuma, R. E. and Best, A. and Brune, C. R. and Fields, C. E. and Jones, S. and Pignatari, M. and Sayre, D. and Smith, K. and Timmes, F. X. and Uberseder, E.},
	month = jul,
	year = {2017},
	note = {Publisher: APS
ADS Bibcode: 2017RvMP...89c5007D},
	pages = {035007},
}

@article{cassisiUpdatedElectronConductionOpacities2007,
	title = {Updated {Electron}-{Conduction} {Opacities}: {The} {Impact} on {Low}-{Mass} {Stellar} {Models}},
	volume = {661},
	issn = {0004-637X},
	shorttitle = {Updated {Electron}-{Conduction} {Opacities}},
	url = {https://ui.adsabs.harvard.edu/abs/2007ApJ...661.1094C},
	doi = {10.1086/516819},
	urldate = {2025-03-11},
	journal = {\apj},
	author = {Cassisi, S. and Potekhin, A. Y. and Pietrinferni, A. and Catelan, M. and Salaris, M.},
	month = jun,
	year = {2007},
	note = {Publisher: IOP
ADS Bibcode: 2007ApJ...661.1094C},
	pages = {1094--1104},
}

@article{cassisiElectronConductionOpacities2021,
	title = {Electron conduction opacities at the transition between moderate and strong degeneracy: {Uncertainties} and impacts on stellar models},
	volume = {654},
	issn = {0004-6361},
	shorttitle = {Electron conduction opacities at the transition between moderate and strong degeneracy},
	url = {https://ui.adsabs.harvard.edu/abs/2021A&A...654A.149C},
	doi = {10.1051/0004-6361/202141425},
	urldate = {2025-06-18},
	journal = {\aap},
	author = {Cassisi, Santi and Potekhin, Alexander Y. and Salaris, Maurizio and Pietrinferni, Adriano},
	month = oct,
	year = {2021},
	note = {ADS Bibcode: 2021A\&A...654A.149C},
	pages = {A149},
}

@article{zhouCoupling1DStellar2025,
	title = {Coupling {1D} stellar evolution with {3D}-hydrodynamical simulations on-the-fly {III}: stellar evolution at different metallicities},
	issn = {0035-8711},
	shorttitle = {Coupling {1D} stellar evolution with {3D}-hydrodynamical simulations on-the-fly {III}},
	url = {https://doi.org/10.1093/mnras/staf937},
	doi = {10.1093/mnras/staf937},
	urldate = {2025-06-18},
	journal = {\mnras},
	author = {Zhou, Yixiao and Rørsted, Jakob L and Weiss, Achim and Jørgensen, Andreas C S and Lagae, Cis and Díaz, Luisa F Rodríguez and Li, Yaguang and Winther, Mark L and Larsen, Jens R and Christensen-Dalsgaard, Jørgen and Kochukhov, Oleg and Pollard, Karen R and Li, Tanda},
	month = jun,
	year = {2025},
	pages = {staf937},
}

@article{rizzuti3DStellarEvolution2023,
	title = {{3D} stellar evolution: hydrodynamic simulations of a complete burning phase in a massive star},
	volume = {523},
	issn = {0035-8711},
	shorttitle = {{3D} stellar evolution},
	url = {https://doi.org/10.1093/mnras/stad1572},
	doi = {10.1093/mnras/stad1572},
	number = {2},
	urldate = {2025-06-18},
	journal = {\mnras},
	author = {Rizzuti, F and Hirschi, R and Arnett, W D and Georgy, C and Meakin, C and Murphy, A StJ and Rauscher, T and Varma, V},
	month = aug,
	year = {2023},
	pages = {2317--2328},
}

@article{chaboyerHeavyElementDiffusionMetalpoor2001,
	title = {Heavy-{Element} {Diffusion} in {Metal}-poor {Stars}},
	volume = {562},
	issn = {0004-637X},
	url = {https://iopscience.iop.org/article/10.1086/323872/meta},
	doi = {10.1086/323872},
	language = {en},
	number = {1},
	urldate = {2025-06-18},
	journal = {\apj},
	author = {Chaboyer, Brian and Fenton, W. H. and Nelan, Jenica E. and Patnaude, D. J. and Simon, Francesca E.},
	month = nov,
	year = {2001},
	note = {Publisher: IOP Publishing},
	pages = {521},
}

@article{asplundChemicalCompositionSun2009,
	title = {The {Chemical} {Composition} of the {Sun}},
	volume = {47},
	issn = {0066-4146, 1545-4282},
	url = {https://www.annualreviews.org/content/journals/10.1146/annurev.astro.46.060407.145222},
	doi = {10.1146/annurev.astro.46.060407.145222},
	language = {en},
	number = {Volume 47, 2009},
	urldate = {2025-06-17},
	journal = {\araa},
	author = {Asplund, Martin and Grevesse, Nicolas and Sauval, A. Jacques and Scott, Pat},
	month = sep,
	year = {2009},
	note = {Publisher: Annual Reviews},
	pages = {481--522},
}

@article{colganNEWGENERATIONALAMOS2016,
	title = {A {NEW} {GENERATION} {OF} {LOS} {ALAMOS} {OPACITY} {TABLES}},
	volume = {817},
	issn = {0004-637X},
	url = {https://dx.doi.org/10.3847/0004-637X/817/2/116},
	doi = {10.3847/0004-637X/817/2/116},
	language = {en},
	number = {2},
	urldate = {2025-06-17},
	journal = {\apj},
	author = {Colgan, J. and Kilcrease, D. P. and Magee, N. H. and Sherrill, M. E. and Abdallah Jr., J. and Hakel, P. and Fontes, C. J. and Guzik, J. A. and Mussack, K. A.},
	month = jan,
	year = {2016},
	note = {Publisher: The American Astronomical Society},
	pages = {116},
}

@article{faragExpandedSetAlamos2024,
	title = {An {Expanded} {Set} of {Los} {Alamos} {OPLIB} {Tables} in {MESA}: {Type}-1 {Rosseland}-mean {Opacities} and {Solar} {Models}},
	volume = {968},
	issn = {0004-637X},
	shorttitle = {An {Expanded} {Set} of {Los} {Alamos} {OPLIB} {Tables} in {MESA}},
	url = {https://ui.adsabs.harvard.edu/abs/2024ApJ...968...56F},
	doi = {10.3847/1538-4357/ad4355},
	urldate = {2025-06-17},
	journal = {\apj},
	author = {Farag, Ebraheem and Fontes, Christopher J. and Timmes, F. X. and Bellinger, Earl P. and Guzik, Joyce A. and Bauer, Evan B. and Wood, Suzannah R. and Mussack, Katie and Hakel, Peter and Colgan, James and Kilcrease, David P. and Sherrill, Manolo E. and Raecke, Tryston C. and Chidester, Morgan T.},
	month = jun,
	year = {2024},
	note = {Publisher: IOP
ADS Bibcode: 2024ApJ...968...56F},
	pages = {56},
}

@article{seatonOpacitiesStellarEnvelopes1994,
	title = {Opacities for stellar envelopes},
	volume = {266},
	issn = {0035-8711},
	url = {https://doi.org/10.1093/mnras/266.4.805},
	doi = {10.1093/mnras/266.4.805},
	number = {4},
	urldate = {2025-06-17},
	journal = {\mnras},
	author = {Seaton, M. J. and Yan, Yu and Mihalas, D. and Pradhan, Anil K.},
	month = feb,
	year = {1994},
	pages = {805--828},
}

@article{alexanderLowTemperatureRosselandOpacities1994,
	title = {Low-{Temperature} {Rosseland} {Opacities}},
	volume = {437},
	issn = {0004-637X},
	url = {https://ui.adsabs.harvard.edu/abs/1994ApJ...437..879A},
	doi = {10.1086/175039},
	urldate = {2025-06-17},
	journal = {\apj},
	author = {Alexander, D. R. and Ferguson, J. W.},
	month = dec,
	year = {1994},
	note = {Publisher: IOP
ADS Bibcode: 1994ApJ...437..879A},
	pages = {879},
}

@book{jorgensenMoleculesStellarEnvironment1994,
	title = {Molecules in the {Stellar} {Environment}},
	volume = {428},
	url = {https://ui.adsabs.harvard.edu/abs/1994LNP...428.....J},
	urldate = {2025-06-17},
	author = {Jorgensen, Uffe G.},
	month = jan,
	year = {1994},
	doi = {10.1007/3-540-57747-5},
	note = {Publication Title: IAU Colloq. 146: Molecules in the Stellar Environment
ADS Bibcode: 1994LNP...428.....J},
}

@article{blouinNewConductiveOpacities2020,
	title = {New {Conductive} {Opacities} for {White} {Dwarf} {Envelopes}},
	volume = {899},
	issn = {0004-637X},
	url = {https://ui.adsabs.harvard.edu/abs/2020ApJ...899...46B},
	doi = {10.3847/1538-4357/ab9e75},
	urldate = {2025-03-11},
	journal = {\apj},
	author = {Blouin, Simon and Shaffer, Nathaniel R. and Saumon, Didier and Starrett, Charles E.},
	month = aug,
	year = {2020},
	note = {Publisher: IOP
ADS Bibcode: 2020ApJ...899...46B},
	pages = {46},
}

@article{kunzAstrophysicalReactionRate2002,
	title = {Astrophysical {Reaction} {Rate} of {12C}(α, γ){16O}},
	volume = {567},
	issn = {0004-637X},
	url = {https://iopscience.iop.org/article/10.1086/338384/meta},
	doi = {10.1086/338384},
	language = {en},
	number = {1},
	urldate = {2025-06-16},
	journal = {\apj},
	author = {Kunz, R. and Fey, M. and Jaeger, M. and Mayer, A. and Hammer, J. W. and Staudt, G. and Harissopulos, S. and Paradellis, T.},
	month = mar,
	year = {2002},
	note = {Publisher: IOP Publishing},
	pages = {643},
}

@article{sunoPreciseCalculationTripleAlpha2016,
	title = {Precise calculation of the triple-\${\textbackslash}ensuremath\{{\textbackslash}alpha\}\$ reaction rates using the transmission-free complex absorbing potential method},
	volume = {94},
	url = {https://link.aps.org/doi/10.1103/PhysRevC.94.054607},
	doi = {10.1103/PhysRevC.94.054607},
	number = {5},
	urldate = {2025-06-16},
	journal = {\prc},
	author = {Suno, Hiroya and Suzuki, Yasuyuki and Descouvemont, Pierre},
	month = nov,
	year = {2016},
	note = {Publisher: American Physical Society},
	pages = {054607},
}

@article{anguloCompilationChargedparticleInduced1999,
	title = {A compilation of charged-particle induced thermonuclear reaction rates},
	volume = {656},
	issn = {0375-9474},
	url = {https://ui.adsabs.harvard.edu/abs/1999NuPhA.656....3A},
	doi = {10.1016/S0375-9474(99)00030-5},
	urldate = {2025-06-16},
	journal = {Nuclear Physics A},
	author = {Angulo, C. and Arnould, M. and Rayet, M. and Descouvemont, P. and Baye, D. and Leclercq-Willain, C. and Coc, A. and Barhoumi, S. and Aguer, P. and Rolfs, C. and Kunz, R. and Hammer, J. W. and Mayer, A. and Paradellis, T. and Kossionides, S. and Chronidou, C. and Spyrou, K. and degl'Innocenti, S. and Fiorentini, G. and Ricci, B. and Zavatarelli, S. and Providencia, C. and Wolters, H. and Soares, J. and Grama, C. and Rahighi, J. and Shotter, A. and Lamehi Rachti, M.},
	month = aug,
	year = {1999},
	note = {ADS Bibcode: 1999NuPhA.656....3A},
	pages = {3--183},
}

@book{collinsFundamentalsStellarAstrophysics1989,
	title = {The fundamentals of stellar astrophysics},
	url = {https://ui.adsabs.harvard.edu/abs/1989fsa..book.....C},
	urldate = {2025-06-16},
	author = {Collins, II, George W.},
	month = jan,
	year = {1989},
	note = {Publication Title: The fundamentals of stellar astrophysics
ADS Bibcode: 1989fsa..book.....C},
}

@article{eggenObservationalAspectsStellar1965,
	title = {Some {Observational} {Aspects} of {Stellar} {Evolution}},
	volume = {3},
	issn = {0066-4146, 1545-4282},
	url = {https://www.annualreviews.org/content/journals/10.1146/annurev.aa.03.090165.001315},
	doi = {10.1146/annurev.aa.03.090165.001315},
	language = {en},
	number = {Volume 3, 1965},
	urldate = {2025-06-12},
	journal = {\araa},
	author = {Eggen, Olin J.},
	month = sep,
	year = {1965},
	note = {Publisher: Annual Reviews},
	pages = {235--274},
}

@article{boudreauxUpdatedHightemperatureOpacities2023,
	title = {Updated {High}-temperature {Opacities} for the {Dartmouth} {Stellar} {Evolution} {Program} and {Their} {Effect} on the {Jao} {Gap} {Location}},
	volume = {944},
	issn = {0004-637X},
	url = {https://ui.adsabs.harvard.edu/abs/2023ApJ...944..129B},
	doi = {10.3847/1538-4357/acb685},
	urldate = {2025-03-12},
	journal = {\apj},
	author = {Boudreaux, Emily M. and Chaboyer, Brian C.},
	month = feb,
	year = {2023},
	note = {Publisher: IOP
ADS Bibcode: 2023ApJ...944..129B},
	pages = {129},
}

@article{joyceReviewMixingLength2023,
	title = {A {Review} of the {Mixing} {Length} {Theory} of {Convection} in {1D} {Stellar} {Modeling}},
	volume = {11},
	url = {https://ui.adsabs.harvard.edu/abs/2023Galax..11...75J},
	doi = {10.3390/galaxies11030075},
	urldate = {2023-10-12},
	journal = {Galaxies},
	author = {Joyce, Meridith and Tayar, Jamie},
	month = jun,
	year = {2023},
	note = {ADS Bibcode: 2023Galax..11...75J},
	pages = {75},
}

@article{averEffectsHeTextbackslashuplambda108302015,
	title = {The effects of {He} {I} \{\vphantom{\}}{\textbackslash}textbackslashuplambda{\textbackslash}10830 on helium abundance determinations},
	volume = {2015},
	issn = {1475-7516},
	url = {https://doi.org/10.1088/1475-7516/2015/07/011},
	doi = {10.1088/1475-7516/2015/07/011},
	language = {en},
	number = {07},
	urldate = {2022-03-23},
	journal = {\jcap},
	author = {Aver, Erik and Olive, Keith A. and Skillman, Evan D.},
	month = jul,
	year = {2015},
	pages = {011--011},
}

@article{dotterDartmouthStellarEvolution2008,
	title = {The {Dartmouth} {Stellar} {Evolution} {Database}},
	volume = {178},
	issn = {0067-0049},
	url = {https://ui.adsabs.harvard.edu/abs/2008ApJS..178...89D},
	doi = {10.1086/589654},
	urldate = {2022-03-08},
	journal = {\apjs},
	author = {Dotter, Aaron and Chaboyer, Brian and Jevremović, Darko and Kostov, Veselin and Baron, E. and Ferguson, Jason W.},
	year = {2008},
	pages = {89--101},
}

@article{casagrandeHeliumAbundanceDY2007,
	title = {The helium abundance and Δ{Y}/Δ{Z} in lower main-sequence stars},
	volume = {382},
	issn = {0035-8711},
	url = {https://doi.org/10.1111/j.1365-2966.2007.12512.x},
	doi = {10.1111/j.1365-2966.2007.12512.x},
	number = {4},
	journal = {\mnras},
	author = {Casagrande, Luca and Flynn, Chris and Portinari, Laura and Girardi, Leo and Jimenez, Raul},
	month = dec,
	year = {2007},
	note = {\_eprint: https://academic.oup.com/mnras/article-pdf/382/4/1516/3944799/mnras0382-1516.pdf},
	pages = {1516--1540},
}

@article{peimbertPrimordialHeliumAbundance2016,
	title = {The primordial helium abundance and the number of neutrino families},
	volume = {52},
	issn = {0185-1101},
	url = {https://ui.adsabs.harvard.edu/abs/2016RMxAA..52..419P},
	doi = {10.48550/arXiv.1608.02062},
	urldate = {2023-11-14},
	journal = {Revista Mexicana de Astronomia y Astrofisica},
	author = {Peimbert, A. and Peimbert, M. and Luridiana, V.},
	month = oct,
	year = {2016},
	note = {ADS Bibcode: 2016RMxAA..52..419P},
	pages = {419},
}

@article{hauschildtNextGenModelAtmosphere1999,
	title = {The {NextGen} {Model} {Atmosphere} {Grid} for 3000{\textbackslash}textless={Teff}{\textbackslash}textless=10,000 {K}},
	volume = {512},
	issn = {0004-637X},
	url = {https://ui.adsabs.harvard.edu/abs/1999ApJ...512..377H},
	doi = {10.1086/306745},
	urldate = {2022-03-23},
	journal = {\apj},
	author = {Hauschildt, Peter H. and Allard, France and Baron, E.},
	month = feb,
	year = {1999},
	pages = {377--385},
}

@article{martaN14pensuremathgammaO15ReactionStudied2011,
	title = {The {N14}(p,{\textbackslash}ensuremath{\textbackslash}gamma){O15} reaction studied with a composite germanium detector},
	volume = {83},
	doi = {10.1103/PhysRevC.83.045804},
	number = {4},
	journal = {\prc},
	author = {Marta, M. and Formicola, A. and Bemmerer, D. and Broggini, C. and Caciolli, A. and Corvisiero, P. and Costantini, H. and Elekes, Z. and Fülöp, Zs. and Gervino, G. and Guglielmetti, A. and Gustavino, C. and Gyürky, Gy. and Imbriani, G. and Junker, M. and Lemut, A. and Limata, B. and Mazzocchi, C. and Menegazzo, R. and Prati, P. and Roca, V. and Rolfs, C. and Rossi Alvarez, C. and Somorjai, E. and Straniero, O. and Strieder, F. and Terrasi, F. and Trautvetter, H. P. and Vomiero, A.},
	month = apr,
	year = {2011},
	note = {\_eprint: 1103.5393},
	pages = {045804},
}

@article{acharyaUncertaintyQuantificationProtonproton2016,
	title = {Uncertainty quantification for proton-proton fusion in chiral effective field theory},
	volume = {760},
	doi = {10.1016/j.physletb.2016.07.032},
	journal = {Physics Letters B},
	author = {Acharya, B. and Carlsson, B. D. and Ekström, A. and Forssén, C. and Platter, L.},
	month = sep,
	year = {2016},
	note = {\_eprint: 1603.01593},
	pages = {584--589},
}

@article{hubbardThermalConductionElectrons1969,
	title = {Thermal {Conduction} by {Electrons} in {Stellar} {Matter}},
	volume = {18},
	issn = {0067-0049},
	url = {https://ui.adsabs.harvard.edu/abs/1969ApJS...18..297H},
	doi = {10.1086/190192},
	urldate = {2022-03-23},
	journal = {\apjs},
	author = {Hubbard, W. B. and Lampe, Martin},
	month = jul,
	year = {1969},
	pages = {297},
}

@article{demarqueY2IsochronesImproved2004,
	title = {Y2 {Isochrones} with an {Improved} {Core} {Overshoot} {Treatment}},
	volume = {155},
	issn = {0067-0049},
	url = {https://ui.adsabs.harvard.edu/abs/2004ApJS..155..667D},
	doi = {10.1086/424966},
	urldate = {2023-03-16},
	journal = {\apjs},
	author = {Demarque, Pierre and Woo, Jong-Hak and Kim, Yong-Cheol and Yi, Sukyoung K.},
	month = dec,
	year = {2004},
	pages = {667--674},
}

@article{iglesiasUpdatedOpalOpacities1996,
	title = {Updated {Opal} {Opacities}},
	volume = {464},
	issn = {0004-637X},
	url = {https://ui.adsabs.harvard.edu/abs/1996ApJ...464..943I},
	doi = {10.1086/177381},
	urldate = {2022-03-23},
	journal = {\apj},
	author = {Iglesias, Carlos A. and Rogers, Forrest J.},
	month = jun,
	year = {1996},
	pages = {943},
}

@article{planckcollaborationPlanck2018Results2020,
	title = {Planck 2018 results. {VI}. {Cosmological} parameters},
	volume = {641},
	doi = {10.1051/0004-6361/201833910},
	journal = {åp},
	author = {{Planck Collaboration} and Aghanim, N. and Akrami, Y. and Ashdown, M. and Aumont, J. and Baccigalupi, C. and Ballardini, M. and Banday, A. J. and Barreiro, R. B. and Bartolo, N. and Basak, S. and Battye, R. and Benabed, K. and Bernard, J. -P. and Bersanelli, M. and Bielewicz, P. and Bock, J. J. and Bond, J. R. and Borrill, J. and Bouchet, F. R. and Boulanger, F. and Bucher, M. and Burigana, C. and Butler, R. C. and Calabrese, E. and Cardoso, J. -F. and Carron, J. and Challinor, A. and Chiang, H. C. and Chluba, J. and Colombo, L. P. L. and Combet, C. and Contreras, D. and Crill, B. P. and Cuttaia, F. and de Bernardis, P. and de Zotti, G. and Delabrouille, J. and Delouis, J. -M. and Di Valentino, E. and Diego, J. M. and Doré, O. and Douspis, M. and Ducout, A. and Dupac, X. and Dusini, S. and Efstathiou, G. and Elsner, F. and Enßlin, T. A. and Eriksen, H. K. and Fantaye, Y. and Farhang, M. and Fergusson, J. and Fernandez-Cobos, R. and Finelli, F. and Forastieri, F. and Frailis, M. and Fraisse, A. A. and Franceschi, E. and Frolov, A. and Galeotta, S. and Galli, S. and Ganga, K. and Génova-Santos, R. T. and Gerbino, M. and Ghosh, T. and González-Nuevo, J. and Górski, K. M. and Gratton, S. and Gruppuso, A. and Gudmundsson, J. E. and Hamann, J. and Handley, W. and Hansen, F. K. and Herranz, D. and Hildebrandt, S. R. and Hivon, E. and Huang, Z. and Jaffe, A. H. and Jones, W. C. and Karakci, A. and Keihänen, E. and Keskitalo, R. and Kiiveri, K. and Kim, J. and Kisner, T. S. and Knox, L. and Krachmalnicoff, N. and Kunz, M. and Kurki-Suonio, H. and Lagache, G. and Lamarre, J. -M. and Lasenby, A. and Lattanzi, M. and Lawrence, C. R. and Le Jeune, M. and Lemos, P. and Lesgourgues, J. and Levrier, F. and Lewis, A. and Liguori, M. and Lilje, P. B. and Lilley, M. and Lindholm, V. and López-Caniego, M. and Lubin, P. M. and Ma, Y. -Z. and Macías-Pérez, J. F. and Maggio, G. and Maino, D. and Mandolesi, N. and Mangilli, A. and Marcos-Caballero, A. and Maris, M. and Martin, P. G. and Martinelli, M. and Martínez-González, E. and Matarrese, S. and Mauri, N. and McEwen, J. D. and Meinhold, P. R. and Melchiorri, A. and Mennella, A. and Migliaccio, M. and Millea, M. and Mitra, S. and Miville-Deschênes, M. -A. and Molinari, D. and Montier, L. and Morgante, G. and Moss, A. and Natoli, P. and Nørgaard-Nielsen, H. U. and Pagano, L. and Paoletti, D. and Partridge, B. and Patanchon, G. and Peiris, H. V. and Perrotta, F. and Pettorino, V. and Piacentini, F. and Polastri, L. and Polenta, G. and Puget, J. -L. and Rachen, J. P. and Reinecke, M. and Remazeilles, M. and Renzi, A. and Rocha, G. and Rosset, C. and Roudier, G. and Rubiño-Martín, J. A. and Ruiz-Granados, B. and Salvati, L. and Sandri, M. and Savelainen, M. and Scott, D. and Shellard, E. P. S. and Sirignano, C. and Sirri, G. and Spencer, L. D. and Sunyaev, R. and Suur-Uski, A. -S. and Tauber, J. A. and Tavagnacco, D. and Tenti, M. and Toffolatti, L. and Tomasi, M. and Trombetti, T. and Valenziano, L. and Valiviita, J. and Van Tent, B. and Vibert, L. and Vielva, P. and Villa, F. and Vittorio, N. and Wandelt, B. D. and Wehus, I. K. and White, M. and White, S. D. M. and Zacchei, A. and Zonca, A.},
	month = sep,
	year = {2020},
	note = {\_eprint: 1807.06209},
	pages = {A6},
}

@article{pietrinferniUpdatedBaSTIStellar2024,
	title = {The updated {BaSTI} stellar evolution models and isochrones - {IV}. α-{Depleted} calculations},
	volume = {527},
	issn = {0035-8711},
	url = {https://ui.adsabs.harvard.edu/abs/2024MNRAS.527.2065P},
	doi = {10.1093/mnras/stad3267},
	urldate = {2024-10-09},
	journal = {\mnras},
	author = {Pietrinferni, Adriano and Salaris, Maurizio and Cassisi, Santi and Savino, Alessandro and Mucciarelli, Alessio and Hyder, David and Hidalgo, Sebastian},
	month = jan,
	year = {2024},
	note = {Publisher: OUP
ADS Bibcode: 2024MNRAS.527.2065P},
	pages = {2065--2070},
}

@misc{honFlowBasedGenerativeEmulation2024,
	title = {Flow-{Based} {Generative} {Emulation} of {Grids} of {Stellar} {Evolutionary} {Models}},
	url = {https://ui.adsabs.harvard.edu/abs/2024arXiv240709427H},
	doi = {10.48550/arXiv.2407.09427},
	urldate = {2024-10-09},
	author = {Hon, Marc and Li, Yaguang and Ong, Joel},
	month = jul,
	year = {2024},
	note = {Publication Title: arXiv e-prints
ADS Bibcode: 2024arXiv240709427H},
}

@book{eddingtonInternalConstitutionStars1926,
	title = {The {Internal} {Constitution} of the {Stars}},
	url = {https://ui.adsabs.harvard.edu/abs/1926ics..book.....E},
	urldate = {2025-03-07},
	author = {Eddington, A. S.},
	month = jan,
	year = {1926},
	note = {Publication Title: The Internal Constitution of the Stars
ADS Bibcode: 1926ics..book.....E},
}

@article{thoulElementDiffusionSolar1994,
	title = {Element {Diffusion} in the {Solar} {Interior}},
	volume = {421},
	issn = {0004-637X},
	url = {https://ui.adsabs.harvard.edu/abs/1994ApJ...421..828T},
	doi = {10.1086/173695},
	urldate = {2022-03-23},
	journal = {\apj},
	author = {Thoul, Anne A. and Bahcall, John N. and Loeb, Abraham},
	month = feb,
	year = {1994},
	pages = {828},
}

@article{canutoElectricalConductivityConductive1970,
	title = {Electrical {Conductivity} and {Conductive} {Opacity} of a {Relativistic} {Electron} {Gas}},
	volume = {159},
	issn = {0004-637X},
	url = {https://ui.adsabs.harvard.edu/abs/1970ApJ...159..641C},
	doi = {10.1086/150338},
	urldate = {2022-03-23},
	journal = {\apj},
	author = {Canuto, Vittorio},
	month = feb,
	year = {1970},
	pages = {641},
}

@article{fergusonLowTemperatureOpacities2005,
	title = {Low-{Temperature} {Opacities}},
	volume = {623},
	issn = {0004-637X},
	url = {https://ui.adsabs.harvard.edu/abs/2005ApJ...623..585F},
	doi = {10.1086/428642},
	urldate = {2022-03-23},
	journal = {\apj},
	author = {Ferguson, Jason W. and Alexander, David R. and Allard, France and Barman, Travis and Bodnarik, Julia G. and Hauschildt, Peter H. and Heffner-Wong, Amanda and Tamanai, Akemi},
	month = apr,
	year = {2005},
	pages = {585--596},
}

@article{choiMesaIsochronesStellar2016,
	title = {Mesa {Isochrones} and {Stellar} {Tracks} ({MIST}). {I}. {Solar}-scaled {Models}},
	volume = {823},
	doi = {10.3847/0004-637X/823/2/102},
	journal = {\apj},
	author = {Choi, J. and Dotter, A. and Conroy, C. and Cantiello, M. and Paxton, B. and Johnson, B. D.},
	month = jun,
	year = {2016},
	note = {\_eprint: 1604.08592},
	pages = {102},
}

@article{deboerMonteCarloUncertainty2014,
	title = {Monte {Carlo} uncertainty of the {\textasciicircum}3{\textbackslash}{mathrmHe}({\textbackslash}ensuremath{\textbackslash}alpha,{\textbackslash}ensuremath{\textbackslash}gamma){\textasciicircum}7{\textbackslash}{mathrmBe} reaction rate},
	volume = {90},
	url = {https://link.aps.org/doi/10.1103/PhysRevC.90.035804},
	doi = {10.1103/PhysRevC.90.035804},
	number = {3},
	journal = {\prc},
	author = {deBoer, R. J. and Görres, J. and Smith, K. and Uberseder, E. and Wiescher, M. and Kontos, A. and Imbriani, G. and Di Leva, A. and Strieder, F.},
	month = sep,
	year = {2014},
	note = {Publisher: American Physical Society},
	pages = {035804},
}

@article{xuNACREIIUpdate2013,
	title = {{NACRE} {II}: an update of the {NACRE} compilation of charged-particle-induced thermonuclear reaction rates for nuclei with mass number {A}{\textbackslash}textless16},
	volume = {918},
	issn = {0375-9474},
	shorttitle = {{NACRE} {II}},
	url = {https://ui.adsabs.harvard.edu/abs/2013NuPhA.918...61X},
	doi = {10.1016/j.nuclphysa.2013.09.007},
	urldate = {2022-03-23},
	journal = {Nuclear Physics A},
	author = {Xu, Y. and Takahashi, K. and Goriely, S. and Arnould, M. and Ohta, M. and Utsunomiya, H.},
	month = nov,
	year = {2013},
	pages = {61--169},
}

@article{joyceNotAllStars2018,
	title = {Not {All} {Stars} {Are} the {Sun}: {Empirical} {Calibration} of the {Mixing} {Length} for {Metal}-poor {Stars} {Using} {One}-dimensional {Stellar} {Evolution} {Models}},
	volume = {856},
	issn = {0004-637X},
	shorttitle = {Not {All} {Stars} {Are} the {Sun}},
	url = {https://dx.doi.org/10.3847/1538-4357/aab200},
	doi = {10.3847/1538-4357/aab200},
	language = {en},
	number = {1},
	urldate = {2023-10-12},
	journal = {\apj},
	author = {Joyce, M. and Chaboyer, B.},
	month = mar,
	year = {2018},
	note = {Publisher: The American Astronomical Society},
	pages = {10},
}

@article{marcucciProtonProtonWeakCapture2013,
	title = {Proton-{Proton} {Weak} {Capture} in {Chiral} {Effective} {Field} {Theory}},
	volume = {110},
	doi = {10.1103/PhysRevLett.110.192503},
	number = {19},
	journal = {\prl},
	author = {Marcucci, L. E. and Schiavilla, R. and Viviani, M.},
	month = may,
	year = {2013},
	note = {\_eprint: 1303.3124},
	pages = {192503},
}

@article{krishnaswamyProfilesStrongLines1966,
	title = {Profiles of {Strong} {Lines} in {K}-{Dwarfs}},
	volume = {145},
	issn = {0004-637X},
	url = {https://ui.adsabs.harvard.edu/abs/1966ApJ...145..174K},
	doi = {10.1086/148752},
	urldate = {2022-03-23},
	journal = {\apj},
	author = {Krishna Swamy, K. S.},
	month = jul,
	year = {1966},
	pages = {174},
}

@article{dotterMESAISOCHRONESStelLAR2016,
	title = {{MESA} {ISOCHRONES} {AND} {S}℡{LAR} {TRACKS} ({MIST}) 0: {METHODS} {FOR} {THE} {CONSTRUCTION} {OF} {S}℡{LAR} {ISOCHRONES}},
	volume = {222},
	issn = {0067-0049},
	shorttitle = {{MESA} {ISOCHRONES} {AND} {S}℡{LAR} {TRACKS} ({MIST}) 0},
	url = {https://dx.doi.org/10.3847/0067-0049/222/1/8},
	doi = {10.3847/0067-0049/222/1/8},
	language = {en},
	number = {1},
	urldate = {2023-10-31},
	journal = {\apjs},
	author = {Dotter, Aaron},
	month = jan,
	year = {2016},
	note = {Publisher: The American Astronomical Society},
	pages = {8},
}

@article{claretNewGridsStellar2004,
	title = {New grids of stellar models including tidal-evolution constants up to carbon burning: {I}. {From} 0.8 to 125 \textit{ \textbf{{M}}} \_{\textbackslash}textbackslashodot{\textbackslash} at \textit{ \textbf{{Z}}} = 0.02},
	volume = {424},
	issn = {0004-6361, 1432-0746},
	shorttitle = {New grids of stellar models including tidal-evolution constants up to carbon burning},
	url = {http://www.aanda.org/10.1051/0004-6361:20040470},
	doi = {10.1051/0004-6361:20040470},
	number = {3},
	urldate = {2023-03-16},
	journal = {\aap},
	author = {Claret, A.},
	month = sep,
	year = {2004},
	pages = {919--925},
}

@article{haftStandardNonstandardPlasma1994,
	title = {Standard and {Nonstandard} {Plasma} {Neutrino} {Emission} {Revisited}},
	volume = {425},
	issn = {0004-637X},
	url = {https://ui.adsabs.harvard.edu/abs/1994ApJ...425..222H},
	doi = {10.1086/173978},
	urldate = {2022-03-23},
	journal = {\apj},
	author = {Haft, Martin and Raffelt, Georg and Weiss, Achim},
	month = apr,
	year = {1994},
	pages = {222},
}

@article{adelbergerSolarFusionCross2011,
	title = {Solar fusion cross sections. {II}. {The} pp chain and {CNO} cycles},
	volume = {83},
	doi = {10.1103/RevModPhys.83.195},
	number = {1},
	journal = {Reviews of Modern Physics},
	author = {Adelberger, E. G. and García, A. and Robertson, R. G. Hamish and Snover, K. A. and Balantekin, A. B. and Heeger, K. and Ramsey-Musolf, M. J. and Bemmerer, D. and Junghans, A. and Bertulani, C. A. and Chen, J. -W. and Costantini, H. and Prati, P. and Couder, M. and Uberseder, E. and Wiescher, M. and Cyburt, R. and Davids, B. and Freedman, S. J. and Gai, M. and Gazit, D. and Gialanella, L. and Imbriani, G. and Greife, U. and Hass, M. and Haxton, W. C. and Itahashi, T. and Kubodera, K. and Langanke, K. and Leitner, D. and Leitner, M. and Vetter, P. and Winslow, L. and Marcucci, L. E. and Motobayashi, T. and Mukhamedzhanov, A. and Tribble, R. E. and Nollett, Kenneth M. and Nunes, F. M. and Park, T. -S. and Parker, P. D. and Schiavilla, R. and Simpson, E. C. and Spitaleri, C. and Strieder, F. and Trautvetter, H. -P. and Suemmerer, K. and Typel, S.},
	month = jan,
	year = {2011},
	note = {\_eprint: 1004.2318},
	pages = {195--246},
}

@article{mowlaviStellarMassAge2012,
	title = {Stellar mass and age determinations . {I}. {Grids} of stellar models from {Z} = 0.006 to 0.04 and {M} = 0.5 to 3.5 {M}\_{\textbackslash}ensuremathødot},
	volume = {541},
	doi = {10.1051/0004-6361/201117749},
	journal = {åp},
	author = {Mowlavi, N. and Eggenberger, P. and Meynet, G. and Ekström, S. and Georgy, C. and Maeder, A. and Charbonnel, C. and Eyer, L.},
	month = may,
	year = {2012},
	note = {\_eprint: 1201.3628},
	pages = {A41},
}

@article{chaboyerTestingMetalPoorStellar2017,
	title = {Testing {Metal}-{Poor} {Stellar} {Models} and {Isochrones} with {HST} {Parallaxes} of {Metal}-{Poor} {Stars}},
	volume = {835},
	doi = {10.3847/1538-4357/835/2/152},
	number = {2},
	journal = {\apj},
	author = {Chaboyer, B. and McArthur, B. E. and O'Malley, E. and Benedict, G. F. and Feiden, G. A. and Harrison, T. E. and McWilliam, A. and Nelan, E. P. and Patterson, R. J. and Sarajedini, A.},
	month = feb,
	year = {2017},
	note = {\_eprint: 1702.00803},
	pages = {152},
}

@article{chakrabortySystematicRmatrixAnalysis2015,
	title = {Systematic {R}-matrix analysis of the {C} 13 (p, γ) {N} 14 capture reaction},
	volume = {91},
	number = {4},
	journal = {\prc},
	author = {Chakraborty, Suprita and deBoer, Richard and Mukherjee, Avijit and Roy, Subinit},
	year = {2015},
	note = {Publisher: APS},
	pages = {045801},
}

@article{husserNewExtensiveLibrary2013,
	title = {A new extensive library of {PHOENIX} stellar atmospheres and synthetic spectra},
	volume = {553},
	doi = {10.1051/0004-6361/201219058},
	journal = {\aap},
	author = {Husser, T. -O. and Wende-von Berg, S. and Dreizler, S. and Homeier, D. and Reiners, A. and Barman, T. and Hauschildt, P. H.},
	month = may,
	year = {2013},
	note = {\_eprint: 1303.5632},
	pages = {A6},
}

@article{chaboyerLowerLimitAge1996,
	title = {A {Lower} {Limit} on the {Age} of the {Universe}},
	volume = {271},
	doi = {10.1126/science.271.5251.957},
	number = {5251},
	journal = {Science},
	author = {Chaboyer, Brian and Demarque, Pierre and Kernan, Peter J. and Krauss, Lawrence M.},
	month = feb,
	year = {1996},
	note = {\_eprint: astro-ph/9509115},
	pages = {957--961},
}

@article{yingAbsoluteAgeM922023,
	title = {The {Absolute} {Age} of {M92}},
	volume = {166},
	issn = {1538-3881},
	url = {https://dx.doi.org/10.3847/1538-3881/acd9b1},
	doi = {10.3847/1538-3881/acd9b1},
	language = {en},
	number = {1},
	urldate = {2023-07-24},
	journal = {\aj},
	author = {Ying, Jiaqi (Martin) and Chaboyer, Brian and Boudreaux, Emily M. and Slaughter, Catherine and Boylan-Kolchin, Michael and Weisz, Daniel},
	month = jun,
	year = {2023},
	note = {Publisher: The American Astronomical Society},
	pages = {18},
}

@article{yingAbsoluteAgeNGC2024,
	title = {The {Absolute} {Age} of {NGC} 3201 {Derived} from {Detached} {Eclipsing} {Binaries} and the {Hess} {Diagram}},
	volume = {970},
	issn = {0004-637X},
	url = {https://ui.adsabs.harvard.edu/abs/2024ApJ...970..184Y},
	doi = {10.3847/1538-4357/ad59a9},
	urldate = {2024-10-06},
	journal = {\apj},
	author = {Ying, Jiaqi (Martin) and Chaboyer, Brian and Du, Wenxin},
	month = aug,
	year = {2024},
	note = {Publisher: IOP
ADS Bibcode: 2024ApJ...970..184Y},
	pages = {184},
}
\bibliographystyle{aasjournal}
\end{document}